\documentclass[
a4paper]{iopart}
\usepackage{graphicx}

\usepackage{iopams}

\def\aa{$\rm a{}\!\!\!^\circ$} 

\eqnobysec

\def\9#1{{\bf #1}}
\def\0{\over } 
\def\2{{\textstyle{1\over2}}} \def\4{{\textstyle{1\over4}}}
\def\5{\hat } \def\6{\partial }
\def\8#1{{\textstyle{#1}}}
 
\let\a=\alpha \let\b=\beta \let\g=\gamma \let\d=\delta
\let\e=\epsilon \let\z=\zeta  
  \let\l=\lambda 
 \let\s=\sigma
 \let\w=\omega \let\om=\omega 
\let\ph=\varphi   
\let\S=\Sigma 
  \let\D=\Delta 
\def\mn{{\mu\nu}}

\let\I=\rmi
\let\E=\rme
\def\CL{\mathcal L}

\def\({\left(} \def\){\right)} \def\<{\langle } \def\>{\rangle }
\def\[{\left[} \def\]{\right]}  

\newcommand{\bea}{\begin{eqnarray}}
\newcommand{\eea}{\end{eqnarray}}
\newcommand{\be}{\begin{equation}}
\newcommand{\ee}{\end{equation}}
\newcommand{\nn}{\nonumber\\ }

\def\Im{{\,\rm Im\,}}
\def\Re{{\,\rm Re\,}}

\begin{document}
\begin{flushright}
TUW-03-32
\end{flushright}
\jl{19} 
\title[Thermal field theory]{%
Advances in perturbative thermal field theory
\\{{\footnotesize\em Review article 
submitted to REPORTS ON PROGRESS IN PHYSICS}}}
\author{U Kraemmer and A Rebhan}
\address{Institut f\"ur Theoretische Physik,
         Technische Universit\"at Wien,\\
         Wiedner Hauptstra\ss e 8-10/136,
         A-1040 Vienna, Austria}
\begin{abstract}
The progress of the last decade
in perturbative quantum field theory at high temperature and density
made possible by the use of effective field theories and
hard-thermal/dense-loop resummations in ultrarelativistic gauge
theories is reviewed.
The relevant methods are
discussed
in field theoretical models from simple scalar theories to 
non-Abelian gauge
theories including gravity. In the simpler models, the aim
is to give a pedagogical account of some of the relevant problems
and their resolution, while in the more complicated but also more
interesting models such as quantum chromodynamics, 
a summary of the results obtained
so far are given together with references to a few most recent
developments and open problems.
\end{abstract}
\tableofcontents
\title[Thermal field theory]{}

\maketitle

\section{Introduction}

Ultrarelativistically hot and dense matter needs to be understood
in many problems ranging from early-universe cosmology 
\cite{Rubakov:1996vz,Schwarz:2003du}
as well as the astrophysics
of compact stars \cite{Gle:CS}
to the current and prospected experimental programs at
relativistic heavy-ion colliders in the USA (RHIC at Brookhaven) and
in Europe (LHC at CERN), which seek to
produce and investigate the new state of matter called the quark-gluon plasma
\cite{HwaW:QGP3}.
The theoretical framework for analysing equilibrium and
near-equilibrium properties is thermal field theory,
which at a fundamental level involves
Abelian and non-Abelian gauge theories.

In this review we shall focus on the technical difficulties
to derive results by analytical, weak-coupling techniques.
Because of asymptotic freedom of non-Abelian gauge theories,
this is a valid approach at least at sufficiently high temperatures
and densities. Much of the present activities is concerned
with quantum chromodynamics (QCD), which is however strongly
interacting at all temperatures of experimental interest.
The perturbative calculations that have been carried out
to high loop orders for thermodynamic potentials seem
to signal that perturbative QCD at finite temperature
is loosing any predictive power below ridiculously high
temperatures $\sim 10^5$ GeV, where the coupling is so small
that everything is described well enough by an ideal gas of quarks and
gluons without much need of perturbative refinements.
However, recent studies indicate that this conclusion is much too pessimistic,
and it appears that even
in strongly coupled QCD it is indeed possible (at least to some extent)
to employ weak-coupling methods if proper use is made of
effective field theories and the fact that the nontrivial
spectral data of quasiparticles can absorb much of the strong interactions
\cite{Blaizot:2003tw}. 

In the case of static quantities at high temperature and small chemical
potential the relevant techniques are effective field theories
which take advantage of the phenomenon of dimensional reduction;
for dynamical quantities the required techniques involve
the effective theory produced by hard thermal loops
\cite{Blaizot:2001nr}.
At high chemical potential and low temperature, the
resummation of hard dense loops becomes important even
in static quantities and
also for describing the astrophysically interesting
case of colour superconductivity \cite{
Rischke:2003mt}.

The present review may to some extent be viewed as an extension
of the material presented in the textbooks of
Kapusta \cite{Kap:FTFT}, Das \cite{Das:FTFT} 
and LeBellac \cite{LeB:TFT}, with a certain
amount of overlap with the latter which already covers
the most important aspects of hard thermal loop resummations.
Of course, numerous interesting new developments could only
be touched upon, but hopefully enough references are provided
as pointers to the recent literature;
a lot of topics had to be omitted by lack of
space or competence.

\subsection{Outline}

In section \ref{secbasics}, after a brief recapitulation of the
fundamental formulae of quantum statistical mechanics, the
imaginary-time and real-time formalisms in thermal field theory
are discussed with an overview of some
more recently developed alternative approaches
in the real-time formalism and in the treatment of gauge theories.

In section \ref{sec3} we then exhibit some of the particular
issues that arise in
finite-temperature field theory in a simple scalar model: the
appearance of thermal masses, the phenomenon of
dimensional reduction at high temperatures, the need for a resummation
of naive perturbation theory, and the problem
of a deterioration of apparent convergence
after resummation and possibilities for its improvement.
After a short discussion of the phenomenon of restoration of
spontaneously broken symmetries at high temperatures at the
end of section \ref{sec3},
section \ref{secQCDtd} describes the progress made
during the last decade in the (resummed) perturbative evaluation
of the thermodynamic potential of unbroken nonabelian gauge theory.
At high temperature and not too high chemical
potential, dimensional reduction provides the most
efficient means for calculating perturbatively the thermodynamic potential, and
this calculation has been carried to its limits, which are
given by the inherent nonperturbative nature of nonabelian magnetostatic
fields. It is argued that initial pessimism regarding the
utility of weak-coupling methods in hot QCD is unnecessary, and
that the results rather indicate that perturbative methods,
when properly improved, may work already at temperatures that
are only a few times higher than the deconfinement transition temperature.
The recent progress made for the case of
a nonvanishing fermion chemical potential is also reviewed,
which includes quark number susceptibilities and non-Fermi-liquid
contributions to the low-temperature entropy and specific heat
of QCD and QED.

Section \ref{secqpgt} discusses the structure of the propagators
of a gauge theory with fermions, the leading-order results
for the respective quasiparticles and the issue of
gauge (in)dependence, both in unbroken nonabelian gauge theory
and in the presence of colour superconductivity.

Section \ref{secHTLres} then introduces the concept of
hard-thermal/dense-loop resummation which is typically necessary
(though not always sufficient) to calculate the effects
of collective phenomena such as dynamical screening and
propagating plasmons. 
Two recent approaches to improve the problem of the poor apparent convergence
observed in the thermodynamic potential are discussed, which
suggests that already the lowest order calculations in terms of
quasiparticles may capture the most
important contributions even in the strongly coupled
quark-gluon plasma at temperatures that are only a few times
higher than the deconfinement temperature.

The known results concerning next-to-leading order corrections to the
quasi-particle spectrum in gauge theories are then reviewed in
section \ref{sublsect}. In this section a few simpler cases
such as the leading infrared-sensitive contributions to the Debye mass
and to damping rates and dynamical screening lengths are shown 
in more detail. Other cases of special interest are
enhancements from collinear physics, which include non-Fermi-liquid
corrections to the fermion self-energy in the vicinity of the
Fermi surface and a modification of longitudinal plasmons
for nearly light-like momenta.

Section \ref{secbey} briefly discusses some of the progress
made recently in the case of ultrasoft scales, where
nonabelian gauge fields have nonperturbative dynamics, and in
the case of
observables that are sensitive to collinear physics, where also
infinitely many loops contribute even after hard-thermal-loop
resummation. 

Section \ref{secgrav} finally considers
thermal field theory in a curved background geometry and
in particular the hard thermal loop contribution to
the gravitational polarization tensor. The latter
encodes the physics of cosmological perturbations
in the presence of nearly collisionless ultrarelativistic matter,
for which analytical solutions can be obtained
in the physically relevant case of a conformally flat geometry.

\section{Basics}
\label{secbasics}

In a relativistic quantum theory where interactions typically
imply the destruction or creation of particles it is appropriate
to formulate a statistical description by means of the grand canonical
ensemble. 

A system in thermodynamical equilibrium for which only
mean values of energy and any conserved charges are prescribed
is characterized by a density matrix $\hat\rho$ such that
$[\hat\rho,\hat H]=0$ with $\hat H$ the Hamiltonian operator and
the requirement of maximal entropy
\be\label{Slnrho}
S = \<-\ln\hat\rho\> \equiv - \Tr \hat\rho \ln\hat\rho.
\ee
The temperature $T$ and the chemical potentials $\mu_i$ appear as Lagrange
multipliers $\b=T^{-1}$ and $\a_i=-\beta\mu_i$ determining the
mean energy $\<\hat H\>$ and mean charges $\<\hat N_i\>$, respectively,
with $[\hat H,\hat N_i]=0=[\hat N_i,\hat N_j]$.
This singles out
\be\label{rhoeq}
\hat\rho = Z^{-1} \exp \{ -\beta \hat H - \sum_i \a_i \hat N_i \},
\ee
where the normalization factor $Z$ is the grand canonical partition
function
\be\label{partfct}
Z(V,\beta,\mu_i)=\Tr \exp \{ -\beta \hat H - \sum_i \a_i \hat N_i \}.
\ee

The partition function,
$Z$, determines all of the other conventional thermodynamic 
(or rather thermo-static)
quantities such as
pressure, entropy, energy, and charge
densities, which are denoted by
$P$, $\mathcal S$, $\mathcal E$, and $\mathcal N$,
respectively. In the thermodynamic infinite-volume 
limit ($V\to\infty$) these are given by
\bea
P &=& T{\6\ln Z\0\6V}={T\0V}\ln Z, \label{P} \\
{\cal S} &\equiv& S/V = {\6 P \0\6 T}, \label{SdPdT}\\
{\cal E} &\equiv& \<\hat H\>/V = -{1\0V} {\6 \ln Z\0\6\beta}, \\
{\cal N}_i &\equiv& \<\hat N_i\>/V = {\6 P\0\6\mu_i}.
\eea

Combining \eref{Slnrho} and \eref{rhoeq} one obtains
the Gibbs-Duhem relation in the form
\be\fl
{\cal S}=\<-\ln\hat\rho\>={1\over V}\ln Z+\beta{\cal E}+\sum_i \alpha_i
{\cal N}_i=\beta(P+{\cal E}-\sum_i \mu_i{\cal N}_i)
\ee
or
\be
E=-PV+TS+\mu_i N_i,
\ee
which explains why $P$ was introduced as the (thermodynamic) pressure.
A priori, the hydrodynamic pressure,
which is defined through the spatial components of the
energy-momentum tensor through ${1\over 3}\<{T^{ii}\>}$,
is a separate object. In equilibrium, it can be identified with the
thermodynamic one through scaling arguments \cite{Landsman:1987uw},
which however do not allow for the possibility of scale (or ``trace'')
anomalies that occur in all quantum field theories with
non-zero $\beta$-function (such as QCD). In \cite{Drummond:1999si}
it has been shown recently that the very presence of the 
trace anomaly
can be used to prove the equivalence of the two in equilibrium.

All these formulae pertain to the rest frame of the heat bath.
A manifestly covariant formulation can be obtained by explicitly
introducing the 4-velocity vector $u^\mu$ of this rest frame.
In thermal equilibrium the energy momentum tensor takes the
form
\be
T^\mn={\cal E}u^\mu u^\nu + P (u^\mu u^\nu - \eta^\mn),
\ee
where $\eta^\mn={\rm diag}(+,-,-,-)$ is the Minkowski metric.
Introducing the 4-vectors $\beta^\mu = \beta u^\mu$ and $j^\mu_i = n_i
u_\mu$, the grand canonical partition function can be written 
covariantly as \cite{
Israel:1981}
\be
Z=\Tr \exp \int d\Sigma_\mu \{-\beta_\nu \hat T^\mn-\sum_i \a_i \hat j^\mu_i
\}
\ee
with reference to a hypersurface $\Sigma$ with normal vector $u$.

It is of course more convenient to stick to the rest frame of the
heat bath. Lorentz invariance appears to be broken then, but it is
usually a trivial matter to switch to Lorentz covariant
expressions using $u^\mu$.

Information about the dynamics of a thermal system can be obtained
by considering thermal expectation values of the generally time-dependent
observables (in the Heisenberg picture). In linear response theory,
the time evolution of small disturbances of an equilibrium system
is determined by the correlation functions of pairs of observables
\cite{FetW:Q}.

For both the purpose of calculating the
partition function $Z$ and correlation functions there exist
two equivalent but rather differently looking formalisms to set up
perturbation theory. These correspond to the two most popular
choices of a complex time path in the path integral formula
\be\label{greenfcts}
 \langle \mathrm T_c \hat \ph_1 \cdots \hat \ph_n \rangle
= {\cal N} \int {\cal D}\ph \, \ph_1 \cdots \ph_n 
\exp{\I\int\limits_{\cal C}\! dt\! \int\! d^3x \, {\cal L}}\,,
\ee
where $\mathrm T_c$ denotes {contour ordering} along 
the complex time path $ {\cal C}$ from $ t_0$ to $ t_0-\I\beta$
such that $t_i \in {\cal C}$, and $t_1 \succeq t_2 \succeq \cdots \succeq t_n$
with respect to a monotonically increasing contour parameter.
$\cal L$ is the Lagrangian, which in the presence of a
chemical potential $\mu\not=0$ may be replaced by
${\cal L} \to \bar{\cal L}= {\cal L}+\mu {\cal N}$, provided
$\cal N$ does not contain time derivatives.
Analyticity requires that along the complex time path
the imaginary part of $t$ is monotonically decreasing \cite{Landsman:1987uw};
in the limiting case of a constant imaginary part of $t$ along
(parts of) the contour, distributional quantities (generalized functions)
arise.

Perturbation theory is set up in the usual fashion.
Using the interaction-picture
representation one can derive
\be
\langle  \mathrm T_c \mathcal O_1 \cdots \mathcal O_n \rangle  = {Z_0\over Z}
\langle  \mathrm T_c \mathcal O_1 \cdots \mathcal O_n\, \E^{
\I\!\int_{\mathcal C} \mathcal L_I} \rangle _0,
\ee
where $\mathcal L_I$ is the interaction part of $\cal L$,
and the correlators on the right-hand-side can be evaluated
by a  Wick(-Bloch-DeDominicis) theorem:
\index{Wick theorem} 
\be
\langle \mathrm T_c \E^{\I\!\int_{\mathcal C}d^4x\, j\ph}\rangle _0=
\exp\{-{1\over2}\int_{\mathcal C}\int_{\mathcal C}d^4x\,d^4x'
j(x)D^c(x-x')j(x')\},
\ee
where $D^c$ is the 2-point function and this is the only
building block of Feynman graphs with an explicit $T$ and $\mu$
dependence.
It satisfies the {KMS} (Kubo-Martin-Schwinger)
condition
\be\label{DKMS}
D^c(t-\I{\beta})=\pm \E^{-{\mu}{\beta}}D^c(t),
\ee
stating that $\E^{\I\mu t}D(t)$ is periodic (anti-periodic) for
bosons (fermions).

A KMS condition can be formulated for all correlation functions
in thermal equilibrium, and can in turn be viewed as a general
criterion for equilibrium \cite{Haag:LQP}. There exists also
a relativistic version of the KMS condition \cite{Bros:1994ua,Bros:1996mw},
which encodes the stronger analyticity requirements of
relativistic quantum field theories. The relativistic spectrum
condition $H\ge|\9P|$, where $\9P$ is the total
three-momentum, implies analyticity involving all
space-time variables $x\to z\in \mathbb C^4$ in tube domains
$|\Im\9z|<\Im z_0<\beta-|\Im\9z|$, whereas the usual KMS condition
corresponds to restricting this to $\Im\9z=0$.

\subsection{Imaginary-time (Matsubara) formalism}

The simplest possibility for choosing the complex {time path} 
in \eref{greenfcts} is
the straight line from $t_0$ to $t_0-\I\beta$, which is named
after Matsubara \cite{Matsubara:1955ws} who first formulated
perturbation theory based on this contour. It is also referred
to as {imaginary-time formalism} (ITF), because for $t_0=0$
one is exclusively dealing with imaginary times. 

Because of the (quasi-)periodicity (\ref{DKMS}), the propagator
is given by a Fourier series
\be
D^c(t) = {1\over -\I{\beta}}\sum_\nu \tilde D(z_\nu) \E^{-\I z_\nu t},
\quad \tilde D(z_\nu) = \int_0^{-\I {\beta}}\!\!\!dt\, D^c(t) \E^{\I z_\nu t}
\ee
with discrete complex (Matsubara) frequencies \index{Matsubara frequencies}
\be
z_\nu=2\pi \I \nu/{\beta}+{\mu}, \quad
\nu\in \left\{ \begin{array}{ll}\mathbb Z & \mbox {bos.} \\
                              \mathbb Z -{1\over2} &  \mbox {ferm.} \\
               \end{array} \right. 
\ee

Since $\tilde D(z_\nu)$ is defined only for a discrete set of complex
number, the analytic continuation to arbitrary frequencies is unique
only when one requires that $|\tilde D(z)|\to 0$ for $|z|\to \infty$
and that $\tilde D(z)$ is analytic off the real axis \cite{Baym:1961}. 
Then the
analytic continuation is provided by the spectral representation
\be
\tilde D(z)=\int_{-\infty}^\infty {dk_0\02\pi}{\rho(k_0)\0k_0-z}.
\ee

The transition to Fourier space turns the integrands of Feynman
diagrams from convolutions to products as usually, with the
difference that there is no longer an integral but a discrete
sum over the frequencies, and compared to standard momentum-space
Feynman rules one has
\be\fl
\int {d^4k\over \I(2\pi)^4} \to 
\b^{-1}\sum_\nu \int {d^3k\over (2\pi)^3},\qquad
\I(2\pi)^4 \d^4(k) \to\b (2\pi)^3 \d_{\nu,0} \d^3(k).
\ee

However, all Green functions that one can calculate in this
formalism are initially defined only for times on $\cal C$, so
all time arguments have the same real part. The analytic continuation
to several different times on the real axis is, however, frequently a highly
involved task \cite{Landsman:1987uw}, so that it can be
advantageous to use a formalism that supports real time arguments from
the start.

\subsection{Real-time (Keldysh) formalism}

\begin{figure}
\unitlength6mm
\begin{picture}(12,4)(-5.5,0)
\put(0,3){\vector(1,0){13}}
\put(6.5,0){\vector(0,1){4}}
\put(12.5,2.3){Re $t$}
\put(6.8,3.9){Im $t$}
\put(-1,3.2){\small $-\infty \gets t_0$}
\put(11.8,3.2){\small $t_f \to + \infty$}
\put(1,0.2){\small $t_0-\I\beta$}
\put(5.5,3.2){${\cal C}_1$}
\put(5.5,2.2){${\cal C}_2$}
\put(0.3,1.6){${\cal C}_3$}
\thicklines
\put(1,3){\vector(1,0){5.5}}
\put(1,3){\line(1,0){11}}
\put(12,3){\line(0,-1){0.2}}
\put(1,2.8){\line(1,0){11}}
\put(12,2.8){\vector(-1,0){5.5}}
\put(1,2.8){\vector(0,-1){2.2}}

\end{picture}
\caption{Complex {time path} in the Schwinger-Keldysh real-time
formalism \label{FigRTF}}
\end{figure}
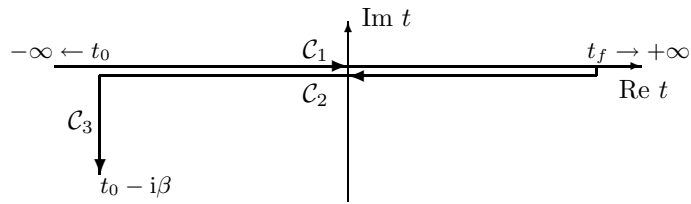

In the so-called real-time formalism(s), the complex {time path} $\cal C$ is
chosen such as to include the real-time axis from an initial time $t_0$
to a final time $t_f$. This
requires further parts of the contour to run backward in
real time \cite{Schwinger:1961qe,Bakshi:1963}
and to end up at $t_0-\I\beta$.
There are a couple of
paths $\cal C$ that have been proposed in the literature. The
oldest one due to Keldysh \cite{Keldysh:1964ud} is shown in figure~\ref{FigRTF},
where the first part of the contour ${\cal C}_1$ is on the real
axis, and a second part runs from $t_f-i\delta$ to $t_0-i\delta$
with $\delta\to 0$.
For some time, the more symmetric choice where the backward-running
contour is placed such as to have imaginary part $-\delta=-\beta/2$
with two vertical contour pieces at $t_0$ and $t_f$
of equal length has enjoyed some popularity 
\cite{Niemi:1984ea,Niemi:1984nf,Landsman:1987uw}
in particular in the axiomatic thermo-field-dynamics (TFD) operator
formalism \cite{Takahasi:1975zn,UmeM:TF,Matsumoto:1983ry,Matsumoto:1984au}
but by now the original Keldysh contour seems to be the one
most widely employed, above all because of its close relationship
to  the so-called closed time-path formalism 
\cite{Keldysh:1964ud,Chou:1985es,Calzetta:1988cq} of nonequilibrium thermodynamics.

With the Keldysh contour, 
if none of the field operators in (\ref{greenfcts}) has
time argument on $\mathcal C_1$ or $\mathcal C_2$, the contributions
from these parts of the contour simply cancel and one is back to
the ITF.
On the other hand, 
if all operators have finite real time arguments,
the contribution from contour $\mathcal C_3$ decouples.
The standard 
argument for this assertion
\cite{Niemi:1984ea,Niemi:1984nf,Landsman:1987uw} relies on
the limit $t_0\to-\infty$ and the fact that
the propagator connecting contour $\mathcal C_1$ and
$\mathcal C_3$ decays by virtue of the Riemann-Lebesgue theorem
\be\fl
D^{13}(t-(t_0-\I \lambda),k)= 
\int\limits_{-\infty}^\infty d\omega\,  \E^{-\I \omega(t-t_0)}
{
\E^{\l\omega}
\over \E^{\b \omega}-1}
\rho(\omega,k) \stackrel{t_0\to-\infty}{\longrightarrow} 0 
\ee
for $\lambda\in(0,\beta)$.
However, as argued by Ni\'egawa\cite{Niegawa:1989dr} there are cases where
this line of reasoning breaks down. The decoupling
of the vertical part(s) of the contour in RTF does however take place
provided the statistical distribution function in
the free RTF propagator defined below in (\ref{DcRTF1}) does
have as its argument $|k_0|$ and not the seemingly equivalent
$\omega_k=(\mathbf k^2+m^2)^{1/2}$. Reference~\cite{Niegawa:1989dr} also
relied on the limit $t_0\to-\infty$, but Gelis \cite{Gelis:1994dp,Gelis:1999nx}
pointed out that the decoupling of $\mathcal C_3$
should take place regardless of the magnitude of $t_0$,
and showed this indeed to be the case provided the ``$n(|k_0|)$ prescription''
is used (see also \cite{LeBellac:1996at,Mabilat:1997rx}).

With only $\mathcal C_1$ and $\mathcal C_2$ contributing, the
action in the path integral decomposes according to
\be\label{C12split}
\int\limits_{\mathcal C_1 \cup\, \mathcal C_2} \mathcal L(\ph)=
\int_{-\infty}^\infty dt\, \mathcal L(\ph^{(1)})-
\int_{-\infty}^\infty dt\, \mathcal L(\ph^{(2)})
\ee
where one has to distinguish between fields of type 1 (those
from contour $\mathcal C_1$) and of type 2 (those
from contour $\mathcal C_2$) because of the prescription
of {contour ordering} in (\ref{greenfcts}).\footnote{Type-2
fields are sometimes called ``thermal ghosts'', which misleadingly
suggests that type-1 fields are physical and type-2 fields
unphysical. In fact, they differ only with respect to the
time-ordering prescriptions they give rise to.} From 
(\ref{C12split}) it follows that
type-1 fields have vertices only among themselves,
and the same holds true for the type-2 fields. However,
the two types of fields are coupled through the propagator,
which is a $2\times2$ matrix with non-vanishing off-diagonal
elements:
\be
{\bf D}^c(t,{ t'})=
\pmatrix{ \<{{\rm T} \ph(t){\ph(t')}}\> & {\s} 
\<{ {\ph(t')}\ph(t)}\> \cr
\<{ \ph(t){\ph(t')}}\> & \<{{\tilde{\rm T}} \ph(t){\ph(t')}}\> }.
\ee
Here $\tilde{\rm T}$ denotes anti-time-ordering for the
2-2 propagator and $\sigma$ is a sign which is positive
for bosons and negative for fermions. 
The off-diagonal elements do not need a time-ordering
symbol because type-2 is by definition always later (on the contour) than
type-1.

In particular, for a massive scalar field one obtains
\bea\label{DcRTF1}\fl
{\bf D}^c(k)&=\pmatrix{ {\I\over k^2-m^2+\I \e} & 2\pi\delta^-(k^2-m^2) \cr
             2\pi\delta^+(k^2-m^2) & {-\I \over k^2-m^2-\I \e} } +
2\pi{\delta(k^2-m^2)}n(|k_0|) \pmatrix{1&1\cr1&1},\nonumber\\ \fl
&\equiv  {\bf D}^c_0(k) + {\bf D}^c_T(k)
\eea
where $n(\omega)=[\E^{\beta \omega}-1]^{-1}$ and
$\delta^\pm(k^2-m^2)=\theta(\pm k_0)\delta(k^2-m^2)$.
The specifically thermal contribution ${\bf D}^c_T$ is 
a homogeneous Green function, as it is proportional to $\delta(k^2-m^2)$. 
Physically,
this part corresponds to Bose-Einstein-distributed, 
real particles on mass-shell.

The matrix propagator (\ref{DcRTF1}) can also be written
in a diagonalized form \cite{Matsumoto:1983gk,Matsumoto:1984au,Henning:1994aa}
\be\label{DcdiagF}
{\bf D}^c(k)=\mathbf M(k_0) { \pmatrix{ {\I G_F} & 0 \cr
             0 & {-\I G_F^*} } } \mathbf M(k_0)
\ee
with $G_F\equiv 1/(k^2-m^2+\I \e)$ and
\be\label{Mk0}
\mathbf M(k_0)= {1\over \sqrt{\E^{\beta|k_0|}-1}}
\pmatrix{\E^{{1\over2}\beta|k_0|} & \E^{(\delta-{1\over2}\beta) k_0} \cr
         \E^{({1\over2}\beta-\delta) k_0} & \E^{{1\over2}\beta|k_0|}}
\ee
where $\delta\to0$ for the Keldysh contour.

For complex time paths with $\delta\propto\beta$, the
$T\to0$ limit decouples type-1 and type-2 fields completely.
For $\delta\to0$, the limit $T\to0$ leads to
\be\label{Mk00}{\mathbf M(k_0)}
\stackrel{ \beta\to\infty}{\longrightarrow}\mathbf M_0(k_0)
=\pmatrix{1&\theta(-k_0)\cr
          \theta(k_0)&1}
\ee
so that one still has propagators connecting fields of
different type. However,
if all the external lines of a diagram are of the same type,
then also all the internal lines are,
because
$\prod_i \theta(k_{(i)}^0)=0$
when $\sum_i k_{(i)}^0 = 0$ and any connected region of
the other field-type leads to a factor of zero.

The matrix structure \eref{DcdiagF} also applies to the
full propagator, and consequently the self energy
$\I\mathbf\Pi=\mathbf D^{-1}-\mathbf D^{-1}_0$ has the
analogous form with $\mathbf M^{-1}$ in place of $\mathbf M$,
\be\label{PidiagF}
{\mathbf \Pi}(k)=\mathbf M(k_0) { \pmatrix{ {\Pi_F} & 0 \cr
             0 & {-\Pi_F^*} } } \mathbf M(k_0).
\ee

It is also possible to diagonalize in terms of retarded and
advanced quantities according to 
\be\label{DcdiagRA}
{\bf D}^c(k)=\mathbf U(k_0) { \pmatrix{ {\I G_R} & 0 \cr
             0 & {\I G_A^*} } } \mathbf V(k_0)
\ee
with
\be\fl
\mathbf U(k_0)=\pmatrix{1 & -n(k_0) \cr 1 & -(1+n(k_0)) \cr},\quad
\mathbf V(k_0)=\pmatrix{1+n(k_0) & n(k_0) \cr 1 & 1 \cr}
\ee
and to include the matrices $\mathbf U$ and $\mathbf V$ in the
vertices
\cite{Aurenche:1992hi,Guerin:1994ik}. 
This has the advantage of leading to $n$-point
Green functions with well-defined causal properties, which
correspond directly to the various analytic continuations of
ITF Green functions.
On the other hand, the type-1/type-2 basis
leads to rather involved relations 
\cite{Kobes:1990kr,Kobes:1991ua,Evans:1992ky}.

Because of
\be
\mathbf U(k) \tau_1 = \mathbf V^T(-k),\quad \tau_1=\pmatrix{0&1\cr1&0\cr},
\ee
more symmetric retarded/advanced Feynman rules can be formulated
by including a factor $\tau_1$ in $\mathbf U$ and putting
$G_R$ and $G_A$ in the off-diagonal entries of \eref{DcdiagRA}
\cite{vanEijck:1992mq,vanEijck:1994rw}.

A precursor of this transformation is in fact given by
the so-called Keldysh basis \cite{Keldysh:1964ud,Chou:1985es,vanEijck:1994rw}
\be
\ph_+=\2(\ph_1+\ph_2),\quad \ph_-=(\ph_1-\ph_2)
\ee
(sometimes labelled by indices $r,a$ instead). This also has
the advantage of a rather direct relationship to retarded/advanced
$n$-point Green functions, and because the transformation
does not involve $n(k_0)$, it is of use also in
the nonequilibrium closed-time-path formalism
\cite{Wang:1998wg}.

Another economical method to derive retarded/advanced quantities
in the real time formalism is provided by the use of the
outer products of 2-component
column vectors \cite{Chu:1993nc,Henning:1993gh,Henning:1995sm} as
worked out in
\cite{Carrington:1998rx,Hou:1998yc}.

\subsection{Extension to gauge theories}

In the partition function \eref{partfct} and in thermal averages
$\<\hat{\cal Q}\>\equiv \Tr \hat\rho \hat{\cal Q}$, the trace is
taken over the physical Hilbert space. But covariant formulations
of gauge theories are built in larger spaces containing unphysical
states, while the definitions following
\eref{partfct} are true only in the physical
Hilbert subspace. The standard solution is to extend the trace to
the larger unphysical space and to cancel unphysical contributions
by Faddeev-Popov ghosts.

In the path integral formalism, the Faddeev-Popov ghost fields arise from
a functional determinant in the configuration space of the bosonic
gauge fields. This requires that although Faddeev-Popov fields
behave as fermions with respect to the diagrammatical combinatorics,
they are subject to the same boundary conditions as the gauge bosons
and therefore have the same statistical distribution functions,
namely Bose-Einstein ones \cite{Bernard:1974bq}.

In operator language which starts from a BRS invariant theory involving
fermionic Faddeev-Popov field operators \cite{Kugo:1978zq}, 
this prescription can be
understood through the observation \cite{Hata:1980yr} that the
operator $\exp{i\pi \hat N_c}$, where $\hat N_c$ is the ghost-number
operator, is equivalent, under the trace, to a projection operator
onto the physical Hilbert space. This means that the fermionic Faddeev-Popov
fields are given an imaginary chemical potential $\mu_c=i\pi/\beta$.
But a Fermi-Dirac distribution with such a chemical potential is
nothing other than a Bose-Einstein distribution.

There is however an alternative approach, developed in 
\cite{Landshoff:1992ne,Landshoff:1993ag},
which avoids assigning thermal distributions to unphysical degrees of
freedom altogether. 
In the real-time formalism, one may switch off the gauge coupling
adiabatically as the beginning of the time contour is moved to $-\infty$.
Then the condition to choose physical states can be the same as in
Abelian gauge theory. The unphysical states 
are identified as those which are due to the
Faddeev-Popov fields and the temporal and longitudinal polarizations of the
gauge fields. Because the free Hamiltonian is a sum of commuting parts
containing respectively only physical and unphysical operators, and
because the unphysical part has zero eigenvalue on the physical states,
all unphysical contributions factor out such that only the transverse
polarizations of the gauge fields acquire thermal parts in their
propagators.

In Feynman gauge for instance the gauge propagator, which
usually is simply ${\mathbf D}^\mn=-\eta^\mn {\mathbf D}^c$, 
with ${\mathbf D}^c$ the matrix propagator \eref{DcRTF1}, becomes
\be\label{fgpropLR}
{\mathbf D}^\mn=-A^\mn {\mathbf D}^c - (\eta^\mn-A^\mn)  {\mathbf D}_0^c
\ee
with
\be\label{Amn}
A^{0\mu}=0,\quad A^{ij}=-(\d^{ij}-{k^i k^j \0 {\bf k}^2 })
\ee
and the Faddeev-Popov ghost propagator remains non-thermal, ${\mathbf D}_{\rm
FP} = {\mathbf D}_0^c $.

In a general linear gauge with a quadratic gauge breaking term in
momentum space according to
\be\label{gbrL}
\tilde {\cal L}_{\rm g.br.}=-{1\02\a}\tilde A^\mu(-k) 
\tilde f_\mu \tilde f_\nu \tilde A^\nu(k)
\ee
the vacuum piece generalizes
by replacing 
\be
\eta^\mn\to \eta^\mn-{k^\mu \tilde f^\nu + \tilde f^\mu k^\nu \0 \tilde f
\cdot k} + (\tilde f^2 + \a k^2){k^\mu k^\nu \0 (\tilde f
\cdot k)^2}.
\ee
The ghost
propagator is replaced by $(\tilde f\cdot k)^{-1}$ with real-time
propagator matrix analogous to the vacuum part of the gauge propagator.

At finite temperature, where manifest Lorentz invariance has been lost
anyway, the modification \eref{fgpropLR} does not introduce additional
non-covariance. In fact, it simplifies calculations of thermal
contributions in general gauges \cite{Landshoff:1992ne}, but it makes it more
intricate to investigate resummation effects \cite{Landshoff:1993ag}.

In gauges where the ghost degrees of freedom are non-thermal anyway,
such as Coulomb gauge or axial gauges, the above Feynman rules
are identical to those of the conventional approach. In particular,
it reproduces the real-time
Feynman rules for temporal axial gauge of \cite{James:1990fd} 
which presents major difficulties in the imaginary-time formalism
\cite{James:1990it}. 


\section{Resummation issues in scalar $\phi^4$-theory}
\label{sec3}

Before discussing gauge theories further, we shall consider
perturbation theory at finite temperature 
in a scalar field theory with quartic coupling
and address the necessity
for resummations of the perturbative series.

\subsection{Daisy and foam resummation}

A particularly simple ``solvable'' model is given by the
the large $N$ limit of
a massless O($N$) scalar field theory with quartic interactions as given
by the Lagrangian \cite{Dolan:1974qd,Baym:1977qb,Bardeen:1983st,Zin:QFT,Drummond:1997cw}
\be\label{scLagr}
{\cal L}(x)=\2(\6_\mu\bphi)^2- 
{3\0N+2} g^2_0 (\bphi^2)^2
\ee
where there are $N$ scalar fields $\phi_i$ and
$\bphi^2=\phi_1^2+\ldots\phi_N^2$.

\begin{figure}
\centerline{\includegraphics[viewport=0 30 300 300,scale=0.25]{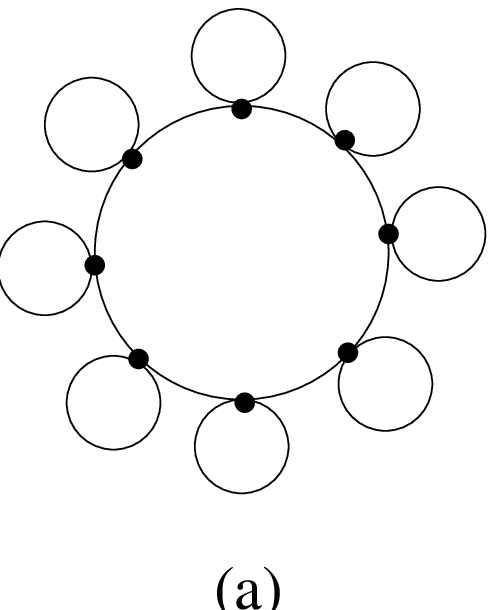}
\includegraphics[viewport=0 30 300 300,scale=0.25]{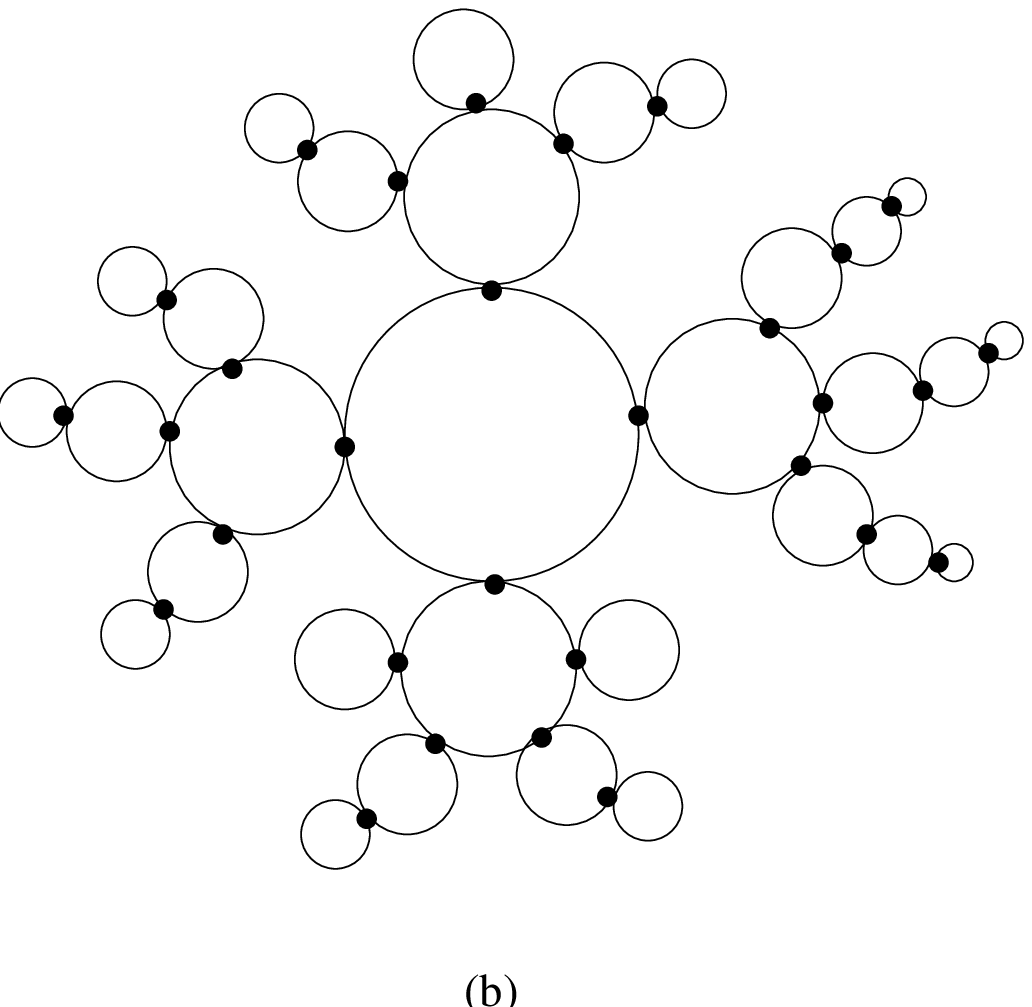}}
\caption{Ring and extended ring or ``foam'' diagrams}
\label{figfoam}
\end{figure}

In the limit of $N\to\infty$ the only Feynman diagrams that survive are
those that derive from ring (``daisy'') diagrams or
nested rings (``superdaisies'', ``cactus'', or
``foam'' diagrams) as shown in
figure~\ref{figfoam}.

In dimensional regularization, the zero-temperature scalar theory
is massless provided the bare mass is zero. The coupling however
receives an infinite renormalization in $n\to4$ dimensions. In the
large-$N$ limit this is determined in modified minimal subtraction
($\overline{\hbox{MS}}$) by
\be\label{scg2ren}
{1\0g^2}={1\0g_0^2} 
+{3\02\pi^2} {1\04-n} 
\,.
\ee

The sign of the counterterm in \eref{scg2ren} in fact hints at
the well-known problem of triviality of $\phi^4$ theories \cite{Zin:QFT}.
If the bare coupling $g_0^2$ is positive and $n$ approaches 4 from
below, the renormalized coupling $g$ goes to zero.
As discussed in reference~\cite{Drummond:1997cw}, 
if one insists on a non-trivial
theory with $g^2>0$ (which is only possible when $g_0^2$ is negative,
and divergent for $n\to4$), 
one finds that there is a tachyon (Landau pole) with
mass
\be\label{mtachyon}
m_{\rm tachyon}^2=-\bar\mu^2\exp\({8\pi^2\03g^2}+2\)
\ee
which appears to disqualify this model completely. 
Here $\bar\mu^2=4\pi e^{-\g}\mu^2$ and $\mu$ is a mass scale
introduced to make the coupling dimensionless in $n\not=4$
dimensions. 

However, at
small renormalized coupling, $g^2\ll1$, the tachyon's mass is exponentially
large. If everything is restricted to momentum scales smaller than
(\ref{mtachyon}), e.g. by a slightly smaller but still
exponentially large cutoff, the $n\!=\!4$ theory seems perfectly
acceptable. For our purposes we shall just have to restrict ourselves
to temperature scales smaller than (\ref{mtachyon})
when considering the finite-temperature effects in this scalar theory.

\subsection{Thermal masses}
\label{sec:scthm}

\begin{figure}[t]
\centerline{\includegraphics[scale=0.3]{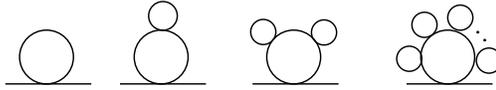}}
\caption{One-loop correction to the self-energy in scalar
$\phi^4$ theory, and some-higher-loop diagrams with increasing
degree of infrared singularity when the propagators are massless}
\label{figringPI}
\end{figure}

To one-loop order, the scalar self-energy diagram is the simple
tadpole shown in the first diagram of \fref{figringPI}, which
is quadratically divergent in cutoff regularization, but
strictly zero in dimensional regularization.
The Bose distribution function occurring at nonzero temperature
provides a cutoff at the scale of the temperature which gives
\be\label{msconeloop}\fl
\Pi=(m_{\rm th}^{(1)})^2=4!\,g^2\int{d^3q\0(2\pi)^3}\theta(q^0)
\d(q^0{}^2-\bi{q}{}^2)\{n(q^0)+\2\} 
=g^2T^2
\ee
so the initially massless scalar fields acquire a temperature-dependent
mass. As we shall see, in more complicated theories 
the thermal self-energy will generally
be a complicated function of frequencies and momenta, but the appearance
of a thermal mass scale $\sim gT$ is generic.

It should be noted, however, that
thermal masses are qualitatively different from ordinary
Lorentz-invariant mass terms. In particular they do not contribute
to the trace of the energy-momentum tensor as $m^2T^2$, as an ordinary
zero-temperature mass would do \cite{Nachbagauer:1996wn}.
So while the dispersion law of excitations is changed by the
thermal medium, the theory itself retains its massless nature.

At higher orders in perturbation theory, the thermal
contributions to the scalar self-energy
become nontrivial functions of frequency and momentum which
is complex-valued, implying a finite but parametrically
small width of thermal (quasi) particles
\cite{Parwani:1992gq,Wang:1996qf}, so that the latter
concept makes sense perturbatively. 

In the large-$N$ limit the self-energy of
the scalar field remains a momentum-independent
real mass term also beyond one-loop order
and is given by the Dyson equation
\be
\Pi=4!\,g_0^2\, \mu^{4-n} \int {d^nq\0(2\pi)^{n-1}} \{n(q^0)+\2\} 
\theta(q^0)\d(q^2-\Pi).
\ee
(Note that in (\ref{msconeloop}) we had simply replaced $g_0^2$ by $g^2$,
disregarding the difference as being of higher than one-loop order.)
The appearance of a thermal mass introduces quadratic ultraviolet
divergences in $\Pi$, which are however of exactly the form required
by the renormalization of the coupling according to (\ref{scg2ren}).
Including the latter, one finds a closed equation for the
thermal mass of the form
\bea
\fl m_{\rm th}^2=24g^2 \Bigl\{ \int {d^nq\0(2\pi)^{n-1}} n(q^0)
\theta(q^0)\d(q^2-m_{\rm th}^2)
+{1\032\pi^2}m_{\rm th}^2\(\ln{m_{\rm th}^2\0\bar\mu^2}-1\) \Bigr\} \nn
\fl \phantom{m_{\rm th}^2}=:
24g^2 \left\{ I_T(m_{\rm th}) + I_0^f(m_{\rm th},\bar\mu) \right\}.
\label{mth2sc}
\eea

The last term in the braces in \eref{mth2sc}, which has been neglected in
\cite{Dolan:1974qd,Altherr:1990rk,LeB:TFT}, is responsible for 
a non-trivial interplay between thermal
and vacuum contributions. Its explicit dependence on $\bar\mu$
is such that it cancels the one implicit in $g^2=g^2(\bar\mu)$,
\be
{d g\0d\ln(\bar\mu)}={3\04\pi^2}g^3,
\ee
which is exact for $N\!\to\!\infty$. 

More complicated effects of
zero-temperature renormalization on reorganized thermal perturbation
theories have been discussed in a scalar $\phi^3_6$ model
in \cite{Bodeker:1998an}. 
Another solvable toy model
is given by the large-$N_f$ limit of QCD or QED, which
has been worked out in 
a thermal field theory context
in \cite{Moore:2001fg,Moore:2002md,Ipp:2003zr,Ipp:2003jy}.
These theories also require the introduction of a cutoff to
avoid the Landau singularity and triviality. In contrast to
the O($N\to\infty$) $\phi^4$ theory, they involve complicated
momentum-dependent dispersion laws as well as damping effects.
While the large-$N$ $\phi^3_6$ theory has unphysical instabilities 
above a certain temperature (aside from the Landau singularity),
large-$N_f$ QED and QCD is well-behaved and very useful as
a benchmark for approximations to real QED and QCD.

\subsection{Perturbation series}\label{sectscpth}

The integral appearing in
\eref{mth2sc} can be evaluated e.\ g.\ using Mellin transformation
techniques \cite{Braden:1982we} to obtain a series expansion of $m_{\rm th}$
whose first few terms read \cite{Drummond:1997cw}
\be
\fl
{m_{\rm th}^2\0T^2} = g^2 - {3g^3\0\pi} + {3g^4\02\pi^2}
\(3-\g-\ln{\bar\mu\04\pi T}\)
+ {27g^5\08\pi^3} \(2\g+2\ln{\bar\mu\04\pi T}-1\)+\Or(g^6).
\label{msc}\ee

This result shows that ordinary perturbation theory is unable to
go beyond the one-loop result \eref{msconeloop}, for ordinary
perturbation theory is an expansion in powers of $g^2$. \Eref{msc}
however involves odd powers of $g$. Indeed, because of the
masslessness of the scalar theory, ordinary perturbation
theory encounters infrared divergences starting at two-loop
order, which are exacerbated by the Bose distribution function
behaving as $n(q^0) \sim T/q^0$ for $q^0 \to 0$. 
For example, the second diagram in \fref{figringPI} involves
two massless propagators with equal momentum. The inserted tadpole
does not vanish at zero momentum, but is given by the
constant term \eref{msconeloop}. At zero temperature, such
an insertion (if nonzero), would make this diagram logarithmically
infrared divergent; at finite temperature it is instead
linearly infrared divergent. The higher-loop diagrams shown in
\fref{figringPI} are even more infrared divergent.
But, as we have seen, the full propagator contains the thermal mass,
so all these divergences are spurious---they just
signal the need for using a resummed i.e.\ massive propagator.

A systematic method to perform the required resummation of ordinary
perturbation
theory is to add a thermal mass term $-1/2m^2_{\rm th}\bphi^2$
to the Lagrangian \eref{scLagr} and to subtract it as a counter-term
which is treated as a one-loop-order quantity.
The corresponding calculation at $N=1$, where \eref{mth2sc} is
only part of the full result, has been performed up to and
including order $g^5$ in \cite{Parwani:1992gq,Andersen:1998zx}.

\subsection{Dimensional reduction}
\label{subsecscdr}

An important technical concept for studying static quantities
such as thermodynamic potentials and (static) screening masses
is that of ``dimensional reduction''
\cite{Ginsparg:1980ef,Appelquist:1981vg,Nadkarni:1983kb,Nadkarni:1988fh,%
Landsman:1989be,Braaten:1995na,Kajantie:1996dw}, which in
the case of scalar field theories has been worked out
in \cite{Braaten:1995cm}.
In this approach one separates hard ($k \sim T$) from soft ($k \lesssim gT$)
modes and integrates out the former. Since in the
Matsubara formalism all non-static modes are necessarily hard,
this yields a three-dimensional effective theory containing
zero-modes only whose parameters (masses and coupling constants)
are to be determined by perturbative matching to the full theory.
To lowest order in scalar $\phi^4$ theory, this yields
(now for $N=1$)
\be
{\mathcal L}_3= \2 (\nabla \phi_0)^2+\2 m_3^2 \phi_0^2+g_3^2 \phi_0^4 +
\ldots
\ee
where $\phi_0=\sqrt{T}\int_0^\beta d\tau \phi(\tau,\mathbf x)$ and,
to lowest order, $m_3^2=g^2T^2$, $g_3^2=g^2 T$.

Calculating now also soft one-loop corrections, one obtains
in dimensional regularization
\be
\delta m^2=12g_3^2 \int{d^3k\0(2\pi)^3}{1\0k^2+m_3^2}=
-{3\0\pi}g_3^2 m_3=-{3\0\pi}g^3 T^2,
\ee
which is exactly the correction given in \eref{msc}.
(In cutoff regularization there would be an extra term $\propto g_3^2\Lambda$,
cancelling a contribution $\propto g^2T\Lambda$ to $m_3^2$.)
Using this method, the perturbative expansion of $m_{th}$ for
$N=1$ has
been worked out to order $g^5$ in \cite{Andersen:1998zx}.

\subsection{Apparent convergence}
\label{subsecscappconv}

The large-$N$ limit 
provides an 
instructive opportunity to study
the convergence properties of a resummed perturbation theory
with the exact result obtained by simply solving \eref{mth2sc}
numerically. This has been carried out in great detail 
in \cite{Drummond:1997cw}
with the result that only for $g$ sufficiently smaller than 1
there is quick convergence, which further deteriorates
if the renormalization scale $\bar\mu$ is very different
from $T$. An optimal value turned out to be $\bar\mu\approx 2\pi T$,
the scale of the bosonic Matsubara frequencies,
which has been argued previously in \cite{Braaten:1995cm}
to be a natural choice.

However, the convergence of the resummed thermal
perturbation series seems to be surprisingly poor given that
the exact result following from \eref{mth2sc}
is a rather unspectacular function. Naturally, a perturbative result 
is a truncated polynomial and thus bound to diverge more and
more rapidly at large coupling as the order is increased.
This may be the case even when the physical effects described by the
lowest-order terms 
are still predominant.

In \cite{Bender:1990ks} it has been shown that the alternative
so-called nonlinear
$\d$-expansion scheme \cite{Bender:1988rq} yields approximations to
$m_{\rm th}^2$ that converge
almost uniformly in $g$, but this scheme has not yet found applications
in more complicated (gauge) field theories.

\label{scalarPade}

\begin{figure}
\centerline{\includegraphics[viewport=70 200 540 550,scale=0.45]{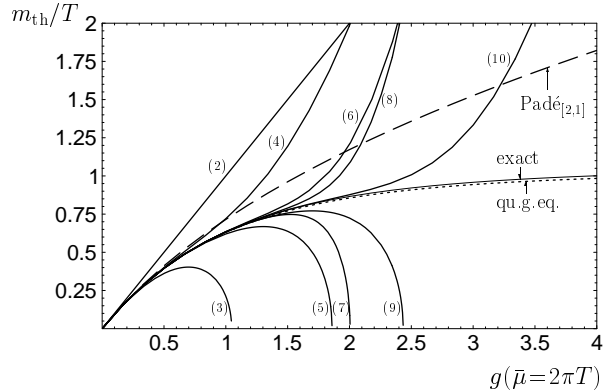}}
\caption{Thermal mass in large-$N$ $\phi^4$-theory as a function
of $g(\bar\mu=2\pi T)$ together with
the perturbative results \eref{msc}
accurate to order $g^2$, $g^3$, \ldots, $g^{10}$. 
The $g^3$ result is the one reaching zero at $g\approx 1$; its
Pad\'e-improved version \eref{mthscPade} is given by the
long-dashed line. The short-dashed line just below the exact result
is obtained by solving the quadratic gap equation \eref{mtruncgap},
which is also perturbatively equivalent to the order $g^3$ result.}
\label{figmthsc}
\end{figure}

A simpler 
alternative is provided by Pad\'e approximants \cite{Bak:Pade},
that is rewriting a given perturbative
result as a perturbatively equivalent
rational function in $g$ by replacing
\be\fl
F_n(g)=c_0+c_1g^1+\ldots+c_n g^n \to
F_{[p,n-p]}(g)={c_0+a_1g^1+\ldots+a_p g^p \0
1+a_{p+1}g^1+\ldots+a_n g^{n-p}}.
\ee
For example,
truncating \eref{msc} above order $g^3$ gives an approximation $F_3$ to
$m_{\rm th}^2$ that stops growing as a function of $g$ at $g=2\pi/9\approx 0.7$
and goes back to zero and then to negative values for $g\ge \pi/3
\approx 1$ (line labelled ``(3)'' in \fref{figmthsc}).
The true (large-$N$) thermal mass from \eref{mth2sc} however is a monotonically
growing function (line labelled ``exact'' in \fref{figmthsc}).
The result $F_3$ in fact ceases to be an improvement over the
leading-order result $F_2=g^2T^2$ roughly where it stops growing.
On the other hand, the simplest possibility for a
perturbatively equivalent Pad\'e approximant, $F_{[2,1]}$, is
a monotonic function in $g$,
\be\label{mthscPade}
m_{\rm th}^2/T^2 = {g^2\01+3g/\pi} + \Or(g^4)
\ee
(long-dashed line in \fref{figmthsc}), and it
gives a substantial improvement for $g\gtrsim 1$.


Higher-order Pad\'e approximants converge rather well
in the simple scalar model \cite{Drummond:1997cw} (except
when they happen to have a pole of the denominator at positive
coupling) and
they have
been proposed as a possibility to improve also the unsatisfying
convergence of perturbative results for the thermodynamical
potential in finite-temperature QCD 
\cite{Kastening:1997rg,Hatsuda:1997wf,Drummond:1997cw},
which will be discussed in more detail in Sect.~\ref{QCDappconv}.
There it does increase the apparent convergence of the resummed
perturbation series for the first few orders, but at higher orders it looks
less convincing. In fact, at these higher orders the perturbation
series also involves $\ln(g)$-contributions, which make a simple
Pad\'e improvement appear less natural. Indeed,
in the above large-$N$ 
scalar model, where the higher Pad\'e
approximants converge rather quickly \cite{Drummond:1997cw},
no $\ln(g)$-terms arise.


Already the simplest Pad\'e resummation \eref{mthscPade}
suggests that the low quality of standard perturbative results
is due to the fact that the latter are polynomials in $g$
which inevitably blow up at larger values of $g$.
In particular those contributions which can be traced to
a resummation of the screening mass involve large coefficients.
Since this resummation is a priori nonperturbative in that it
involves arbitrarily high powers of $g$, this signals the need for
a more complete treatment of such resummation effects.

In the above scalar toy model, one can in fact easily obtain
an efficient resummation of the term involving $g^3$ in \eref{msc}
which is responsible for the poor convergence of the perturbative
results displayed in figure \ref{figmthsc}. If one just retains
the first two terms in a $(m/T)$ expansion of the
one-loop gap equation \eref{mth2sc}, one ends up with
a simple quadratic gap equation
\be
\label{mtruncgap}
m_{\rm th}^2=g^2T^2-{3\0\pi}g^2Tm_{\rm th}
\ee
which is perturbatively equivalent to \eref{msc} to order $g^3$,
but the solution of \eref{mtruncgap} turns out to be
extremely close to that of the full gap equation
if the $\overline{\mbox{MS}}$ renormalization scales 
$\bar\mu \approx 2\pi T$.


There exist formalisms which at a given order
of approximations perform a complete
propagator resummation: the so-called 
self-consistent $\Phi$-derivable approximations \cite{Baym:1962} 
(in particle physics also known in connection with
the composite-operator effective action or
Cornwall-Jackiw-Tomboulis formalism \cite{Cornwall:1974vz}).

In the Luttinger-Ward representation \cite{Luttinger:1960}
the thermodynamic potential $\Omega=-PV$ 
is expressed as a functional of full propagators $D$ and
two-particle irreducible (2PI) diagrams. Considering
a scalar field theory with both cubic and quartic vertices
for the moment, $\Omega[D]$ has the form
\bea
\label{LW}\fl
\Omega[D]&=&-T \log Z=\2 T \,\Tr \log D^{-1}-\2 T \,\Tr\, \Pi D
+T \Phi[D]\nonumber\\
\fl
&=&\int\!{d^4k\0(2\pi)^4}n(\omega) \Im \left[
\log D^{-1}(\omega,k)-\Pi(\omega,k) D(\omega,k) \right]+T\Phi[D],\;\;
\eea
where $\Tr$ denotes the trace in configuration space,
and
$\Phi[D]$ is the sum of the 2-particle-irreducible ``skeleton''
diagrams
\be\label{skeleton}
-\Phi[D]= 
\includegraphics[bb = 50 390 550 440,width=5cm]{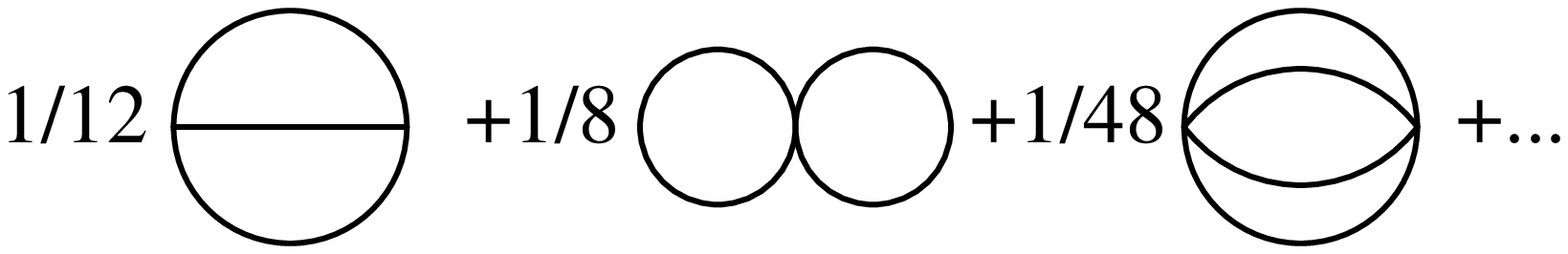}
\ee
\vspace{2mm}
The self energy $\Pi=D^{-1}-D^{-1}_0$, 
where $D_0$ is the bare propagator, is
related to $\Phi[D]$ by 
\be\label{PhiPi}
\delta\Phi[D]/\delta D=\2\Pi.
\label{Pi}
\ee
An important property of the functional  $\Omega[D]$, which is easily verified
using  (\ref{PhiPi}), is that it is stationary under variations of
$D$:
\be\label{selfcons}
{\delta\Omega[D] / \delta D}=0.
\ee
Self-consistent (``$\Phi$-derivable'')
\cite{Baym:1962} approximations
are  obtained by selecting a class of skeletons in 
$\Phi[D]$ and calculating $\Pi$ from equation~(\ref{Pi})
above, preserving the stationarity condition.

$\Phi$-derivable approximations have
been worked out in scalar theory to 3-loop order
\cite{Braaten:2001en,Braaten:2001vr} and it has recently
been shown that in this model
these approximations can be nonperturbatively renormalized in a
self-consistent manner
\cite{vanHees:2001ik,VanHees:2001pf,vanHees:2002bv,Blaizot:2003br}.

In the present large-$N$ $\phi^4$ toy model, the 2-loop
$\Phi$-derivable approximation is in fact exact,
so that one may use this example for studying the
quality of further approximations on top of the former \cite{Blaizot:2003tw},
which are typically unavoidable to make the formalism tractable
in more complicated theories.

In particular,
a resummation of quasi-particle propagators in thermodynamic quantities
is conveniently performed to 2-loop order in the entropy density
$\mathcal S=dP/dT$, which turns out to have the simple representation
\cite{Blaizot:1999ip}
\be\fl
{\cal S}=-\int\!\!{d^4k\0(2\pi)^4}{\6n(\om)\0\6T} \left\{ \Im 
\log D^{-1}(\om,k) - \Im\Pi(\om,k) \Re D(\om,k) \right\},
\label{Ssc}\ee
where $\Pi$ is the one-loop self-energy calculated self-consistently
with dressed propagator $D$. At this order all fundamental interactions
can be completely absorbed in the spectral properties of
the quasi-particles, whose residual interactions enter only
at three-loop order.
For a real momentum-independent
self-energy $\Pi=m^2$, \eref{Ssc}
even coincides with the free (Stefan-Boltzmann)
expression for the entropy density, ${\cal S}_{\rm free}(m)$.
The more general form \eref{Ssc} is also applicable in
gauge theories and for fermions 
\cite{Vanderheyden:1998ph,Blaizot:1999ip,Blaizot:1999ap,Blaizot:2000fc}
and we shall return to that
in Sect. \ref{secBIR}.


In references\ \cite{Karsch:1997gj,Chiku:1998kd,Pinto:1999py,Andersen:2000yj} 
a different reorganization of thermal perturbation theory
has been proposed,
``screened perturbation theory'' (SPT), 
which amounts to adding and subtracting a mass term
to the Lagrangian \eref{scLagr}
according to
\be\label{scrpth}
{\cal L}\to {\cal L} - \2m^2 \bphi^2 + \d \2m^2 \bphi^2
\ee
and to treating $\d$ as a one-loop quantity prior to putting it eventually
to $\d=1$. The difference to conventional resummation techniques
is to refrain from identifying and expanding out the coupling $g$
implicit in $m$, which in the end is chosen as some approximation to the
thermal mass and thus proportional to $g$ for small coupling.

Starting from two-loop order, it is possible to determine $m$ by
a principle of minimal sensitivity, which makes SPT a variant
of the so-called variational perturbation theories
(see e.g.\ \cite{Duncan:1988hw,Hamprecht:2003vh}).

This scheme has been applied with apparent success to
scalar field theory \cite{Karsch:1997gj,Chiku:1998kd,Pinto:1999py,Andersen:2000yj}.
It has also been generalized to gauge theories, where
a local mass term is insufficient but needs to be replaced
by a nonlocal gauge invariant extension of a thermal mass
term (see sect.\ \ref{secHTLPT}) 
\cite{Andersen:1999fw,Andersen:1999sf,Andersen:2002ey,Andersen:2003zk}.
A special difficulty of SPT is that at any finite order of
the new perturbative expansion it gives rise to
new ultra-violet divergences and corresponding new
scheme dependences, which need to be fixed in some way or other
\cite{Rebhan:2000uc}.

\subsection{Restoration of spontaneously broken symmetry}

A physically important case ignored so far is spontaneous
symmetry breaking, the simplest example of which is provided
by 
adding a wrong-sign mass term to
the $\phi^4$ theory considered above
\be
{\mathcal L}=\2(\6_\mu \phi)^2-V_{\rm cl}(\phi) \quad \hbox{with
$V_{\rm cl}(\phi)=-\2\nu^2+{\lambda\04!} \phi^4$},
\ee 
where we have switched to the more conventional notation of $\lambda/4!=g^2$.
While ${\mathcal L}$ is symmetric under
$\phi\to-\phi$, this symmetry is ``spontaneously broken''
by choosing one of the minima of $V_{\rm cl}$, which are given by $\phi_{\rm
min}=
\pm \sqrt{3!\nu^2/\lambda}$.

At high temperatures, however, there is symmetry restoration
\cite{Kirzhnits:1972ut,Weinberg:1974hy,Dolan:1974qd,Kirzhnits:1976ts,Arnold:1993rz}:
When $T\gg\nu$, the scalar field receives a contribution of
$\hat m^2_{\rm th}=\l T^2/4!$ to its (initially negative) mass squared:
$-\nu^2\to -\nu^2+\l T^2/4!$ or, equivalently, 
\be\label{Veff2nd}
V_{\rm cl}\to V_{\rm eff}(T)=V_{\rm cl}+
\2 \hat m^2_{\rm th}\phi^2.
\ee 
As a result, the minimum of $V_{\rm cl}$
becomes $\phi_{\rm min}=0$ for $T\ge T_c=\sqrt{4!\nu^2/\l}$.
Since 
\be
P=-V_{\rm eff}(T)\Big|_{\phi_{\rm min}} ={\l\0384}\theta(T_c-T)\times
(T_c^2-T^2)^2,\ee 
the phase transition is
of second order, i.e., there is no discontinuity in the first derivative
of the pressure (in the entropy), but only in its second derivative,
the specific heat.

The above {\em effective potential} can be derived more directly and
systematically from the partition function evaluated at spatially
constant field configurations $\bar\phi$,
\be
V_{\rm eff}(\bar\phi;T)=-{1\0\beta V} \ln Z \Big|_{\bar\phi}.
\ee
At one-loop order the temperature-dependent
contribution of a scalar field with (field-dependent)
mass $m^2(\bar\phi)=V_{\rm cl}''(\bar\phi)$ is
\bea\label{Veff1loop}
\fl
V_{\rm eff}^{(1)}(\bar\phi;T)
&=V_{\rm cl}+T\int{d^3p\0(2\pi)^3} \ln \(1-e^{-\omega/T}\) \quad
\hbox{with $\omega=\sqrt{p^2+m^2(\bar\phi)}$} \nonumber\\ \fl
&=V_{\rm cl}-{\pi^2 T^4\090}+{1\024}m^2(\bar\phi)T^2-{1\012\pi}m^3(\bar\phi)T+\ldots
\eea

The first field-dependent correction term reproduces the effective potential
of \eref{Veff2nd}. Subsequent terms in the high-T expansion as well as
higher order loop contributions require a resummation of the 
thermal mass of the
scalar field: 
\be\label{m2efflo}
m^2(\bar\phi)\to m^2_{\rm eff}(\bar\phi;T)=m^2(\bar\phi)+\hat m^2_{\rm th}.
\ee
To avoid over-counting, this resummation has to take place either
through the explicit introduction of thermal counterterms \cite{Arnold:1993rz}
or by using a self-consistent formalism like that
of $\Phi$-derivable approximation or the CJT effective action
\cite{Amelino-Camelia:1993nc}. The latter allows to include
superdaisy diagrams, which treated naively would give a spurious
and in fact completely misleading
contribution to the effective potential $\propto m(\bar\phi)T^3$.

Insertion of \eref{m2efflo} in \eref{Veff1loop} 
gives an effective potential which at the critical
temperature has two degenerate minima, $\bar\phi=0$ and $\bar\phi\sim
\sqrt\l T_c$. However, at the second non-trivial minimum,
$m_{\rm eff}(\bar\phi) \sim \l T$ and the loop expansion parameter of
the perturbation series, which is $\l T/m_{\rm eff}$, ceases to be small.
Hence, perturbation theory cannot decide whether there is a first-order
phase transition as implied by a second degenerate minimum in the potential
or not. In fact, universality arguments show that the phase transition
must be second order for a $\phi^4$ theory with $Z_2$ symmetry
\cite{Zin:QFT}.

The situation is different in gauge theories. Again, resummed perturbation
theory signals a first order phase transition, but when
the Higgs mass is small compared to the massive vector-boson mass, it
is sufficiently strongly first-order so that the perturbation series
does not break down where the degenerate minima appear. 
The (resummed) effective potential has been calculated perturbatively
to one-loop \cite{Anderson:1992zb,Carrington:1992hz,Dine:1992wr}
and two-loop order \cite{Arnold:1993rz,Fodor:1994bs,Buchmuller:1995sf}
(see also \cite{Gynther:2003za}).
Of particular interest is the question whether the electroweak
sector of the Standard Model (or extensions thereof) admits
a first order phase transition, which would be of great interest
for baryogenesis in the Early Universe \cite{Kuzmin:1985mm}.
Nonperturbative studies indeed confirmed a first-order transition
for sufficiently small Higgs masses \cite{Fodor:1995sj,Kajantie:1996kf}, but
found an endpoint \cite{Kajantie:1996mn,Rummukainen:1998as,Csikor:1998eu} 
at Standard Model Higgs mass $m_H\lesssim 80$ GeV, so that
a first-order phase transition in the Standard Model is excluded
by the current experimental bounds on the Higgs mass (though not
in possible (supersymmetric)
extensions of the Standard Model \cite{Riotto:1999yt,Csikor:2000sq}).



\section{QCD thermodynamics}
\label{secQCDtd}

The (resummed) perturbative evaluation of the thermodynamic potential
of QCD at high temperature has been
pushed in 
recent years up to
the order $g^{6}\ln g$
\cite{Shuryak:1978ut,Kapusta:1979fh,Toimela:1983hv,Arnold:1995eb,Zhai:1995ac,Braaten:1996ju,Braaten:1996jr,Kajantie:2002wa}.
At higher orders in $g$ this is much facilitated by the possibility to
employ effective field theory methods which in this case
lead to a dimensional reduction to a 3-dimensional
Yang-Mills plus adjoint Higgs theory \cite{Appelquist:1981vg}.
Completion of these results at order $g^6$ is in fact impossible
without inherently nonperturbative input, but further
progress has been made most recently by the extension to nonzero quark
chemical potential within dimensional reduction
\cite{Vuorinen:2003fs}.

At zero temperature and high chemical potential the pressure
is known to order $g^4$ \cite{Freedman:1977xs,Freedman:1977dm,Freedman:1977ub,Baluni:1978ms,Vuorinen:2003fs}, while the low-temperature
expansion of the pressure leads to the phenomenon
of non-Fermi-liquid behaviour of entropy and specific heat
\cite{Ipp:2003cj}.

\subsection{Dimensional reduction}

Dimensional reduction in hot QCD 
leads to an effective three-dimensional
Lagrangian \cite{%
Appelquist:1981vg,Nadkarni:1983kb,Nadkarni:1988fh}
\be\label{LQCDdr}
\mathcal L_E=\2 \tr F_{ij}^2 + \tr [D_i,A_0]^2
+m_E^2 \tr A_0^2 + \2 \lambda_E (\tr A_0^2)^2 +\ldots
\ee
where the parameters are determined perturbatively by matching
\cite{Braaten:1996jr,Kajantie:1997tt}. In lowest order\footnote{%
Some higher-dimension terms in the effective theory
\eref{LQCDdr} have been determined in 
references~\cite{Landsman:1989be,Chapman:1992wk,Diakonov:2003yy},
and, including the electroweak sector, in references 
\cite{Moore:1996jv,Kajantie:1998ky}.}
and at zero chemical potential one has a dimensionful
coupling $g_E^2 = g^2T$ and \cite{Nadkarni:1988fh}
\be\label{LQCDparam}
m_E^2=(1+N_f/6)g^2 T^2,\qquad
\lambda_E={9-N_f\012\pi^2}g^4T,
\ee
though $\lambda_E$ starts to contribute to the
pressure only at order $g^6$. At this order, however,
the self-interactions of the massless magnetostatic gluons
start to contribute, and these contributions are
inherently non-perturbative because the three-dimensional
theory for the zero modes $A_i(\vec x)$ is
a confining theory \cite{Polyakov:1978vu,Linde:1980ts,Gross:1981br}.

The thermal pressure of the 4-dimensional theory can be
decomposed into contributions from the hard modes $\sim T$,
calculable by standard perturbation theory, and soft contributions
governed by (\ref{LQCDdr}) which involves both perturbatively
calculable contributions up to order $g^5 T^4$ and the
nonperturbative ones starting at order $g^6 T^4$.

In reference~\cite{Braaten:1996jr}
the effective theory based on (\ref{LQCDdr}) has been used
to organize and reproduce the perturbative calculation of
the thermal pressure to order $g^5$ of references~\cite{Arnold:1995eb,Zhai:1995ac}.
This turns out to be particularly elegant when dimensional
regularization is used to provide both the UV and IR cutoffs
of the original and effective field theories.

To order $g^4$, the contribution of the hard modes can then be
written as \cite{Braaten:1996jr}
\bea\label{P3h}
\fl P_{\rm hard}&=\8{8\pi^2\045}T^4 \biggl\{\left(1+\8{21\032}N_f\right)-
\8{15\04}\left(1+\8{5\012}N_f\right){\alpha_s\0\pi}\nn\fl&\quad
+\Bigl\{244.9+17.24N_f-0.415N_f^2
+{135}\left(1+\8{1\06} N_f\right)\ln{\bar\mu\02\pi T}\nn\fl
&\qquad-\8{165\08}\left(1+\8{5\012}N_f\right)\left(1-\8{2\033}N_f\right)
\ln{\bar\mu\02\pi T}
\Bigr\}\left({\alpha_s\0\pi}\right)^2\biggr\}.
\eea
In the first logarithm the dimensional regularization scale $\bar\mu$ 
is associated with regularization in the infrared and thus has to
match a similar logarithm in the effective theory, whereas the second
logarithm is from UV and involves the first coefficient of the beta
function.

Indeed, calculating the pressure contribution of the soft sector
described by (\ref{LQCDdr}) in dimensional regularization
gives, to three-loop order (neglecting $\lambda_E$-contributions)
\bea\label{P3s}
P_{\rm soft}/T &=& {2\03\pi}m_E^3-{3\08\pi^2}\left(
4\ln{\bar\mu\02m_E}+3\right)g_E^2 m_E^2\nn&&
-{9\08\pi^3}\left({89\024}-{11\06}\ln2+{1\06}\pi^2\right)
g_E^4\,m_E^{\phantom4}.
\eea
All the contributions to the pressure involving odd powers of $g$
in (\ref{Fpt})
(as well as part of those involving even powers) are coming
from the soft sector. Inserting the leading-order value (\ref{LQCDparam})
for $m_E$ gives the QCD pressure up to and including order $g^4\ln g$;
to obtain all the terms to order $g^5$, next-to-leading order
corrections to the $m_E$-parameter have to be obtained
by a matching calculation as given in reference~\cite{Braaten:1996jr}.
The result is known in closed form
\cite{Arnold:1995eb,Zhai:1995ac,Braaten:1996jr}, but we shall
quote here only the case of SU(3) with $N_f$ quark flavours
and numerical values for the various coefficients:
\bea\label{Fpt}
\fl P &= \8{8\pi^2\045}T^4 \biggl\{
\left(1+\8{21\032}N_f\right)-
\8{15\04}\left(1+\8{5\012}N_f\right){\alpha_s\0\pi}
+30\left[\left(1+\8{1\06} N_f\right)\left({\alpha_s\0\pi}\right)\right]^{3/2}\nn
\fl&\qquad\quad+\Bigl\{237.2+15.97N_f-0.413N_f^2+
\8{135\02}\left(1+\8{1\06} N_f\right)\ln\left[{\alpha_s\0\pi}(1+\8{1\06} N_f)\right]\nn
\fl&\qquad\qquad
-\8{165\08}\left(1+\8{5\012}N_f\right)\left(1-\8{2\033}N_f\right)
\ln{\bar\mu\02\pi T}
\Bigr\}\left({\alpha_s\0\pi}\right)^2\nn
\fl&\qquad\quad+\left(1+\8{1\06} N_f\right)^{1/2}\biggl[-799.2-21.96 N_f - 1.926 N_f^2\nn
\fl&\qquad\qquad+
\8{495\02}\left(1+\8{1\06} N_f\right)\left(1-\8{2\033}N_f\right)
\ln{\bar\mu\02\pi T}\biggr]\left({\alpha_s\0\pi}\right)^{5/2}
+\mathcal O(\alpha_s^3\ln\alpha_s) \biggr\}.
\eea
Here $\bar\mu$ is the renormalization scale parameter of
the $\overline{{\rm MS}}$ scheme and $\alpha_s(\bar\mu)$ is the
corresponding running coupling.


The coefficient of the $\alpha_s^3\ln\alpha_s$ term, the last
in the pressure at high $T$ and vanishing chemical potential
that is calculable completely within perturbation theory, has
recently been determined as \cite{Kajantie:2002wa,Kajantie:2003ax}
\bea\label{Pg6}
P\Big|_{g^6\ln g}&=
\8{8\pi^2\045}T^4 \biggl[1134.8+65.89 N_f+7.653 N_f^2\nn&
-\8{1485\02}\left(1+\8{1\06} N_f\right)\left(1-\8{2\033}N_f\right)
\ln{\bar\mu\02\pi T}\biggr]\left({\alpha_s\0\pi}\right)^{3}
\ln{1\0\alpha_s}\,.
\eea

In order to obtain the $g^6\ln g$-contribution (\ref{Pg6}) one
also needs $g_E^2$ to order $g^4$ (given in reference~\cite{Kajantie:1997tt})
and above all
the four-loop contribution of the effective theory (\ref{LQCDdr})
which has recently been calculated analytically as
\cite{Kajantie:2002wa}
\be\label{P4s}
\fl P_{\rm soft}^{(4)}/T = N_g {(N g_E^2)^3 \0(4\pi)^4}
\left[ \left( \8{43\012}-\8{157\pi^2\0768} \right) \ln{\bar\mu\0g_E^2}
+ \left( \8{43\04}-\8{491\pi^2\0768} \right) \ln{\bar\mu\0m_E}+c \right]
\ee
up to a constant $c$ that
is strictly nonperturbative and needs to
be determined by three-dimensional lattice calculations. 
Such calculations have been undertaken in reference~\cite{Kajantie:2000iz},
but they depend
on an as yet undetermined 4-loop matching coefficient. At the moment the
conclusion is that it is at least
not excluded that the lattice results
based on dimensional reduction can be matched to the full four-dimensional
results at temperatures of a few times the transition temperature.
For this reason, the most reliable results on the thermodynamics
of hot QCD (particularly for pure-glue QCD) remain to date the
four-dimensional lattice data. However, since inclusion of
fermions is particularly easy in the dimensional reduction
method, but computationally expensive in lattice gauge theory,
a full three-dimensional prediction would
clearly be most desirable.

\subsection{Apparent convergence}
\label{QCDappconv}

\begin{figure}[ht]
\centerline{\includegraphics[viewport=70 200 540 550,scale=0.45]{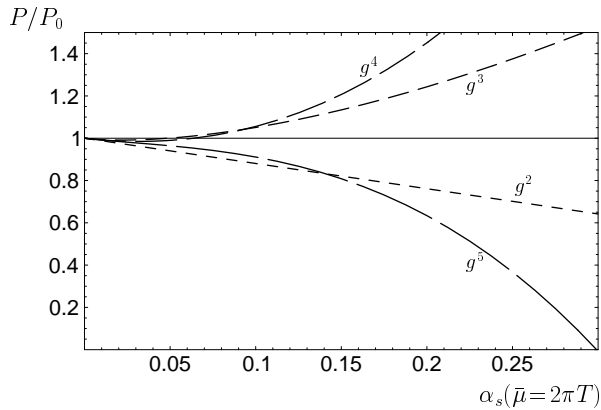}}
\caption{\label{figpFal}
Strictly perturbative results for the thermal pressure
of pure glue QCD normalized to the ideal-gas value
as a function of
$\alpha_s(\bar\mu=2\pi T)$.}
\end{figure}

\begin{figure}[t]
\centerline{\includegraphics[viewport=70 200 540 550,scale=0.45]{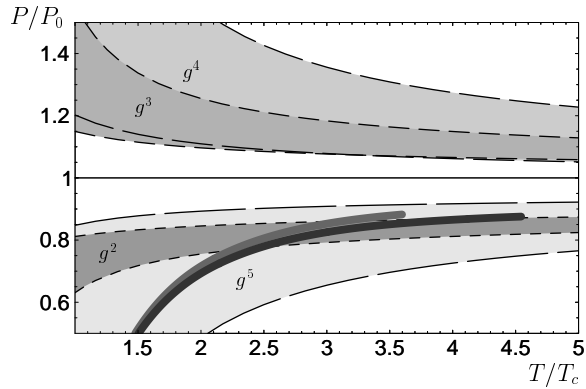}
}
\caption{Strictly perturbative results for the thermal pressure
of pure glue QCD as a function
of $T/T_c$ (assuming $T_c/\Lambda_{\overline{\hbox{\scriptsize MS}}}=1.14$).
The various gray bands bounded by differently
dashed lines show the perturbative results
to order $g^2$, $g^3$, $g^4$, and $g^5$, 
using a 2-loop running coupling with $\overline{\hbox{MS}}$
renormalization
point $\bar\mu$ varied between $\pi T$ and $4\pi T$. The thick
dark-grey line shows the continuum-extrapolated
lattice results from reference~\protect\cite{Boyd:1996bx};
the lighter one behind that of a lattice calculation
using an RG-improved action \cite{Okamoto:1999hi}.
\label{fig:qcd}}
\end{figure}

Figure \ref{figpFal} shows the outcome of
evaluating the perturbative result \eref{Fpt} for the thermodynamic pressure
at $N_f=0$
to order $\alpha_s$, $\alpha_s^{3/2}$, $\alpha_s^2$, and
$\alpha_s^{5/2}$, respectively, with a choice of $\bar\mu=2\pi T$.
Apparently, there is no convergence for $\alpha_s \gtrsim 0.05$ which
in QCD corresponds to $T\lesssim 10^5 T_c$, where $T_c$ is
the deconfinement phase transition temperature.
What is more, the numerical dependence on the renormalization scale $\bar\mu$
does not diminish as the order of the perturbative result
is increased, but becomes more and more severe,
as shown in figure~\ref{fig:qcd}, where 
$\alpha_s(\bar\mu)$ is determined by a 2-loop renormalization
group equation and
$\bar\mu$ is varied
between $\pi T$ and $4\pi T$. 
So there seems to be a complete loss of predictive power at any temperature of
interest \cite{Arnold:1995eb,Zhai:1995ac,Braaten:1996jr,Braaten:1996ju}.

To alleviate this situation,
various mathematical extrapolation techniques  have been tried,
such as  Pad\'e approximants
\cite{Kastening:1997rg,Hatsuda:1997wf,Cvetic:2002ju},
self-similar approximants \cite{Yukalov:2000zr}, and
Borel resummation \cite{Parwani:2000rr,Parwani:2000am},
however with limited success. 
While in the scalar toy model of
Sect.~\ref{scalarPade}, Pad\'e approximants work remarkably well,
in the QCD case there is a problem how to handle
logarithms of the coupling and, perhaps related to that, the
numerical results obtained so far appear less satisfactory.


When compared with the perturbative results, it is 
however remarkable
that the next-to-leading result to order $g^2$ performs rather
well at temperatures $\gtrsim 2T_c$, though the higher-order results
prove that perturbation theory is inconclusive. Moreover, 
simple quasiparticle models which describe the effective
gluonic degrees of freedom by $2N_g$ ($N_g=N^2-1$) scalar
degrees of freedom with asymptotic masses taken from a
HTL approximation can be used quite successfully to model
the lattice data by fitting the running coupling
\cite{Peshier:1996ty,Levai:1997yx,Peshier:1999ww,Schneider:2001nf}.

In fact, we have seen in section \ref{subsecscappconv} that
already in the simplest scalar model resummed perturbation theory
gives rather poorly convergent results, and that fairly
simple reorganizations as in \eref{mtruncgap} lead to
dramatic improvements of the situation.

\begin{figure}[t]
\centerline{\includegraphics[viewport=50 200 540 555,scale=0.43]{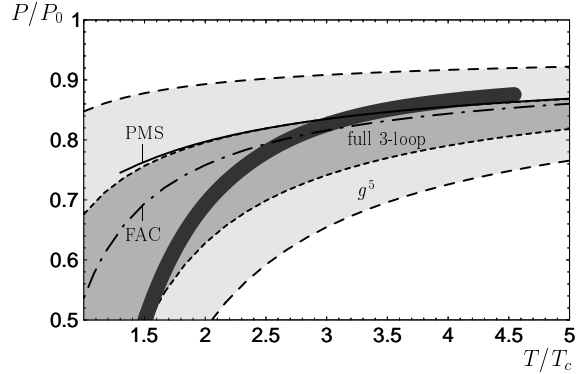}}
\caption{\label{fig:3loop}
Three-loop pressure in pure-glue QCD with 
unexpanded effective-field-theory parameters 
when $\bar\mu$ is varied between $\pi T$ and $4\pi T$ (medium-gray band); the
dotted lines indicate the position of this band when only the leading-order
result for $m_E$ is used.
The broad light-gray band underneath is the strictly perturbative result to
order $g^5$ with the same scale variations. The full line gives the
result upon extremalization (PMS) with respect to $\bar\mu$ (which does
not have solutions below $\sim 1.3T_c$); the dash-dotted line corresponds
to fastest apparent convergence (FAC) in $m_E^2$, which sets
$\bar\mu\approx 1.79\pi T$. (Taken from \cite{Blaizot:2003iq})}
\end{figure}
\begin{figure}[t]
\centerline{\includegraphics[viewport=50 200 540 555,scale=0.45]{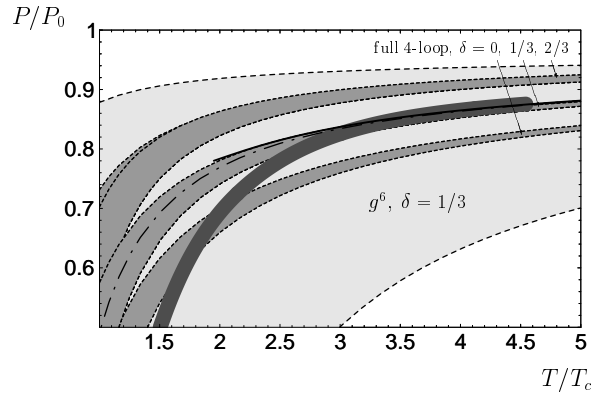}}
\caption{\label{fig:4loop}
Like figure \ref{fig:3loop}, but extended to four-loop order by
including the recently determined $g^6\ln(1/g)$ contribution of
\cite{Kajantie:2002wa} together with three values for
the undetermined constant $\delta$ in $[g^6\ln(1/g)+\delta]$.
The broad light-gray band underneath is the strictly perturbative result to
order $g^6$ corresponding to the central value $\delta=1/3$,
which has a larger scale dependence than the order $g^5$ result
in figure \ref{fig:3loop}; the untruncated results on the other
hand show rather small scale dependence.
The full line gives the untruncated
result with $\delta=1/3$ and $\bar\mu$ fixed by PMS (which does
not have solutions below $\sim 1.9T_c$); the dash-dotted line corresponds
to fastest apparent convergence (FAC) in $m_E^2$, which sets
again to $\bar\mu\approx 1.79\pi T$. (Taken from \cite{Blaizot:2003iq})}
\end{figure}

As has been noted in references \cite{Kajantie:2002wa,Blaizot:2003iq},
in the calculation as
organized through the dimensionally reduced effective theory
(\ref{LQCDdr}), the large scale dependence of strict
perturbation theory can be significantly reduced when the 
perturbative values of the effective parameters
are kept as they appear in (\ref{P3s}), 
without expanding and truncating the final result to the considered order
in the coupling.
What is more, the unexpanded two-loop and
three-loop contributions from the soft sector lead to results which no
longer exceed the ideal-gas result (as the strictly perturbative
results to order $g^3$ and $g^4$ do), and their (sizable) scale dependence
diminishes by going from two to three-loop order \cite{Blaizot:2003iq}.

At three-loop order, it is in fact possible to eliminate the
scale dependence altogether by a principle of minimal sensitivity.
The result in fact agrees
remarkably well with the 4-d lattice results down to $\sim 2.5T_c$
as shown in \fref{fig:3loop}.

The four-loop order result depends on the unknown constant
$c$ in (\ref{P4s}). However, it is at least not excluded
that $c$ could be such that the four-loop result
is also close to the 4-d lattice results \cite{Kajantie:2002wa,Laine:2003ay}.
In fact, while a strictly perturbative treatment leads
to increased scale dependence compared to 3-loop order,
keeping the soft contributions unexpanded in $g$ further
diminishes the scale dependence \cite{Blaizot:2003iq}
as shown in \fref{fig:4loop}.

While this goes only minimally beyond a strictly perturbative treatment,
it strongly suggests that perturbative QCD at high temperature, when
supplemented by appropriate resummation of soft physics, is 
{\em not} limited to $T\gg 10^5 T_c$
as previously thought \cite{Arnold:1995eb,Braaten:1996ju}, but
seems capable of 
quantitative predictions at temperatures of possibly
only a few times the transition temperature. 

\subsection{Finite chemical potential}

The calculation of the thermodynamical potential using
dimensional reduction can in fact be extended to nonvanishing
quark chemical potential $\mu_q$, provided however that $T\gg m_E$.
This has been carried out up to and including order $g^6 \ln g$
in reference~\cite{Vuorinen:2003fs}, and a discussion of
the issue of apparent convergence can be found in reference~\cite{Ipp:2003yz}.

A non-zero chemical potential modifies the parameters
of the effective theory, in particular the mass parameter $m_E$,
which is altered already at leading order according to
\eref{mD} below%
. In addition there are new, $C$-odd
terms in the effective Lagrangian. The one with smallest
dimension in nonabelian theories reads 
\cite{KorthalsAltes:1999cp,Hart:2000ha,Bodeker:2001fs}
\be\label{muA03}
\mathcal L_E^{(\mu)}=i{g^3\03\pi^2}\sum_q \mu_q \tr A_0^3.
\ee
In Abelian theories there is also a linear term involving
$\tr A_0$ which has been discussed e.g.~in \cite{Khlebnikov:1996vj}.

In general the effects of these additional $C$-odd terms are
small compared to the $C$-even operators in \eref{LQCDdr} which
depend on the chemical potential through its parameters.
One quantity which is determined to leading order by
the operator \eref{muA03} is the flavour off-diagonal
quark number susceptibility 
at zero chemical potential \cite{Blaizot:2001vr}
\be
\chi_{ij}\equiv{\6^2 {P} \0 \6 \mu_i\6 \mu_j}.
\ee
When quark masses are negligible, all off-diagonal components
are equal at $\mu_i=0$. Denoting them by $\tilde\chi$, the
leading-order term involves a logarithmic term coming from
the exchange of three electrostatic gluons and is given
by \cite{Blaizot:2001vr}
\be\label{tcc0}
{\tilde \chi}\simeq-
{(N^2-1)(N^2-4)\0 384 N}\left({g\0\pi}\right)^6 T^2\ln{1\0g}.
\ee
where $N$ is the number of colours. This vanishes in SU(2)
gauge theory, but not in QED,
where (in the ultrarelativistic limit) \cite{Blaizot:2001vr}
\be
\tilde \chi\Big|_{\mathrm{QED}} \simeq- {e^6\024 \pi^6} T^2\ln{1\0e} . 
\ee

On the other hand, the diagonal quark susceptibilities
have a perturbative expansion whose first few terms are given by
\be
\label{cc0pt}
\fl  {\chi \0 \chi_0}=1-{1\02}{3\0N}{N_g\08}\left(g\0\pi\right)^2+
{3\0N}{N_g\08}\sqrt{{N\03}+{N_f\0 6}}\left(g\0\pi\right)^3 
+{3\04}{N_g\08}\left(g\0\pi\right)^4\log{1\0g}+ {\mathcal O}(g^4),
\ee
with $\chi_0 =NT^2/3$ the ideal gas value and $N_g=N^2-1$.
The higher-order coefficients have been calculated by
Vuorinen \cite{Vuorinen:2002ue} up to and including
order $g^6 \ln(1/g)$. 
The problem with apparent convergence is similar if somewhat
less severe than in the case of the pressure at zero chemical
potential discussed above.

In contrast to the pressure, however,
the coefficient of the order $g^6$ term in $\chi$
is not sensitive to nonperturbative
chromomagnetostatic physics and thus 
calculable in perturbation theory, though not yet available.
Its determination would in fact be of some interest in view of the 
important progress that has recently been made with
the inclusion of small chemical potentials in lattice
gauge theory
\cite{Fodor:2001au,Fodor:2001pe,deForcrand:2002ci,Allton:2002zi,%
Fodor:2002km,D'Elia:2002gd,Gavai:2003mf,Allton:2003vx}.

\subsection{Low temperature and high chemical potential}
\label{seclowT}

At large chemical potential but $T\lesssim g \mu_q$, dimensional
reduction does not occur. In this case a fully four-dimensional
computation has to be performed.

The perturbative result up to and including order $g^4$
at zero temperature
has been calculated by Freedman and McLerran 
\cite{Freedman:1977xs,Freedman:1977dm,Freedman:1977ub}
and by Baluni \cite{Baluni:1978ms} more than a quarter of a century ago.

This result was originally given
in a particular gauge-dependent momentum subtraction scheme.
In order to convert it to the gauge-independent
$\overline{\hbox{MS}}$ scheme, one needs to replace the
scale parameter $\mu_0$
in reference~\cite{Freedman:1977ub} ($M$ in references~\cite{Baluni:1978ms,Kap:FTFT})
according to \cite{Blaizot:2000fc,Fraga:2001id}
\begin{equation}\label{MSbarconversion}
\mu_0=\bar\mu \exp\left\{ [(151
)N-40N_f]/[24(11N-2N_f)] \right\}.
\end{equation}
Furthermore, the order $g^4$ contributions involved
two integrals that were evaluated only numerically with
sizable error bars. This calculation was recently repeated
in reference~\cite{Vuorinen:2003fs}, with one of these integrals
evaluated analytically and the other with very high accuracy.

Specialized to $N=3$ and
uniform chemical potential for
$N_f$ quark flavours, the result for the pressure at zero temperature
to order $g^4$ in
the $\overline{\hbox{MS}}$ scheme reads
\bea\label{FMBP}
\fl
P&=  {N_f \mu_q^4\04\pi^2} \biggl\{1-2{\alpha_s(\bar\mu)\0\pi}- 
\Bigl[ 
18-11 \log 2
- 0.53583 N_f + N_f \log{N_f
        \alpha_s(\bar\mu)\0\pi}  \nonumber\\
\fl&\quad\qquad\qquad\qquad
+ \;(11-{2\03}N_f)\log{\bar\mu\0\mu_q}\; \Bigr]
({\alpha_s(\bar\mu)\0\pi})^2+O(\alpha_s^3 \log \alpha_s)\biggr\},
\eea
where $\mu_q$ is the (common) quark chemical potential, not to be
confused with the renormalization scale $\bar\mu$ of
the $\overline{\hbox{MS}}$ scheme.

At zero temperature and large chemical potential, perturbation
theory is not hampered by a perturbative barrier at order $g^6$
so that higher-order corrections are in principle calculable,
but not yet available (except for one term $\propto
N_f^3 g^6$ extracted numerically from the solvable
large-$N_f$ limit of QCD \cite{Ipp:2003jy}).

However, at small but finite temperature $T\lesssim g\mu$, the
only weakly screened low-frequency 
transverse gauge-boson interactions (see (\ref{dynscr}) below)
lead to a qualitative deviation from the Fermi liquid behaviour
of relativistic systems described in reference
\cite{Baym:1976va}. In particular,
the low-temperature limit of entropy and specific
heat does not vanish linearly with temperature, but
there is a positive contribution proportional to $\alpha T\ln T^{-1}$
\cite{Holstein:1973,Gan:1993,Chakravarty:1995}, which
implies that at sufficiently small temperature the entropy
exceeds the ideal-gas result.
In reference~\cite{Ipp:2003cj} this effect was most recently
calculated beyond the coefficient of the leading log
obtained in \cite{Holstein:1973,Chakravarty:1995}\footnote{The coefficient
of the $\alpha T\ln T^{-1}$ term given in the original paper 
\cite{Holstein:1973} was found to be lacking of a factor of 4.}.
The complete result for the entropy below order $T^3 \ln T$, where
regular Fermi-liquid corrections enter, is given by
(for SU(3) and with numerically evaluated coefficients
which are known in closed form \cite{Ipp:2003cj})
\bea\label{anomS}
\fl\mathcal S=&\mu_q^2 T \biggl\{ {NN_f\03}
+{\alpha_s N_fN_g\018\pi} 
\ln\left(2.2268
\sqrt{\alpha_s N_f\0\pi}{\mu_q\0T}\right) 
-0.17286
N_g\left({\alpha_s N_f\0\pi}\right)^{2\03} 
\left(T\0\mu_q\right)^{2\03}\nonumber\\
\fl&\qquad 
+0.13014
N_g\left({\alpha_s N_f\0\pi}\right)^{1\03} 
\left(T\0\mu_q\right)^{4\03}
\biggr\} + O(T^3 \ln T),\quad T\ll g\mu_q,
\eea
and turns out to involve also fractional powers of the temperature as well
as the coupling.
(The corresponding result in
QED is obtained by replacing $N_g\to1$ and $\alpha_sN_f\to2\alpha$.)
\Eref{anomS} is the beginning of a perturbative expansion
provided $T/\mu_q\ll g$, e.g.\ $T/\mu_q\sim g^{1+\delta}$ with $\delta>0$.
The corresponding contribution to the pressure is then of
the order $g^{4+2\delta}\ln g$ and thus of higher order than
the terms evaluated in \eref{FMBP}. But in the entropy
the zero-temperature limit of the pressure drops out and
the nonanalytic terms found in \eref{anomS} become the
leading interaction effects when $T/\mu_q\ll g$. For exponentially
small $T/(g\mu_q) \sim \exp(-\#/g^2)$,
they eventually become comparable to the ideal-gas part
and the perturbative treatment breaks down (in QCD
one in fact expects a breakdown of perturbation theory already
at the order of $\exp(-\#/g)$ for colour superconducting
quarks, cf.\ section \ref{secCSC}).

Originally, the anomaly in the specific heat was discussed for
a nonrelativistic electron gas with the expectation that this effect
may be too small for experimental detection
\cite{Holstein:1973}. In QCD, however, it is numerically much more
important, not only
because $\alpha_s \gg \alpha_{\rm QED}$, but also because of
the relatively large factor
$N_g=8$. Consequently, it may play some role
in the thermodynamics of (proto-)neutron stars, if those have
a normal (non-superconducting) quark matter component.

For potential phenomenological applications in astrophysical
systems, the specific
heat $C_v$ at constant volume and number density is of
more direct interest, which however
differs from the logarithmic derivative of the entropy
only by subleading terms \cite{LL:V-Cv}:
\be\fl
\mathcal C_{v}
\equiv C_v/V=T\left\{ \left( \frac{\partial \mathcal S}{\partial T}\right) _{\mu_q }-{\left( \frac{\partial \mathcal N}{\partial T}\right) ^{2}_{\mu_q }}{\left( \frac{\partial \mathcal N}{\partial \mu_q }\right)^{-1} _{T}}\right\}=T\left( \frac{\partial \mathcal S}{\partial T}\right) _{\mu_q }+
O(T^3)\,,
\ee
where $\mathcal N$ is the number density.
In QCD this gives
\bea\label{anomCV}
\fl\mathcal C_v=&\mu_q^2 T \biggl\{ {NN_f\03}
+{\alpha_s N_fN_g\018\pi} 
\ln\left(\!
1.9574
\sqrt{\alpha_s N_f\0\pi}{\mu_q\0T}\right) 
-0.288095
N_g\!\left({\alpha_s N_f\0\pi}\right)^{2\03}\!\!
\left(T\0\mu_q\right)^{2\03}\nonumber\\
\fl&\qquad 
+0.3036697
N_g\left({\alpha_s N_f\0\pi}\right)^{1\03} 
\left(T\0\mu_q\right)^{4\03}
\biggr\} + O(T^3 \ln T),\quad T\ll g\mu_q.
\eea

The results \eref{anomS} and \eref{anomCV} 
imply an excess over
their respective ideal-gas values. They depend on having $T\ll g\mu_q$,
and in this regime the low-temperature series
has a small expansion parameter $T/(g\mu_q)$; for $T\gg g\mu_q$ the
standard exchange term gives the leading-order interaction
contribution \cite{Kap:FTFT}
\be\label{normalCvS}
\mathcal C_v \simeq \mathcal S=
\mu_q^2 T \left\{ {NN_f\03}
-N_g N_f {\alpha_s\04\pi} + O(\alpha_s^4)
\right\} + O(T^3),
\ee
so that for larger temperature there is a reduction
compared to the ideal-gas result.
A nonmonotonic behaviour of the entropy as a function of $T$
which interpolates between \eref{anomS} and \eref{normalCvS}
has indeed been found in the numerical evaluation of
the exactly solvable large-$N_f$ limit of QED and QCD
\cite{Ipp:2003zr}, and there the domain where the entropy
has the anomalous feature of exceeding the ideal-gas value
is given by $T/\mu_q \lesssim g \sqrt{N_f}/30$.

Non-Fermi-liquid corrections to $\mathcal C_v$ in the context of
ultrarelativistic QED and QCD
have also been considered
in reference~\cite{Boyanovsky:2000zj}, however the result obtained
therein does not
agree with \eref{anomCV}. When expanding perturbatively
the renormalization-group resummed result of 
reference~\cite{Boyanovsky:2000zj},
it would imply a leading nonanalytic
$\alpha T^3 \ln T$ term,
which is in fact the kind of nonanalytic terms that
appear also in regular Fermi-liquids \cite{Carneiro:1975}.
However, reference~\cite{Boyanovsky:2000zj} did not evaluate all
contributions $\propto \alpha T$, which were considered to be
free of nonanalytic terms.


\section{The quasi-particle spectrum in gauge theories}
\label{secqpgt}

With the exception of static quantities in the high-temperature
limit, where dimensional reduction is applicable, a systematic
calculation of observables in thermal field theory, such
as reaction rates, transport coefficients, and even the
thermodynamic potential at low temperatures and high
chemical potential require the determination and consistent
inclusion of medium effects on the dynamical propagators
in the theory.

As we have seen in the example of scalar field theory in Sect.~\ref{sec:scthm},
the interactions with the particles of the heat bath modify the
spectrum of elementary excitations.
The poles of the propagator, which determines the linear response
of the system under small disturbances, receive thermal corrections
which, in general, 
introduce a mass gap for propagating modes and screening for
non-propagating ones, even when the underlying field theory is massless.
The simple O($N\to\infty$) $\phi^4$ model considered in Sect.~\ref{sec:scthm}
is in fact somewhat misleading in that there the self-energy
is a real quantity to all orders in the coupling, whereas in a nontrivial
quantum field theory a nonvanishing width for propagating
modes is unavoidable \cite{Narnhofer:1983hp}. Nonetheless, the
typical situation in perturbation theory is that the width
is parametrically smaller than the thermal mass at a given momentum,
and that the former can be treated to some extent perturbatively.
It should be noted, however, that such quasiparticle excitations
need not correspond to simple poles on the unphysical sheet.
There could instead be branch points, branch singularities,
essential singularities, or even no singularities at all
\cite{Weldon:2003}.

Moreover,
in the case of gauge theories it is a priori not clear whether
the thermal corrections to the various propagators encode physical
information or not. In Abelian gauge theory, the photon propagator
is linearly related to correlators of the gauge-invariant electromagnetic
field strength and so has a direct physical interpretation; matter
fields on the other hand already transform non-trivially
under gauge transformation and indeed their propagator is a
gauge-fixing dependent quantity. In non-Abelian gauge theories, the
gauge bosons carry colour charge and their propagator is also
gauge dependent. Correlators of the non-Abelian field strength are
not gauge independent either.

Still, even in gauge-dependent quantities there may be gauge-independent
information. Indeed, in \cite{Kobes:1990xf,Kobes:1991dc} (see also
\cite{Rebhan:2001wt} for a more detailed recent review)
it has been shown that the
singularity structure (location of poles and branch singularities)
of certain components of gauge and matter
propagators are gauge independent when all contributions to
a given order of a systematic expansion scheme are taken into account.

Another example for gauge-independent content in
gauge-dependent quantities is provided by the high-temperature limit of 
self-energies and more-point correlation
functions with small external momenta.
Those are related to forward-scattering
amplitudes \cite{Barton:1990fk,Frenkel:1992ts} 
of on-shell plasma constituents and
are therefore completely gauge independent. They can form the building
blocks of an effective theory at soft scales (with respect
to the temperature), as we shall describe further below.
Before doing so, we review the structure of the various
propagators in a gauge theory.

\subsection{Gauge-boson propagator}

The self-energy of gauge bosons is a symmetric tensor, the so-called
polarization tensor, which is defined by
\be
-\Pi^\mn = {G}^{-1\mn}-{G}^{-1\mn}_0.
\ee
(In the real-time formalism, where this should be defined first as
a $2\times2$ matrix relation, we assume that $\Pi$ has been
extracted after diagonalization, for example 
in the $F,\bar F$ basis according to \eref{PidiagF}.)

If the gauge fixing procedure does not break rotational invariance
in the plasma rest frame, this Lorentz tensor can be decomposed in
terms of four independent tensors and associated structure functions,
\be
-\Pi^\mn = \Pi_A A^\mn + \Pi_B B^\mn + \Pi_C C^\mn + \Pi_D D^\mn,
\ee
with, in momentum space, $A^\mn$ being the spatially transverse
tensor \eref{Amn}, and the others chosen as 
\cite{Kapusta:1979fh,Gross:1981br,Kajantie:1985xx,Heinz:1987kz,Landsman:1987uw,Landshoff:1993ag}
\bea
B^\mn(k) &= {\tilde n^\mu \tilde n^\nu \0 \tilde n^2} \equiv
\eta^\mn - {k^\mu k^\nu \0 k^2} - A^\mn(k), \\
C^\mn(k) &= {1\0|{\bf k}|} \left\{ 
\tilde n^\mu k^\nu + k^\mu \tilde n^\nu \right\}, \\
D^\mn(k) &= {k^\mu k^\nu \0 k^2},
\eea
where $\tilde n^\mu = (\eta^{\mu\s}-{k^\mu k^\s / k^2})\d^0_\s$
in the plasma rest frame.

Only $A$ and $B$ are transverse with respect to the four-momentum $k$,
$A^\mn k_\nu=B^\mn k_\nu = 0$. $C$ obeys the weaker relation $C^\mn k_\mu
k_\nu = 0$, and $D$ projects onto $k$. This particular basis is
a convenient choice because $A$, $B$, and $D$ are idempotent
and mutually orthogonal; $C$ is only orthogonal to $A$, but its
product with the other tensors has vanishing trace.

In Abelian gauge theory with linear gauge fixing, the Ward identities
imply transversality of the polarization tensor, $\Pi^\mn k_\nu \equiv 0$.
In non-Abelian gauge theories transversality holds only in 
certain gauges such as axial gauges \cite{Leibbrandt:1987qv}
(with only the temporal
axial gauge respecting rotational invariance) and background-covariant gauges
\cite{
Abbott:1981hw,Abbott:1983zw,Rebhan:1985bg}. 
Contrary to the experience at zero
temperature, at finite temperature the polarization tensor turns
out to be non-transverse already at one-loop order in general covariant
and Coulomb gauges \cite{Heinz:1987kz,Elze:1988rh}, 
with the fortuitous exception
of Feynman gauge (at one-loop order).

For a gauge breaking Lagrangian \eref{gbrL} BRS invariance only requires that
\cite{ItzZ:QFT,Landsman:1987uw,Kobes:1989up,Kunstatter:1992ij}
\be
\tilde f_\mu \tilde f_\nu {G}^\mn = -\alpha\,.
\ee
This entails that \cite{Kobes:1989up,Kunstatter:1992ij}
\be\label{PiD}
\Pi_D \(k^2-\Pi_B\) = \Pi_C^2
\ee
so that generally the polarization tensor contains 3 independent
structure functions. \Eref{PiD} also implies that at one-loop order
$\Pi_D\equiv0$, but not beyond in those gauges where $\Pi_C\not\equiv 0$.

Rotationally invariant gauge fixing vectors can be
written generally as $\tilde f^\mu = \tilde \beta(k) k^\mu+\tilde \gamma(k) \tilde n^\mu$
with $\tilde \beta\not=0$.
This includes covariant gauges ($\tilde \b=1,\tilde \g=0$), Coulomb gauges ($\tilde \b=\tilde n^2,
\tilde \g=-k^0$), or temporal gauges ($\tilde \b=k^0/k^2, \tilde \g=1$).

For these, 
the structure functions in the full propagator
\be
-G^\mn=\D_A A^\mn + \D_B B^\mn + \D_C C^\mn + \D_D D^\mn
\ee
are determined by 
\bea
\D_A &= [ k^2 - \Pi_A ]^{-1} \\
\D_B &= [ k^2 - \Pi_B -
{2\tilde \b\tilde \g |{\bf k}| \Pi_C - \a \Pi_C^2 + \tilde \g^2 {\tilde n}^2 \Pi_D  \0
\tilde \b^2 k^2 - \a \Pi_D } ]^{-1} \\
\D_C &= - {\tilde \b\tilde \g |{\bf k}| - \a \Pi_C \0 \tilde \b^2 k^2 - \a \Pi_D} \D_B \\ \label{DeltaD}
\D_D &= { \tilde \g^2 \tilde n^2 + \a(k^2-\Pi_B) \0 \tilde \b^2 k^2 - \a \Pi_D } \D_B
\eea

\subsubsection{Gauge independence of singularities}\label{gindproof}

In \cite{Kobes:1990xf,Kobes:1991dc} 
it has been shown that under variations of the gauge
fixing parameters (in our case $\a,\tilde \b,\tilde \g$) one has
``gauge dependence identities'' (generalized Nielsen identities 
\cite{Nielsen:1975fs,Aitchison:1984ns,Johnston:1987ib,Fukuda:1976di})
which are of the form
\bea\fl
\d \D_A^{-1}(k) &= { \Delta_A^{-1}}\Bigl[- A^\mu_\nu(k)\delta
X^\nu_{,\mu}(k)\Bigr] \equiv \D_A^{-1}(k) \d Y(k),
\label{dDA}\\ \fl
\d \D_B^{-1}(k) &= { \Delta_B^{-1}}
\left[-{\tilde n^\mu\over \tilde n^2}+
{\tilde\gamma\tilde\beta-\alpha \Pi_C/|\vec k| \over 
\tilde\beta^2 k^2 -\alpha \Pi_D}k^\mu\right]2\tilde n_\nu \delta X^\nu_{,\mu}
\equiv\D_B^{-1}(k) \d Z(k),\label{dDB} 
\eea
where $\delta X^\nu_{,\mu}$ has a diagrammatic expansion which
is one-particle-irreducible except for at most one Faddeev-Popov ghost
line. No such relation exists for $\Delta_C$ or $\Delta_D$.

Now if $\delta Y$ and $\delta Z$ are regular on the two ``mass-shells''
defined by $\Delta_A^{-1}=0$ and $\Delta_B^{-1}=0$, the
relations (\ref{dDA},\ref{dDB}) imply that the locations of these particular
singularities of the gluon propagator are gauge fixing independent,
for if $\Delta_A^{-1}=0=\Delta_B^{-1}$ then also
$\Delta_A^{-1}+\delta\Delta_A^{-1}=0=\Delta_B^{-1}+\delta\Delta_B^{-1}$.

In the case of $\Delta_B$, singularities in $\delta Z(k)$
include a kinematical pole $1/k^2$ hidden in the $\tilde n$'s
and the manifestly gauge-dependent 
$\Delta_D$ (cf.\ (\ref{DeltaD})).
Excluding these obvious gauge artefacts, everything depends on whether 
the possible singularities of $\delta X^\nu_{,\mu}$
could coincide with the expectedly physical dispersion laws
$\Delta_A^{-1}=0$ and $\Delta_B^{-1}=0$. Because
$\delta X^\nu_{,\mu}$ is one-particle reducible with respect
to Faddeev-Popov ghosts, the singularities of the latter have
to be excluded, too. However, these are generically different
from those that define the spatially transverse and longitudinal
gauge-boson quasi-particles. Indeed, in leading-order thermal
perturbation theory the Faddeev-Popov ghost self energy
and the physical self energies
receive contributions carrying different powers of temperature
or chemical potential and therefore
have independent and generically different dispersion laws.

However, $\delta X^\nu_{,\mu}$ may develop singularities also
from one-particle-irreducible subdiagrams, namely when one line
of such a diagram is of the same type as the external one
and the remaining ones are massless. This can give rise
to infrared or mass-shell singularities and seemingly
constitute an obstruction to the gauge-independence proof
\cite{Baier:1992dy,Baier:1992mg}.
But such singularities will be absent as soon as
an overall infrared cut-off is introduced, for example
by restricting everything to a finite volume first. In every finite volume,
this obstruction to the gauge-independence proof is then
avoided, and $\Delta_A^{-1}=0$ and $\Delta_B^{-1}=0$ define
gauge-independent dispersion laws if the infinite-volume limit
is taken last of all \cite{Rebhan:1992ak}. 


\subsubsection{Gauge-field quasi-particles at leading order}
\label{sec:gfqplo}

It turns out that the leading-order contributions to $\Pi^\mn$ for
small frequencies and momenta $k^0,|{\bf k}|\ll T$ are entirely
transverse ($\Pi_C=\Pi_D=0$) and
gauge-parameter independent. They are generated by one-loop
diagrams with hard loop momentum and for this reason they are
termed ``hard-thermal-loops'' (HTL) \cite{Braaten:1990mz}. 
Their form is the same in Abelian \cite{Silin:1960,Fradkin:1965} and
non-Abelian gauge theories \cite{Kalashnikov:1980cy,Weldon:1982aq}, 
and also in the presence of chemical potentials $\mu_f$
for fermions such that ${\rm max}(T,\mu_f) \gg k^0,|{\bf k}|$.
This universal result reads
\bea
\label{PiA} \hat\Pi_A &= \2 (\hat\Pi_\mu{}{}^\mu - \hat\Pi_B) \\
\label{PiB} \hat\Pi_B &= -{k^2\0{\bf k}^2} \hat\Pi_{00}
\eea
with
\bea\label{PiHTL}
\hat\Pi{}_\mu{}{}^\mu=\hat m_D^2,\quad
\hat\Pi_{00}=\hat m_D^2
\(1-{k^0\02|{\bf k}|}\ln{k^0+|{\bf k}|\0k^0-|{\bf k}|}\),\\
\label{mD}\fl \hat m_D^2=\left\{
\begin{array}{ll}
{e^2T^2/3}+{e^2\mu_e^2/\pi^2} & \textrm{for QED,}\\
g^2NT^2/3+\sum_f {g^2\mu_f^2/(2\pi^2)} & \textrm{for
SU($N$) with $N_f$ flavours.}
\end{array} \right.
\eea
The spectral representation of the resulting propagators $\Delta_{A,B}$
involves simple poles and continuous parts for $k^2<0$, which
are given in detail in appendix \ref{App:A}.

\begin{figure}                                            
     \centerline{ \includegraphics[viewport =                           
-40 155 530 530,scale=0.5]{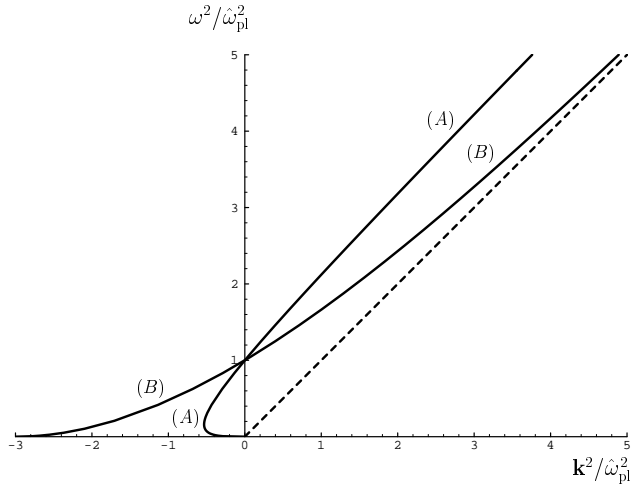} }                    
\caption{Location of poles in $\D_A$ and $\D_B$ of the
hard-thermal-loop gauge propagator. The right part with
$\mathbf k^2\ge0$ corresponds to propagating normal modes,
the left part to (dynamical) screening.} \label{figg}
\end{figure}

In \fref{figg} the location of the poles of the corresponding
propagators $\D_A$ and $\D_B$ are displayed in quadratic scales.
The light-cone is marked by a dashed line. A simple mass hyperboloid
would be given by a line parallel to the latter. Evidently,
the dispersion laws of the HTL quasiparticle excitations are
more complicated---they involve momentum-dependent masses: in
the long-wavelength limit $|\mathbf k|\to0$ 
there is a common lowest (plasma) frequency
$\hat\om_{\rm pl.}=\hat m_D/\sqrt3$ for propagating normal modes. 

For larger frequencies and momenta 
$\w,|\mathbf k| \gg \hat m_D$ it turns out that
mode $A$ approaches asymptotically a mass hyperboloid with
mass $m_\infty=\hat m_D/\sqrt2$. So in this momentum region the physical
spatially transverse polarizations of the gauge bosons acquire
indeed a constant thermal mass. The additional mode $B$, whose
dispersion curve approaches the light-cone exponentially, is found
to have a spectral strength (the residue of the
corresponding pole) that decays exponentially as $k/\hat m_D\to\infty$ 
\cite{Pisarski:1989cs}, showing its exclusively collective nature.

For real $|\mathbf k|$ 
but $\omega^2<\mathbf k^2$, 
$\hat\Pi_\mn$ has an imaginary part $\sim \hat m_D^2$
from the logarithm in \eref{PiHTL}
which prevents the appearance of poles in this region. This
imaginary part corresponds to the possibility of Landau damping,
which is the transfer of energy from soft fields to
hard plasma constituents moving in phase with the field 
\cite{LifP:PK,Blaizot:2001nr}
and is an important part of the spectral density of HTL
propagators. At higher, subleading orders of perturbation theory, it
is, however, not protected against gauge dependences
in nonabelian gauge theories, in contrast to the location
of the singularities which determine the dispersion laws
of quasi-particles. However, at asymptotically large times,
Landau damping is (generically) dominated by the gauge-independent
location of
the branch cuts at $\omega=\pm|\mathbf k|$, resulting in
power-law relaxation of perturbations \cite{Boyanovsky:1998pg}.
There is also an exponential component of Landau damping due to
a pole at purely imaginary $\omega$ and real $|\mathbf k|$,
on the unphysical sheet reached by continuation through the
branch cut between $\omega=\pm|\mathbf k|$
\cite{Rajantie:1999mp}.

For real $\om<\hat\om_{\rm pl.}$,
there are no poles for real $|\mathbf k|$, 
but instead for imaginary $\sqrt{\mathbf k^2}=i\kappa$,
corresponding to exponential (dynamical) screening
of (time-dependent) external sources. These poles,
displayed on the left part of figure \ref{figg}, are in
fact closely related
to the just mentioned poles 
at purely imaginary $\omega$ and real $|\mathbf k|$.

In the static limit $\omega\to 0$,
only mode $B$ is screened with (Debye) screening length $\hat m_D^{-1}$.
This corresponds to exponential screening of (chromo-) electrostatic
fields. 
The transverse mode $A$ on the other hand is only weakly screened
when $\omega \ll m_D$ with a frequency-dependent inverse screening length
\cite{Weldon:1982aq}
\be\label{dynscr}
\kappa_A \simeq \left( \pi m_D^2 \omega\04 \right)^{1/3},
\qquad \omega\ll m_D.
\ee

When $\omega\to0$, mode $A$ describes magnetostatic fields
which are found to be completely unscreened in the HTL approximation. 
In QED the absence of magnetic
screening is intuitively clear and can be proved rigorously to
all orders of perturbation theory \cite{Fradkin:1965,Blaizot:1995kg},
but not in the nonabelian case. 

In QCD, the absence of a magnetostatic screening mass causes
problems for perturbation theory as will be discussed further
in Sect.~\ref{sectmm}.
In fact, lattice simulations
of gauge fixed propagators in nonabelian theories do find
a screening behaviour in the transverse sector, 
though the corresponding singularity is evidently quite
different from a simple pole \cite{Cucchieri:2001tw}.

\subsection{Fermions}

The fermion self-energy at non-zero temperature or density has
one more structure function than usually. In the ultrarelativistic
limit where masses can be neglected, it can be parametrized by
\be\label{Sigma1}
\Sigma(\omega, {\bf  k})\,=\,a(\omega, |\mathbf k|)\,\gamma^0\,+\,b(\omega, |\mathbf k|)
{\hat{\bf  k}}\cdot{\bgamma}.
\ee
(For a massive fermion, this would also include a mass correction,
i.e., $\Sigma = a\gamma^0+b\,
{\hat{\bf  k}}\cdot{\bgamma}+c \mathbf 1$.)
This can be rewritten as:
\be\label{SIGL}
\gamma_0\Sigma(\omega, {\bf  k})\,=\,\Sigma_+(\omega,|\mathbf k|)\,
\Lambda_+(\hat {\bf k})\,-\,\Sigma_-(\omega,|\mathbf k|)\,
\Lambda_-(\hat {\bf k}),
\ee
where $\Sigma_\pm\equiv b \pm a$,
and the spin matrices 
\be
\Lambda_{\pm}(\hat {\bf k})\equiv \frac{1 \pm \gamma^0
 \bgamma\cdot\hat{\bf k}}{2}
\ee
project onto spinors whose chirality is equal ($\Lambda_+$),
or opposite ($\Lambda_-$), to their  helicity.
Dyson's equation $S^{-1}= -{\not\! k} + \Sigma$ then implies
\be\label{SINV}
\gamma_0 S^{-1}=\Delta_+^{-1}
\Lambda_+ \,+\,\Delta_-^{-1}\Lambda_-,
\ee
with $\Delta_\pm^{-1} \equiv - [\omega\mp(|\mathbf k|+\Sigma_\pm)]$.
This is trivially inverted to yield the fermion propagator
\be\label{SFL}
S\gamma_0=
\Delta_+\Lambda_+ +\Delta_-\Lambda_-.
\ee
whose singularities 
are conveniently
summarized by the equation
\be
\det S^{-1}(\omega, {\bf  k}) = 0,
\ee
where the determinant refers to spinor indices.
The potential gauge dependences 
of the singularities of $S$
are described by an identity
of the form \cite{Kobes:1991dc} 
\be
\delta\det S^{-1}(\omega, {\bf  k})=\det S^{-1}(\omega, {\bf  k})\;
\delta\, {\rm tr}\, X(\omega, {\bf  k}),
\ee
where $\delta\, {\rm tr}\, X(\omega, {\bf  k})$ again
has a diagrammatic expansion which is one-particle-irreducible
except for at most one Faddeev-Popov ghost line. The same reasoning
as in Sect.~\ref{gindproof} (with similar qualifications)
leads to the conclusion that
the positions of the singularities of the fermion propagator
are gauge-fixing independent \cite{Kobes:1991dc}.


In the HTL approximation, the fermion self-energies are once again
gauge-independent in their entirety. Explicitly, they
read \cite{Klimov:1981ka,Klimov:1982bv,Weldon:1982bn,Weldon:1989bg,Pisarski:1989wb,Weldon:1989ys}:
\be\label{SIGHTL}
\hat\Sigma_\pm(\omega,|\mathbf k|)\,=\,{\hat M^2\0k}\,\left(1\,-\,
\frac{\omega\mp |\mathbf k|}{2|\mathbf k|}\,\log\,\frac{\omega + |\mathbf k|}{\omega - |\mathbf k|}
\right),\ee
where $\hat M^2$ is the plasma frequency for fermions,
i.e., the frequency of long-wavelength ($k\to 0$) fermionic
excitations:
\be\label{MF}
\hat M^2 
={g^2 C_f\08}\left(T^2+{\mu^2\0\pi^2}\right).\ee
($C_f=(N^2-1)/2N$ in SU($N$) gauge theory, and $g^2C_f\to e^2$ in QED.)

For frequencies $\om<\hat M$, there are, in contrast to the gauge
boson propagator, no solutions with imaginary wave-vectors that
would correspond to screening. Instead, the additional collective
$(-)$ (occasionally dubbed ``plasmino'' \cite{Braaten:1991hg})
branch 
exhibits propagating modes down to 
$\om^{(-)}_{\rm min.} \approx 0.928 
\hat M$
(at $|\mathbf k|_{\rm dip}\approx 0.408 
\hat M$), with a curious dip in the dispersion
curve reminiscent of that of rotons in liquid helium \cite{Feynman:1956}.

For momenta $|\mathbf k|\gg \hat M$,
the normal (+) branch of the poles of the fermion propagator approaches
asymptotically a mass hyperboloid with mass $M_\infty=\sqrt 2 \hat M$
and unit residue, whereas the $(-)$ branch tends to the light-cone
exponentially, with exponentially vanishing residue.

\begin{figure}                                            
     \centerline{ \includegraphics[viewport =                           
0 140 540 600,scale=0.43]{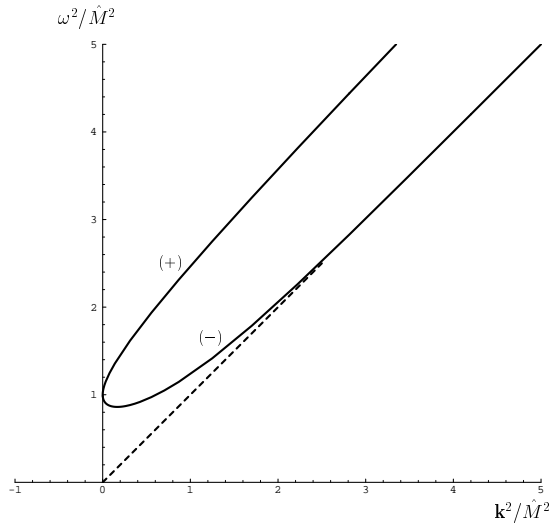} }                    
\caption{Location of poles in the
hard-thermal-loop fermion propagator}\label{figf}
\end{figure}

For space-like momenta, $\omega^2<\mathbf k^2$, there is again a large
imaginary part $\propto \hat M^2$ corresponding to fermionic
Landau damping, which now reflects the possibility of turning
hard fermionic (bosonic) plasma constituents into hard
bosonic (fermionic) ones in the presence of soft fermionic
fields. Beyond lowest (HTL) order, this imaginary part
cannot be expected to be gauge-fixing independent, however, not even
in Abelian theories.

Another case where gauge dependences arise is when the
bare fermion mass cannot be neglected compared to temperature.
Then the HTL approximation is no longer adequate, and
one has to include more than one-loop diagrams for
consistency. At one-loop order, the modifications
of the fermionic dispersion laws have been studied in 
references~\cite{Petitgirard:1992mf,Baym:1992eu}, with the
result that the additional $(-)$ branch disappears gradually
when $m/T \gtrsim 1/3$. This phenomenon has been obtained
both in Coulomb and Feynman gauge, with only weak gauge
dependences in the real parts of the fermion self-energy,
so it is not to be expected that higher-loop corrections
would change this result qualitatively.

There have been also investigations of how the HTL dispersion laws
of fermions as well as gauge bosons
are affected by retaining non-leading powers of temperature
\cite{Peshier:1998dy,Mitra:1999wz} and chemical potential
\cite{Levinson:1985ub,Blaizot:1993bb,OlivaAguero:1998ky,Kalashnikov:1998bj}
which also emphasize the need for higher-loop contributions
to obtain complete, gauge-independent results.

The very existence of the additional $(-)$ branch in the
high-temperature limit has been
confirmed in references~\cite{Peshier:1999dt,Weldon:1999th} 
under rather weak assumption without
use of perturbation theory. 
Possible experimental signatures of the additional fermionic
quasi-particles are the Van Hove singularities in dilepton
production from a quark-gluon plasma
\cite{Braaten:1990wp,Wong:1992be,Peshier:1999dt}
(though smeared out by damping effects to an unknown extent);  
they also play a role in the calculation of the electroweak
baryon genesis of reference~\cite{Farrar:1994hn}
but it has been shown \cite{Gavela:1994dt,Huet:1995jb}
that the fermion damping discussed in
section~\ref{subsecdds} completely swamps their effects in the
standard model (but may be still important in extensions thereof
\cite{Cline:1996dg}). 

\subsection{Diquark condensates and colour superconductivity}
\label{secCSC}

At sufficiently low temperature and high quark chemical potential,
QCD will in fact be in a colour superconducting phase which
modifies the fermion propagator by the appearance of
a diquark condensate. The reason is that 
even an arbitrarily
weak attractive interaction at the Fermi surface leads to
the appearance of Cooper pairs \cite{AbrGD:QFTStPh,FetW:Q},
which form a Bose condensate and give rise to an energy gap leading
to superfluidity, or, in the presence of a gauge symmetry
which is spontaneously broken by the condensate, to
superconductivity.
Whereas in conventional superconductors, the only attractive
interaction is from phonon-mediated interactions, in QCD
already one-gluon exchange in the colour-antitriplet channel
is attractive, so {\em colour}-superconductivity should be
a much less fragile phenomenon with correspondingly large
energy gaps. This was first studied by Barrois \cite{Barrois:1977xd}
in the late seventies and more extensively by Bailin and
Love \cite{Bailin:1984bm} and others \cite{Donoghue:1988ec,Iwasaki:1995ij}.

Over the last few years there has been a renewed flurry of activity 
in this field following
the observation by various groups \cite{Alford:1997zt,Rapp:1998zu} 
that the energy
gap of colour superconductors as well as their
critical temperature may be much larger than previously expected.
Moreover, new symmetry breaking schemes were discovered.
Whereas it was previously \cite{Bailin:1984bm} thought that
only the lightest quark flavours $u$ and $d$ are able to
form Cooper pairs, it was argued in reference~\cite{Alford:1997zt}
and corroborated in reference~\cite{Schafer:1998ef} that a
condensate which locks the breaking of colour and flavour
symmetry (``colour-flavour locking'' (CFL))
is energetically favoured, at least for high
baryon chemical potential. 
In astrophysical applications,
where in contrast to heavy-ion physics the strange-quark
chemical potential is constrained by electric neutrality,
CFL appears to be even more favoured than two-flavour superconductors (2SC)
according to the estimates of 
\cite{Alford:2002kj,Steiner:2002gx}. Alternatively, electric
neutrality may be realized with the so-called gapless 2SC
phase \cite{Shovkovy:2003uu,Huang:2003xd}.

Close to the phase transition to confined nuclear matter,
there is also the possibility for rather involved symmetry
breaking patterns. A particularly fascinating one is the so-called
LOFF phase \cite{Larkin:1964,Fulde:1964,Alford:2000ze}, 
which has a 
gap function that breaks translational invariance,
leading to crystalline structures in a colour superconductor
with possible relevance to neutron star physics \cite{Alford:2002wf}.

Most of the original work has been based on a QCD inspired
phenomenological Nambu-Jona-Lasinio (NJL) model \cite{Nambu:1961tp}.
Starting with the work of Son \cite{Son:1998uk}, the
phenomenon of colour superconductivity has also been investigated
by systematic perturbative techniques based on the fundamental
Lagrangian of QCD. Son established, using renormalization group
techniques, that in QCD the parametric form of the energy gap
is not of the order of $b\mu \exp(-c/g^2)$ as with
point-like interactions, but is modified by long-range
colour magnetic interactions to the parametrically larger
order of $b_1\mu g^{-5}\exp(-c_1/g)$, with $c_1=3\pi^2/\sqrt2$.

Using weak-coupling methods, $b_1$ has been calculated
in \cite{Schafer:1999jg,Pisarski:1999bf,Pisarski:1999tv,Brown:1999aq,Brown:2000eh,Wang:2001aq}, 
and gauge parameter independence of $b_1$
has been verified in \cite{Pisarski:2001af} in
Coulomb-like gauges, though covariant gauges
present a problem \cite{Hong:1998tn,Shovkovy:1999mr,Hong:1999fh,Hong:2003ts}.

The gauge independence identities of the previous section are not
immediately applicable, but require a Nambu-Gor'kov ansatz
\cite{Bailin:1984bm,Wang:2001aq,Manuel:2000nh} for the inverse
propagator of the form
\begin{equation}
  \mathcal S^{-1}=\left(
    \begin{array}{cc} q\!\!\!/+\mu\gamma_0
    +\Sigma & \Phi^- \\
    \Phi^+ & q\!\!\!/-\mu\gamma_0+\bar\Sigma 
    \end{array}
  \right), \label{e1}
\end{equation}
where $\Phi^\pm$ are the gap functions, related by
$\Phi^-(q)=\gamma_0 [\Phi^+(q)]^\dag\gamma_0$, and 
$\bar\Sigma(q)=C[\Sigma(-q)]^T C^{-1}$ with the charge 
conjugation matrix $C$.
Flavour and fundamental colour indices are suppressed in (\ref{e1}). 
This inverse propagator
is the momentum space version of the second derivative of
the effective action,
\begin{equation}
  {\delta^2\Gamma\over\delta\bar\Psi(x)\delta\Psi(y)}\Big|_{\psi=\bar\psi=A_i^a=0,
  A_0^a=\tilde A_0^a},
  \label{e2}
\end{equation}
where $\Psi=(\psi, \psi_c)^T$, $\bar\Psi=(\bar\psi, \bar\psi_c)$, and $\tilde A_0^a$
is the expectation value of $A_0^a$, which is generally
nonvanishing in the colour superconducting phase \cite{Gerhold:2003js}.
The doubling of fermionic fields in
terms of $\Psi$ and $\bar\Psi$ is just a notational convenience here;
the effective action itself should be viewed as depending only
on either $(\psi,\bar \psi)$ or the set $\Psi=(\psi, \psi_c)^T$.

Assuming a spatially homogeneous quark condensate,
one can then derive a gauge dependence identity for
the momentum-space propagator of the form 
\cite{Gerhold:2003js}
\begin{equation}\fl
  \delta\det( S^{-1}_{i\bar j})+\delta\tilde A^{a0}
  {\partial\over\partial \tilde A^{a0}}\det( S^{-1}_{i\bar j})
\equiv \delta_{\rm tot}\det( S^{-1}_{i\bar j})
  =-\det( S^{-1}_{i\bar j})[\delta X^k_{\,,k}
  +\delta X^{\bar k}_{\,,\bar k}], \label{e8}
\end{equation}
where the indices $i$ and $\bar i$ comprise colour, flavour, 
Dirac and Nambu-Gor'kov indices.
Just like in the case of ordinary spontaneous symmetry breaking,
one has to consider a total variation \cite{Nielsen:1975fs,Aitchison:1984ns} 
of the determinant of the inverse
quark propagator, with the first term
corresponding to the explicit variation of the gauge fixing function, 
and the second term coming from the
gauge dependence of $\tilde A_0^a$.

Since the determinant is equal to the product of the eigenvalues, equation (\ref{e8}) implies 
that the location of the singularities of the quark propagator 
is gauge independent, provided the singularities of $\delta X^k_{\,,k}$ do not coincide
with those of the quark propagator. As above, one may argue that $\delta X$ is
1PI up to a full ghost propagator, and up to gluon tadpole insertions,
and the singularities of the ghost propagator are not
correlated to the singularities of the quark propagator. 
Gauge independence of the zeros of the inverse fermion propagator
then follows provided that
also the 1PI parts of $\delta X$ have
no singularities coinciding with the singularities of the propagator.

At leading order, 
when the quark self energy can be neglected, 
this implies that the gap function is gauge independent
on the quasiparticle mass shell (though at higher orders
it becomes necessary to consider the complete dispersion relations
of the quasiparticles).

\section{Hard-thermal-loop effective action and resummation}
\label{secHTLres}

As we have seen in several examples now, complete weak-coupling expansions
in thermal field theory tend to require a reorganization of
the standard loop expansion.
Already the perturbative expansion of the thermal
mass \eref{msc} of a scalar quasi-particle in Sect.~\ref{sectscpth}
has shown that ordinary perturbation theory fails to
determine higher-order corrections, but runs into infra-red
problems. A resummation of the leading-order (HTL) mass
is necessary (and sufficient in this case) to reorganize
the perturbation series \cite{Parwani:1992gq}.
A thermal mass $\propto gT$ introduces an additional (soft
for $g\ll1$) mass scale,
and whenever loop calculations receive important contributions
from this scale, it is clearly mandatory to use propagators
dressed by these masses.

In the general case, it is however equally important to include
vertex corrections. In gauge theories, this is only natural
as Ward identities tie up vertex functions with self-energies. 
But regardless of gauge symmetry considerations,
if there are contributions to $N$-point one-loop vertex functions
that are proportional to $T^2$ like the HTL self-energies,
they are as important as bare vertices when the external
momentum scale is $\sim gT$. One then has
\be
\quad { \Gamma_{,N}^{\rm HTL}} \sim g^N { T^2} { k^{2-N}}
\sim g^{N-2} { k^{4-N}} 
\sim {\partial ^N\mathcal L_{\rm cl}\over \partial A^N}\Big|_{ k\sim gT},
\ee
for bosonic fields $A$.

A one-loop vertex function whose leading contribution
(for soft external momenta) is proportional to a power of
temperature greater than one is called HTL,
as this is again dominated by a hard loop momentum $\sim T$.
In fact, already in spinor QED there are infinitely many
HTL vertex functions, namely those involving two external
fermion lines and an arbitrary number of gauge bosons.
In QCD, there are in addition HTL's with an arbitrary number
of external gluons. These have been first identified in
references~\cite{Frenkel:1990br,Braaten:1990mz,Braaten:1990az} and used to
set up a resummation programme for amplitudes involving
soft external momenta \cite{Braaten:1990mz,Braaten:1990kk}.

\subsection{HTL effective action}

In the case of scalar $\phi^4$ theory, the only HTL is a
mass term corresponding to a local HTL effective Lagrangian
$\mathcal L^{\rm HTL}_{\rm scalar}=-\2\hat m^2_{\rm th}\phi^2$
with $\hat m^2_{\rm th}$ the leading-order term from \eref{msc}.
Remarkably, the infinitely many HTL diagrams of gauge theories
have a comparatively simple and manifestly gauge-invariant
integral representation 
\cite{Taylor:1990ia,Braaten:1992gm,Frenkel:1992ts,Blaizot:1994be}
\bea\fl
\mathcal L^{\rm HTL} &= 
\mathcal L^{\rm HTL}_f+ \mathcal L^{\rm HTL}_g \nonumber\\ \fl
&= \hat M^2 \int {d\Omega_{\9v}\over 4\pi} 
\bar \psi \gamma^\mu
{{ v_\mu}\over {i v}\cdot D(A)} \psi 
 +{\hat m_D^2\over 2}
{\rm tr} \int {d\Omega_{\9v}\over 4\pi} F^{\mu\alpha}
{{ v_\alpha} { v^\beta} \over  ({ v}\cdot D_{adj.}(A))^2} F_{\mu\beta}
\label{HTLeffL}
\eea
where
$v=(1,\9v)$ is a light-like 4-vector, i.e.\ with $ \9v^2=1$, 
and its
spatial components are averaged over by $\int d\Omega_{\9v}\cdots$. 
Here $v$ is the remnant of the hard plasma constituents'
momenta $p^\mu \sim T v^\mu$, namely their light-like
4-velocity, and the overall scale $T$ has 
combined with the coupling constant to form the scale of the
thermal masses, $\hat M, \hat m_D\sim gT$.

The covariant derivatives in the denominators of (\ref{HTLeffL}) 
are responsible for
the fact that there are infinitely many HTL's. Because
in QED one has $D_{adj.}(A)\to\partial$, the only HTL
with exclusively photons as external lines is the
photon self-energy polarization tensor; the other HTL diagrams
of QED have two external fermion lines and an arbitrary number
of photon insertions.

The gauge boson part of \eref{HTLeffL} has in fact been obtained
originally in a form which is not obviously gauge invariant,
namely
\be
\mathcal L^{\rm HTL}_g=\hat m_D^2 \tr\left\{
A_0^2 + \int {d\Omega_{\9v}\04\pi} \mathcal W(v\cdot A) \right\}
\label{HTLeffLTW}
\ee
where the functional $W$
is determined by being a gauge invariant completion of the
Debye mass term \cite{Taylor:1990ia} as
\be\fl
\mathcal W(v\cdot A)=\6_0(v\cdot A) F({1\0v\cdot\6}[iv\cdot A,*])
{1\0v\cdot\6} v\cdot A,\quad F(z)=2\sum_{n=0}^\infty {z^n\0n+2}.
\ee
This has an interesting
interpretation as eikonal of a Chern-Simons gauge
theory \cite{Efraty:1992gk,Efraty:1993pd,Nair:1993rx},
where however parity violations and a quantization of the coefficient
of the action are absent because of the angular average.

Explicit representations of HTL vertices for gauge bosons are in fact
most efficiently obtained from expanding \eref{HTLeffLTW}
rather than the manifestly gauge invariant form \eref{HTLeffL}.
The gauge boson self energy \eref{PiHTL} is then found
to be given by
\be
\hat \Pi_{\mu\nu}(k)=\hat m_D^2 \left[
g_{\mu 0}g_{\nu 0}-k_0 \int {d\Omega_{\9v}\04\pi} {v_\mu v_\nu\0v\cdot k}
\right]
\ee
and an $n$-point vertex function by
\bea\fl
&\hat\Gamma^{a_1\cdots a_n}_{\mu_1\cdots \mu_n}(k_1,\cdots,k_n)
= 2g^{n-2}\hat m_D^2 \int {d\Omega_{\9v}\04\pi} v_{\mu_1} \cdots v_{\mu_n}
\biggl\{ \tr(T^{a_n}[T^{a_{n-1}},[\ldots,T^{a_1}]\cdots])\nonumber\\
\fl &\times
{k_1^0\0v\cdot k_1}{1\0v\cdot(k_1+k_2)}\cdots {1\0v\cdot(k_1+\cdots+k_{n-2})}
+{\rm permutations}(1,\ldots,n-1) \biggr\}
\eea

The effective Lagrangian \eref{HTLeffL} can be understood as an
effective field theory in the sense of Wilson's renormalization
group \cite{Braaten:1991vp}. It arises from integrating out, in leading order,
the effects of the hard momentum modes of the plasma
constituents. Because, unlike the case of effective
field theories in vacuum field theory, these
particles are real rather than virtual, they give rise
to non-localities corresponding to their free, light-like propagation.
The HTL vertex functions encoded
by \eref{HTLeffL} or \eref{HTLeffLTW} 
can thus be understood as forward scattering amplitudes
for hard (collisionless) particles in soft external fields
\cite{Frenkel:1992ts},
thereby explaining their gauge independence which is
less obvious on a purely diagrammatic level (there it
can be understood through the absence of HTL ghost self-energy
and vertex functions \cite{Kobes:1991dc}).

The action provided by
\eref{HTLeffL} or \eref{HTLeffLTW} is Hermitean only in a Euclidean form.
After analytic continuation there are cuts corresponding
to Landau damping, as mentioned above in connection with
the two-point functions. In order to obtain 
the analytic continuation relevant for linear response
theory, this is better performed on the effective
equations of motions \cite{Jackiw:1993zr}, which are also the primary
objects in kinetic theory. 

\subsection{Kinetic theory approach}

While a kinetic-theory derivation of the HTL propagators
generalizing the Abelian case \cite{Silin:1960} to
nonabelian Yang-Mills theory has been given already in
references~\cite{Heinz:1983nx,Heinz:1985yq,Heinz:1986qe,Elze:1989un}, 
a systematic treatment that includes
the HTL vertices has been developed only in references
\cite{Blaizot:1993gn,Blaizot:1993zk,Blaizot:1994be,Kelly:1994ig,Blaizot:1995nr,Kelly:1994dh} (see references~\cite{Blaizot:2001nr,Litim:2001db} for two recent
comprehensive reviews). In a pure-glue nonabelian gauge theory,
the effective equations of motion are given by
\bea\label{DFjW}
&[D_\mu,F^{\mu\nu}]^a= j^{\nu a}=
\hat m_D^2 \int {d\Omega_{\9v}\over 4\pi} v^\nu W^a(x,\9v),\\
\label{cDWvE}
&[v\cdot D,W(x,\9v)]^a=\9v\cdot \9E^a(x),
\eea
where $\9E$ is the chromo-electric field strength, and $W^a(x,\9x)$
describes the fluctuations of the phase space density of hard gluons.
At the expense of introducing $W^a(x,\9x)$ as a new soft degree
of freedom, this formulation permits a completely local
description of the physics of soft gauge boson modes
\cite{Nair:1994xs,Iancu:1998sg}.

The HTL diagrams can be constructed from solving
\eref{DFjW} in terms of $j_\mu^a[A]$ and formally expanding
\be\label{jindexp}
j_\mu^a=-\hat\Pi_{\mu\nu}^{ab} A_b^\nu+{1\02}\hat\Gamma^{abc}_{\mu\nu\rho}
A_b^\nu A_c^\rho + \ldots.
\ee 

The kinetic-theory approach allows also to consider strong deviations
from equilibrium. In references 
\cite{Mrowczynski:2000fp,Romatschke:2003ms,Birse:2003qp} 
the analogue of the HTL self-energies with anisotropic momentum
distributions have been considered, which turn out to lead to
space-like singularities in the gauge boson propagator. The latter are
related to plasma instabilities which could play an important
role in thermalization issues \cite{Randrup:2003cw,Arnold:2003rq}.


\subsection{HTL/HDL resummation}

Since for soft momenta $\sim gT$ HTL self-energies and vertices
are equally important as the tree-level self-energies and vertices,
the former may not be treated perturbatively, but should
rather be combined with the latter to form effective
self-energies and vertices. This can be done formally by replacing
\be\label{resum}
\mathcal L_{\rm cl} \to \mathcal L_{\rm cl} + \mathcal L^{\rm HTL} 
- \delta\times \mathcal L^{\rm HTL}
\ee
where $\delta$ is a parameter that is sent to 1 in the
end, after the last term has been treated as a `thermal counterterm'
by assuming that $\delta$ counts as a one-loop quantity.

Because $\mathcal L^{\rm HTL}$ has been derived under the
assumption of soft external momenta, this prescription is in
fact only to be followed for soft propagators and vertices
\cite{Braaten:1990mz}.
Propagators and vertices 
involving hard momenta (if present) do not require this resummation,
and in fact for obtaining a systematic expansion in the coupling $g$,
one has to expand out all HTL insertions on hard internal lines.
In practice, a separation between hard and soft scales 
may be implemented by introducing an intermediate
scale $\Lambda$ with $gT \ll \Lambda \ll T$, assuming
$g\ll1$, for example $\Lambda\propto \sqrt{g}T$
(see e.g.\ \cite{Braaten:1991dd,Braaten:1991jj}).

The resulting systematic expansions in $g$ typically involve
single powers in $g$ and logarithms of $g$, in contrast to
conventional perturbation theory which would involve only $g^2$
as an expansion parameter. This is because increasing
the loop order by one involves a factor $g^2T$ which in an ultrarelativistic
situation and soft external momenta $\lesssim gT$
is made dimensionless by a thermal mass $m\sim gT$, so
that the effective expansion parameter becomes $g^2T/m\sim g$.

The same resummation scheme arises in the presence of a
chemical potential. In the effective action,
the chemical potential enters only in the mass parameters
$\hat M^2$ and $\hat m_D^2$ according to \eref{MF} and \eref{mD}. 
For $T\approx0$ but
large $\mu_f$, the HTL have also been nicknamed ``hard dense
loops'' (HDL) \cite{Manuel:1996td}. 
At $T\approx0$,
resummation of the HDL effective action
is necessary for soft momenta $\lesssim g\mu$, but
because
of the absence of Bose enhancement this
does not give rise to single powers of $g$
in the perturbations series, but only to logarithms of $g$
in addition to powers of $g^2$,
as found long ago in the ring resummation scheme of Gell-Mann and Brueckner
for an electron gas at high density \cite{Gell-Mann:1957}.
HDL resummation for dynamic quantities
has been considered first in \cite{Altherr:1992mf}
with applications to energy loss rates in astrophysical systems
through axion-like particles, and for energy loss of
heavy quarks at both large $T$ and $\mu$ in \cite{Vija:1995is}.

HDL resummation is also at the basis of the results 
of reference~\cite{Ipp:2003cj} quoted in
Sect.~\ref{seclowT} on non-Fermi-liquid behaviour of the specific
heat at low temperature and high chemical potential, equation~\eref{anomCV}.
In this case the non-analytic terms in $g^2$ involve a
logarithm of $g$ as well as cubic roots of $g^2$, which
are the result of the only weak dynamical screening
of near-static magnetic modes whose screening
lengths involve cubic roots of $m_D^2\omega$, see \eref{dynscr}.

\subsubsection{HTL-screened perturbation theory (HTLPT)}
\label{secHTLPT}

In
\cite{Andersen:1999fw,Andersen:1999sf,Andersen:2002ey,Andersen:2003zk}
a modification of the above scheme has been
suggested, where the mass parameters
$\hat m_D$ and $\hat M$ within $\mathcal L^{\rm HTL}$
are considered as independent of the coupling $g$
and used for a variational improvement of the perturbative series.
This is a generalization of screened perturbation
theory \cite{Karsch:1997gj,Andersen:2000yj}.
It differs from standard HTL/HDL resummation
also in that \eref{resum} is used
for both hard and soft momenta, which results in additional
UV divergences that need to be subtracted at
any finite order of the expansion.

This method has been used to calculate the thermodynamic potential
to two loop order, where it does
improve the apparent convergence of the perturbative
results, but the result deviates significantly from
lattice results even at the highest temperatures that are available
for the latter.

A problem of this approach, at least in quantities that are
dominated by hard excitations like the pressure, seems to be
that the HTL effective action is not a good approximation at
hard momenta. This is also signalled by the fact that
HTL propagators do not satisfy the relativistic KMS condition
at large momenta \cite{Bros:1996mw}.
While this is taken care of eventually by the
counterterms in \eref{resum}, at any finite order of the expansion
there are uncancelled unphysical hard contributions.
In \cite{Blaizot:2003iq} it has been shown that a (simpler)
implementation of a variational perturbation theory in
dimensional reduction which uses just the Debye mass term
(which is the static limit of the HTL effective action)
avoids this problem and indeed
leads to results which are closer to the lattice result
as well as the higher-order calculations in dimensional reduction
when improved as discussed in section \ref{QCDappconv}.
It therefore appears that HTLPT needs to be amended
such that a different treatment of hard and soft modes
is secured.


\subsubsection{HTL resummed thermodynamics
through $\Phi$-derivable approximations}
\label{secBIR}

While HTLPT when applied to the thermodynamical potential
does not work satisfactorily (though it may be of more use
in quantities which depend more dominantly on soft rather
than hard scales), it turns out that a generalization of
the self-consistent expression for the entropy \eref{Ssc}
allows for a resummation of HTL propagators without the
problems of HTLPT and with remarkably good numerical results
when compared to lattice data
\cite{Blaizot:1999ip,Blaizot:1999ap,Blaizot:2000fc}.\footnote{A
similar approach but formulated directly in terms of
the (modified) thermodynamic potential $\Omega$ has been set up
in \cite{Peshier:2000hx}.}

In gauge theories including fermions, 
the self-consistent two-loop expression for
the entropy \eref{Ssc} reads
\bea
\fl
\label{S2loop}
{\cal S}\!&=&\!-\tr \int{d^4k\0(2\pi)^4}{\6n(\omega)\0\6T} \left[ \Im 
\log D^{-1}(\omega,k)-\Im \Pi(\omega,k) \Re D(\omega,k) \right] \nonumber\\
\fl
&&-2\,\tr \int{d^4k\0(2\pi)^4}{\6f(\omega)\0\6T} \left[ \Im
\log S^{-1}(\omega,k)-\Im \Sigma(\omega,k) \Re S(\omega,k) \right],\;\;\;\;
\eea
and a similarly simple expression can be obtained for the
quark number density
\bea
\fl
\label{N2loop}
{\cal N}\!&=&\!-2\,\tr \int{d^4k\0(2\pi)^4}{\6f(\omega)\0\6\mu} \left[ \Im
\log S^{-1}(\omega,k)-\Im \Sigma(\omega,k) \Re S(\omega,k) \right].\;\;\;\;
\eea
Here $n(\omega)=(e^{\beta\omega}-1)^{-1}$,
$f(\omega)=(e^{\beta(\omega-\mu)}+1)^{-1}$, and 
``tr'' refers to all discrete labels,
including spin, colour and flavour when applicable.

In nonabelian gauge theories, the above expressions have to be augmented by
Fad\-deev-Popov ghost contributions which enter like bosonic fields
but with opposite over-all sign, unless a gauge is used where
the ghosts do not propagate such as in axial gauges.
But because $\Phi$-derivable approximations do not generally
respect gauge invariance,\footnote{For this, one would have
to treat vertices on an equal footing with self-energies, which
is in principle possible using the formalism
developed in references~\cite{Norton:1975bm,Freedman:1977xs,Kleinert:1982ki}.}
the self-consistent two-loop approximation
will not be gauge-fixing independent.
It is in fact not even clear that the corresponding gap equations
(\ref{Pi}) have solutions at all or that one can renormalize
these (nonperturbative) equations,
although nonperturbative renormalizability has been
proven in the scalar case
\cite{vanHees:2001ik,VanHees:2001pf,vanHees:2002bv,Blaizot:2003br}.
Concerning gauge fixing dependences it is at least possible
to show that at a stationary point these enter at
twice the order of the truncation \cite{Arrizabalaga:2002hn}.

In \cite{Blaizot:1999ip,Blaizot:1999ap,Blaizot:2000fc} a
manifestly gauge invariant approximation to full self-consistency
has been proposed which maintains equivalence with
conventional perturbation theory up to order $g^3$, which
is the maximum (perturbative) accuracy of a two-loop $\Phi$-derivable
approximation. For these approximations it will be sufficient
to keep only the two transverse structure functions of the
gluon propagator and to neglect ghosts.

For soft momenta, the appropriate leading order propagator
is the HTL one, and indeed there is no HTL ghost self-energy.

For hard momenta, one can identify the contributions to
(\ref{S2loop}) below
order $g^4$ as those linear in the self-energies,
\bea\label{Shard}\fl
\mathcal S^{\rm hard}&=&\mathcal S_0
+2N_g\int\!\!{d^4k\0(2\pi)^4}\,{\6n\0\6T}\,\Re{\Pi_t}\,
\Im\frac{1}{\omega^2-k^2}\nonumber\\
\fl
&-&\!\!4NN_f\int\!\!{d^4k\0(2\pi)^4}\,{\6f\0\6T}\,\Bigl\{
\Re\Sigma_+\Im\frac{-1}{\omega-k}\,-\,
\Re\Sigma_-\Im\frac{-1}{\omega+k}\Bigr\}
\eea
considering now a gauge theory with $N_g$ gluons
and $N_f$ fermion flavours.
Because the imaginary parts of the free propagators restrict their
contribution to the light-cone, only the light-cone projections
of the self-energies
enter. At order $g^2$ this is exactly 
given by the HTL results, without having to assume soft $\omega,k$
\cite{Kraemmer:1990drA,Flechsig:1996ju}
\bea\label{mas2}
&&\Re\Pi_t^{(2)}(\omega^2=k^2)=\hat\Pi_t(\omega^2=k^2)=
\2\hat m_D^2\equiv m_\infty^2,\\
&&2k \,\Re\Sigma_\pm^{(2)}(\omega=\pm k)=2k \,\hat\Sigma_\pm(\omega=\pm k)=
2 \hat M^2 \equiv M_\infty^2,\label{Mas2}
\eea
and without contributions from the other components of $\Pi_{\mu\nu}$
and the Faddeev-Popov self-energy.

There is no contribution $\propto g^2$ from soft momenta
in (\ref{S2loop}) and (\ref{N2loop}) so that one is left with
remarkably simple general formulae for the leading-order interaction
contributions to the thermodynamic potentials expressed
through the asymptotic thermal masses of the bosonic and
fermionic quasiparticles:
\be\label{SNBF2}\fl
{\cal S}^{(2)} =-  T\left\{\sum_B { m_{\infty\,B}^2 \0 12}\,+\,
\sum_F { M_{\infty\,F}^2\0 24}\right\},\quad
{\cal N}^{(2)}=-\,{1\0 8\pi^2}\sum_F \mu_F M_{\infty\,F}^2.\ee
Here the sums run over all the bosonic ($B$) and fermionic ($F$)
degrees of freedom (e.g. 4 for each Dirac fermion), 
which are allowed to have different asymptotic
masses and, in the case of fermions, different chemical potentials.

The result \eref{SNBF2} also makes it clear that relative-order-$g$
corrections to $m_\infty^2$, $M_\infty^2 \sim g^2 T^2$
will contribute to the order-$g^3$ terms in $\mathcal S$ and $\mathcal N$.

The HTL approximation to $\mathcal S$ and $\mathcal N$ thus
includes correctly the leading-order interaction term $\propto g^2$
and only part of the order-$g^3$ terms. Using the
peculiar sum rule \eref{SHTLsumrule} one can in fact show
that in the case of pure-glue QCD the HTL entropy contains
exactly 1/4 of the plasmon term $\sim g^3$.
When treated strictly perturbatively, even 1/4 of the plasmon
term spoils the apparent convergence. However, in the
nonperturbative expression \eref{S2loop} the otherwise
large $g^3$ correction is rendered harmless and leads to
a small correction such that the rough agreement of
the perturbative order-$g^2$ result with lattice results
for $T\gtrsim 3T_c$ is retained and improved.

The plasmon term $\sim g^3$ 
becomes complete only upon inclusion
of the next-to-leading correction to the asymptotic thermal
masses $m_\infty$ and $M_\infty$. These are determined
in standard HTL perturbation theory through
\be\label{dmas}
\begin{array}{l}
\delta m_\infty^2(k)=\Re \delta\Pi_T(\omega=k) \\
=\Re(\begin{picture}(0,0)(0,0)
\put(25,0){\small +}
\put(56,0){\small +}
\put(104,0){\small +}
\put(157,0){$|_{\omega=k}$}
\end{picture}
\!\!\includegraphics[bb=145 430 500 475,width=5.5cm]{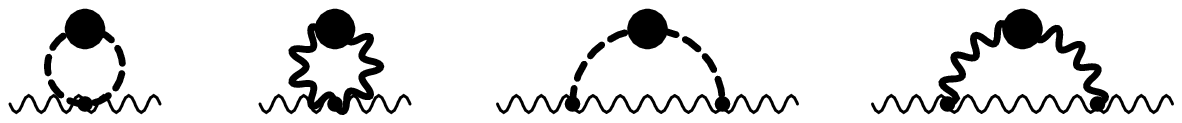})
\end{array}
\ee
where thick dashed and wiggly lines with a blob represent
HTL propagators for longitudinal and transverse polarizations, respectively.
Similarly,
\be\label{dMas}
{1\02k}\delta M_\infty^2(k)=\delta\Sigma_+(\omega=k) 
=\Re(\begin{picture}(0,0)(0,0)
\put(42,0){\small +}
\end{picture}
\includegraphics[bb=75 430 285 475,width=3.2cm]{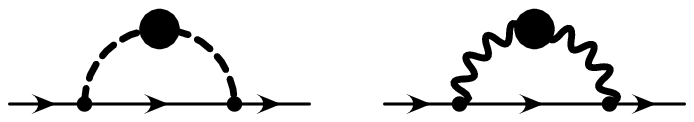})
|_{\omega=k}\;.
\ee
The explicit proof that these contributions indeed restore the
correct plasmon term is given in reference~\cite{Blaizot:2000fc}.

These corrections to the asymptotic thermal masses are, in contrast
to the latter, nontrivial functions of the momentum, which can
be evaluated only numerically. However, as far as the generation
of the plasmon term is concerned, these functions contribute
in the averaged form
\be\label{deltamasav}
\bar\delta m_\infty^2={\int dk\,k\,n_{\rm BE}'(k) \Re \delta\Pi_T(\omega=k) 
\0 \int dk\,k\,n_{\rm BE}'(k)}
\ee
(cf.\ (\ref{Shard})) and similarly
\be\label{deltaMasav}
\bar\delta M_\infty^2={\int dk\,k\,n_{\rm FD}'(k) \Re 
2k \delta\Sigma_+(\omega=k) 
\0 \int dk\,k\,n_{\rm FD}'(k)}\;.
\ee
These averaged asymptotic thermal masses turn out to be given
by the remarkably simple expressions \cite{Blaizot:2000fc}
\be
\label{deltamas}
\bar\delta m_\infty^2=-{1\02\pi}g^2NT\hat m_D,\quad
\bar\delta M_\infty^2=-{1\02\pi}g^2C_fT\hat m_D,
\ee
where $C_f=N_g/(2N)$. Since the integrals in
(\ref{deltamasav}) and (\ref{deltaMasav}) are dominated by hard
momenta, these thermal mass corrections only pertain to hard
excitations. 

Pending a full evaluation of the NLO corrections to $\Re\delta\Pi$
and $\Re\delta\Sigma$, 
references~\cite{Blaizot:1999ip,Blaizot:1999ap,Blaizot:2000fc} 
have proposed
to define a next-to-leading approximation through (for gluons)
\be
{\cal S}_{NLA}={\cal S}_{HTL}\Big|_{\rm soft}+
{\cal S}_{HTL,m_\infty^2\to\bar m_\infty^2}\Big|_{\rm hard},
\ee
where $\bar m_\infty^2$ includes (\ref{deltamas}) and a separation
scale $\sqrt{c_\Lambda 2\pi Tm_D}$ is introduced to make the
distinction between hard and soft domains.

For numerical evaluations, 
a crucial issue here is the definition of the corrected
asymptotic mass $\bar m_\infty$. For the range of coupling constants
of interest ($g\gtrsim 1$), the correction $ |\bar\delta m_\infty^2|$
is greater than the LO value $m_\infty^2$, leading to tachyonic
masses if included in a strictly perturbative manner.

However, this problem is not at all specific to QCD.
In section \ref{subsecscappconv} we have seen that
the perturbative result \eref{msc} for the scalar screening mass
to order $g^3$ also turns tachyonic for $g\gtrsim 1$.
The self-consistent one-loop gap equation \eref{mth2sc},
on the other hand, is a monotonic function in $g$, and
is well approximated by a quadratic equation \eref{mtruncgap} which
keeps just to first two terms in in $m/T$ expansion.
As a model for the rather intractable (nonlocal) gap equations of QCD
in a $\Phi$-derivable approximation,
reference \cite{Blaizot:2000fc} has therefore proposed
to include the corrections \eref{deltamas} in
analogy to \eref{mtruncgap} in order to enforce monotonicity
of $\bar m_\infty$ as function of the coupling.
(In order to be consistent with the large flavour number limit,
this should in fact be done only for the bosonic part; for $N_f\not=0$, the
fermionic gap equation has to remain linear in $\bar M_\infty^2$,
which however poses no problem at small $N_f$
\cite{Rebhan:2003fj}.)

\begin{figure}[t]
\centerline{\includegraphics[viewport=70 190 540 545,width=7cm]{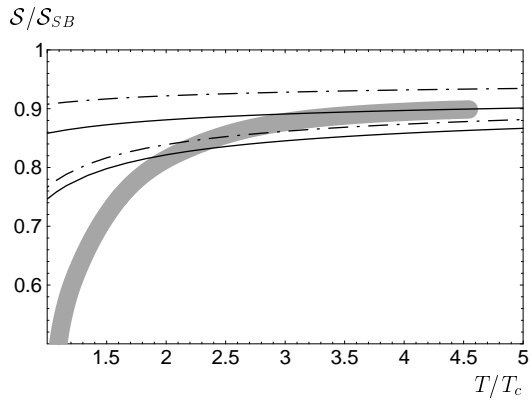}}
\caption{Comparison of the lattice data for
the entropy of pure-glue SU(3) gauge theory of reference~\cite{Boyd:1996bx}
(gray band) with the range of ${\cal S}_{HTL}$ (solid lines)
and ${\cal S}_{NLA}$ (dash-dotted lines) for $\bar\mu=
\pi T\ldots 4\pi T$ and $c_\Lambda= 1/2 \ldots 2$.}\label{figSg}
\end{figure}

In figure~\ref{figSg}, the numerical results for the pure-glue HTL entropy
and the NLA one (using a quadratic gap equation
for $\bar m_\infty$) are given as a function of $T/T_c$ with $T_c$
chosen as $T_c=1.14\Lambda_{\overline{\mathrm MS}}$. The full lines
show the range of results for
${\cal S}_{HTL}$ when the renormalization scale $\bar\mu$
is varied from $\pi T$ to $4\pi T$; the dash-dotted lines mark
the corresponding results for ${\cal S}_{NLA}$ with the
additional variation of $c_\Lambda$ from $1/2$ to 2. The
dark-gray band are lattice data from reference~\cite{Boyd:1996bx}.
Evidently, there is very good agreement for $T\gtrsim 2.5T_c$.

This approach can be generalized 
\cite{Blaizot:1999ap,Blaizot:2000fc,Blaizot:2001vr}
also to nonzero chemical potentials
$\mu_f$, for which lattice results
are available in the form of quark number susceptibilities
\cite{Gottlieb:1987ac,Gavai:2001fr,Gavai:2001ie,Bernard:2002yd,Gavai:2002jt}
and are beginning to become available for finite small chemical potential
\cite{Fodor:2001au,Fodor:2001pe,deForcrand:2002ci,Allton:2002zi,%
Fodor:2002km,D'Elia:2002gd,Gavai:2003mf,Allton:2003vx}.
Simpler quasiparticle models 
\cite{Peshier:1996ty,Levai:1997yx}
have already been used to extrapolate lattice data to finite
chemical potential \cite{Peshier:1999ww} and seem to work
well when compared with the recent lattice results for nonzero
chemical potential \cite{Szabo:2003kg}.
The HTL approach offers a possible refinement, which has been worked out in
\cite{Romatschke:2002pb,Rebhan:2003wn}.

\section{Next-to-leading order
corrections to the quasi-particle spectrum}
\label{sublsect}

In the simple scalar model of Sect.~\ref{sectscpth} and in
the more complicated example of asymptotic (averaged)
gluon and quark masses, eqs.~\eref{deltamasav} and \eref{deltaMasav},
we have seen that HTL/HDL perturbation theory leads
to next-to-leading order corrections of quasi-particle 
dispersion laws which are typically suppressed by 
a single power of $g$ rather than $g^2$.

In the following, we shall review what is at present
known about next-to-leading order corrections to
the HTL quasi-particles in gauge theories.
It will turn out that some corrections, namely screening
lengths for frequencies below the plasma frequency and
damping rates for moving excitations are even more
enhanced by infrared effects, to an extent that they
cannot be determined 
beyond the leading logarithmic term without nonperturbative input.


\subsection{Long-wavelength plasmon damping}

Historically,
the first full-fledged application of HTL resummation techniques
was the calculation of
the damping constant of gluonic plasmons in the long-wavelength
limit.
In fact, the development of the HTL resummation methods
was stimulated by the failure of conventional thermal perturbation
theory to determine this quantity.

In bare perturbation theory, the one-loop long-wavelength plasmon damping
constant had been found
to be gauge-parameter dependent and negative definite in covariant gauges 
\cite{Kalashnikov:1980cy}. Since even
the gauge-independent frameworks of the Vilkovisky-DeWitt
effective action \cite{Vilkovisky:1984st,Rebhan:1987wp}
and the gauge-independent pinch technique \cite{Cornwall:1985eu}
led to negative one-loop damping constants, this was sometimes
interpreted as a signal of an instability of the quark-gluon
plasma \cite{Hansson:1987un,Nadkarni:1988ti}.
Other authors instead argued in favour of ghost-free
``physical'' gauges such as temporal and Coulomb gauge, 
where the result turned out to be positive and seemingly
gauge-independent in this class of gauges
\cite{Kajantie:1985xx,Heinz:1987kz}.

\begin{figure}[t]
\centerline{\includegraphics[width=7truecm]{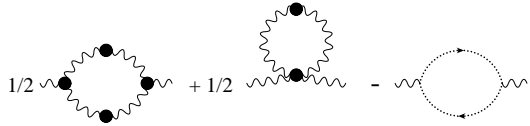}}
\caption{One-loop diagrams in HTL-resummed perturbation theory
contributing to the gauge boson self-energy in pure-glue QCD.
HTL-resummed quantities are marked with a blob.
 \label{Figdpi}}
\end{figure}

It was pointed out in particular by Pisarski \cite{Pisarski:1989vd} 
that all these results at bare one-loop order were incomplete,
as also implied by the arguments for gauge independence
of the singularities of the gluon propagator of reference~\cite{Kobes:1990xf}.
The appropriate resummation scheme was finally developed
by Braaten and Pisarski in 1990 \cite{Braaten:1990kk,Braaten:1990mz}
who first obtained the complete leading term in the plasmon damping
constant by evaluating the diagrams displayed
in figure \ref{Figdpi} \cite{Braaten:1990it} with the result
\bea\label{gamma6}
\gamma(\mathbf k=0)
&={1\02\hat\omega_{\rm pl}}{\rm Im}\,\delta\Pi_A(k_0=\hat \omega_{\rm pl},\mathbf k=0)
\nonumber\\
&=+6.635\ldots{g^2NT\over 24\pi}=
0.264 \sqrt{N}{ g}\,\hat \omega_{\rm pl}. 
\eea

For $g\ll1$ this implies the existence of weakly damped gluonic plasmons.
In QCD ($N=3$), where for all temperatures of interest $g\gtrsim 1$,
the existence of plasmons as identifiable quasi-particles requires
that $g$ is significantly less than about 2.2, so that the situation
is somewhat marginal.

The corresponding quantity for fermionic quasi-particles has been
calcula\-ted in \cite{Kobes:1992ys,Braaten:1992gd}
with a comparable result: weakly damped long-wavelength
fermionic quasi-particles in 2- or 3-flavour QCD require that
$g$ is significantly less than about 2.7.

These results have been obtained in Coulomb gauge with
formal verification of their gauge independence. Actual
calculations, however, later revealed
that in covariant gauges
HTL-resummed perturbation theory
still leads to explicit gauge dependent
contributions to the damping of fermionic \cite{Baier:1992dy}
as well as gluonic \cite{Baier:1992mg}
quasi-particles. But, as was pointed out subsequently
in \cite{Rebhan:1992ak},
these apparent gauge dependences are avoided if the quasi-particle
mass-shell is approached with a general infrared cut-off (such
as finite volume) in place, and this cut-off lifted only in the end.
This procedure defines gauge-independent dispersion laws;
the gauge dependent parts are found to pertain to the residue,
which at finite temperature happens to be linearly infrared
singular in covariant gauges, rather than only logarithmically
as at zero temperature, due to Bose enhancement.

For kinematical reasons, the result \eref{gamma6}, which
has been derived for branch A of the gluon propagator, should
equally hold for branch B, 
since with $\mathbf k\to0$ one can no longer
distinguish between spatially transverse and longitudinal
polarizations. However, as will be discussed
further below, the limit $\mathbf k\to0$
involves infrared problems, and there
are even explicit
calculations \cite{Abada:1997vm,Abada:1998ue}
that claim to find obstructions to this equality,
which are however contradicted by the recent work of \cite{Dirks:1999uc}.

In the simpler $\phi^4$ theory the long-wavelength
plasmon damping constant has been calculated in
\cite{Parwani:1992gq,Wang:1996qf}
and it has been shown in
\cite{Aarts:1997qi,Aarts:1997kp,Buchmuller:1997yw,Buchmuller:1998nw}
that this quantity can also be extracted from classical
field theory after perturbatively matching
to the HTL mass.

\subsection{NLO correction to the plasma frequency}
\index{plasma frequency}

In \cite{Schulz:1994gf}, Schulz has calculated
also the real part of the NLO
contribution to the gluon polarization tensor 
in the limit of $\mathbf k\to0$ which determines the
NLO correction to the gluonic plasma frequency.

The original power-counting arguments of \cite{Braaten:1990mz}
suggested that besides one-loop diagrams with HTL-resummed
propagators and vertices given
in figure \ref{Figdpi}, there could be also contributions
from two-loop diagrams to relative order $g$. 
The explicit (and lengthy) calculation
of \cite{Schulz:1994gf} showed that those contribute only
at order $g^2 \ln(1/g)$ rather than $g$, and
the NLO plasma frequency
in a pure-glue plasma was obtained as
\be\label{NLOmpl}
\omega_{\rm pl.}=\hat\omega_{\rm pl.} \left[ 1-
0.09 \sqrt{N}{ g} \right].
\ee

In this particular result, HTL-resummed perturbation theory turns
out to give a moderate correction to the leading-order HTL value
even for $g\sim 1$.

\subsection{NLO correction to the Debye mass}

Poles of the gauge boson propagator at $\omega<\omega_{\rm pl.}$
and $\mathbf k^2 = -\kappa^2 < 0$
describe exponential 
screening of fields with frequencies below the plasma
frequency and, for $\omega\to0$, of static fields.
The value of $\kappa$ is the (frequency-dependent)
screening mass.

In the static case, branch $A$ of the propagator describes
the screening of (chromo-) magnetostatic fields. 
While there is a finite screening length as long as $\omega>0$,
the $A$-branch of the HTL propagator becomes unscreened in
the static limit.
In QED, a ``{magnetic mass}'' is forbidden by gauge invariance
\cite{Fradkin:1965,Blaizot:1995kg}, but some sort of 
entirely non-perturbative magnetic mass
is expected in non-Abelian gauge theories in view of severe
infrared problems caused by the self-interactions of
magnetostatic gluons \cite{Polyakov:1978vu,Linde:1980ts,Gross:1981br}.

Branch $B$, on the other hand, contains the information about
screening of (chromo) electric fields as generated by
static charges (Debye screening). The {Debye mass} given by
the leading-order HTL propagator is $\hat m_D=\sqrt{3}\omega_{\rm pl.}$.

The evaluation of corrections in thermal perturbation theory
again requires resummation. Even though the nonlocalities of the
HTL effective action do not play a role in the static limit, 
historically
these corrections
have been determined for QCD, as well as for (ultra-relativistic) QED,
only after the more complicated cases of gluonic plasmon damping
and NLO plasma frequency discussed above had been mastered.

Originally, the {Debye mass} (squared) has been {\em defined} as
the infrared limit $\Pi_{00}(\omega=0,k\to0)$, which indeed
is correct at the HTL level, cf.\ \eref{PiHTL}.

In QED, this definition has the advantage of being directly
related to the second derivative of the thermodynamic pressure
with respect to the chemical potential $\mu$ (electric
susceptibility), so that
the higher-order terms known from the latter determine those
of $\Pi_{00}^{\rm QED}(\omega=0,k\to0)$ through \cite{Fradkin:1965,Kap:FTFT}
\be\label{PiQED00}\fl
\Pi
_{00}(0,k\to0)\Big|_{\mu=0}
=e^2{\partial ^2P\over \partial \mu^2}\Big|_{\mu=0}={e^2T^2\over 3}
\left(1-{3e^2\over 8\pi^2}+{\sqrt3 e^3\over 4\pi^3}
+\ldots\right)
\ee
This result is gauge independent because in QED all of $\Pi_\mn$ is.

In the case of QCD, there is no such relation. Moreover,
$\delta m_D^2/\hat m_D^2$ $\sim g$ rather than $g^3$
because of gluonic self-interactions and Bose enhancement.
The calculation of this quantity should be much easier than
the dynamic ones considered above, because in the static
limit the HTL effective action collapses 
to just the local, bilinear HTL {Debye mass} term,
\be\label{HTLeffLstatic}
\CL^{\rm HTL} \stackrel{\rm static}{\longrightarrow}
-{1\over2} \hat m_D^2 {\rm tr} A_0^2.
\ee
This is also
gauge invariant, because $A_0$ behaves like an adjoint scalar under
time-independent gauge transformations.
Resummed perturbation theory for static quantities thus reduces to a resummation
of the HTL Debye mass in the electrostatic propagator
\cite{Gell-Mann:1957,Kap:FTFT,Arnold:1993rz}.

However, in QCD this simple (``ring'') resummation\index{ring resummation}
leads to the gauge dependent result \cite{Toimela:1985ht}
\be\label{Pi0000}
\Pi_{00}(0,0)/\hat m_D^2 = 1+{ \alpha}{N\over 4\pi}\sqrt{6\over 2N+N_f}g 
\ee
where $\alpha$ is the gauge parameter of general covariant gauge (which
coincides with general {Coulomb gauge} in the static limit).

This result was initially interpreted as meaning either
that the non-Abelian Debye
mass could not be obtained in resummed perturbation theory 
\cite{Nadkarni:1986cz} or that one should use a physical gauge
instead \cite{Kajantie:1985xx,Kap:FTFT}. In particular, temporal axial
gauge was put forward, because in this gauge there is, like in
QED, a linear relationship between electric field strength
correlators and the gauge propagator. However, because
static ring resummation clashes with {temporal gauge},
inconclusive and contradicting results were obtained by
different authors \cite{Kajantie:1982hu,Furusawa:1983gb,Kajantie:1985xx}.
A consistent calculation in fact requires
vertex resummations \cite{Baier:1994et,Peigne:1995dn}, but this
does not resolve the gauge dependence issue because the nonabelian
field strength
correlator is gauge variant \cite{Rebhan:1994mx}.

On the other hand, in view of the 
{gauge dependence identities} discussed in
section \ref{gindproof}, the gauge dependence of (\ref{Pi0000}) is no longer
surprising. Gauge independence can only be expected ``on-shell'',
which here means $\omega=0$ but $\mathbf k^2\to-\hat m_D^2$.

Indeed, the exponential fall-off of the electrostatic propagator
is determined by the position of the singularity of $\Delta_B(0,k)$,
and not simply by its infrared limit.
This implies in particular that one should use a different definition
of the {Debye mass} already in QED, despite the gauge independence
of (\ref{PiQED00}), namely \cite{Rebhan:1993az}
\be\label{mDpoledef}
m^2_D=\Pi_{00}(0,k)\Big|_{\mathbf k^2\to-m_D^2}.
\ee

For QED (with massless electrons), the {Debye mass} is thus
not given by (\ref{PiQED00}) but rather as \cite{Rebhan:1993az}
\bea\label{mDMOM}
\fl m^2_D&=&\Pi_{00}(0,k\to0)+\bigl[
\Pi_{00}(0,k)\big|_{k^2=-
m^2_D}-\Pi_{00}(0,k\to0)
\bigr]\nonumber\\
\fl
&=&{e^2T^2\over 3}
\biggl(1-{3e^2\over 8\pi^2}+{\sqrt3 e^3\over 4\pi^3}
+\ldots -{e^2 \over 6\pi^2}[\ln{\tilde\mu\over \pi T}+\g_E-{4\over 3}]
+\ldots \biggr)
\label{mDQED}
\eea
where $\tilde\mu$ is the renormalization scale of
the momentum subtraction scheme i.e.\
$\Pi_\mn(k^2\!=\!-\tilde\mu^2)|_{T=0}=0$.
In the also widely used modified minimal subtraction ($\overline{\hbox{MS}}$)
scheme the last coefficient $-{4\over 3}$ in \eref{mDMOM}
has to be replaced by $-{1\over 2}$.
(The slightly different
numbers in the terms $\propto e^4T^2$ 
quoted in \cite{Blaizot:1995kg,LeB:TFT}
pertain to the minimal subtraction (MS) scheme\footnote{Reference~\cite{LeB:TFT}
erroneously refers to the $\overline{\hbox{MS}}$ scheme when
quoting the MS result of \cite{Blaizot:1995kg}.}.)

Since 
$de/d\ln\tilde\mu=e^3/(12\pi^2)+O(e^5)$,
(\ref{mDQED}) is a renormalization-group invariant result for
the Debye mass in hot QED, which (\ref{PiQED00})
obviously is not. Only the susceptibility $\chi=\partial^2 P/\partial\mu^2$
is renormalization-group invariant, but not $e^2\chi$.
Similarly, one should distinguish between electric susceptibility
and electric Debye mass also in the context of QCD.

In QCD, where gauge independence is not automatic,
the dependence on the gauge fixing parameter $\alpha$
is another indication that (\ref{Pi0000}) is
the wrong definition. For (\ref{mDpoledef})
the full momentum dependence of
the correction $\delta\Pi_{00}(k_0=0,{\mathbf k})$ to
$\hat\Pi_{00}$ is needed.
This is given by \cite{Rebhan:1993az}
\bea\fl
&&\delta\Pi_{00}(k_0=0,{\mathbf k})
=
g\hat m_DN\sqrt{\frac6{2N+N_f}}
\int
\frac{d^{3-2\varepsilon}p}{(2\pi)^{3-2\varepsilon}}
\biggl\{\frac1{{\mathbf p}^2+\hat m_D^2}+\frac1{{\mathbf p}^2}
\nonumber\\
\fl
&&\qquad+\frac{4\hat m_D^2-({\mathbf k}^2+\hat m_D^2)
[3+2{\mathbf p}{\mathbf k}/{\mathbf p}^2]}{{\mathbf p}^2
{[(\mathbf p+\mathbf k)^2+\hat m_D^2]}
}
+{ \alpha} 
{({\mathbf k}^2+\hat m_D^2)}
\frac{{\mathbf p}^2+2{\mathbf p}{\mathbf k}}{{\mathbf p}^4{
{[(\mathbf p+\mathbf k)^2+\hat m_D^2]}
}}
\biggr\}.
\label{dPi00QCD}
\eea

In accordance with the {gauge dependence identities}, the last term
shows that gauge independence holds algebraically
for $\mathbf k^2=-\hat m_D^2$. 
On the other hand, on this ``screening mass shell'', where
the denominator term 
$[(\mathbf p+\mathbf k)^2+\hat m_D^2] \to [\mathbf p^2+2\mathbf p\mathbf k]$, one
encounters IR-singularities. In the $\alpha$-dependent term,
they are such that they produce a factor $1/[\mathbf k^2+m_D^2]$
so that the gauge dependences no longer disappear even on-shell.
This is, however, the very same problem that had to be solved
in the above case of the plasmon damping in {covariant gauges}. Introducing
a temporary infrared cut-off (e.g., finite volume), does not
modify the factor $[\mathbf k^2+m_D^2]$ in the numerator but
removes the infrared divergences that would otherwise
cancel it. Gauge independence thus
holds for all values of this cut-off, which can be sent to zero in the end.
The gauge dependences are thereby identified as belonging to
the (infrared divergent) residue.

The third term in the curly brackets, however, remains
logarithmically singular on-shell when the infrared cut-off is
removed. In contrast to the $\alpha$-dependent term, closer
inspection reveals that these singularities are coming from
the massless magnetostatic modes and not from unphysical massless
gauge modes. 

At HTL level, there is no (chromo) magnetostatic screening, but, as we
have mentioned, one expects some sort of such screening to be generated
non-perturbatively in the static sector of hot QCD at
the scale $g^2T \sim gm_D$
\cite{Polyakov:1978vu,Linde:1980ts,Gross:1981br}.
\index{magnetic mass}

While this singularity prevents evaluating 
$\delta\Pi_{00}$ 
in full,
the fact that this singularity is only logarithmic allows one
to extract the leading term of (\ref{dPi00QCD}) under the
assumption of an effective cut-off at $p \sim g^2T$ as \cite{Rebhan:1993az}
\be\label{mDlng}
{\delta m^2_D\over \hat m^2_D}=
\frac{N}{2\pi}\sqrt{\frac6{2N+N_f}}\,g\,\ln\frac1g+O(g).
\ee

The $O(g)$-contribution, however, is sensitive to the physics
of the magnetostatic sector at scale $g^2T$, and is completely
non-perturbative in that all loop orders $\ge 2$ are
expected to contribute with equal importance
as we shall discuss in section \ref{sectmm}.

Because of the undetermined $O(g)$-term in (\ref{mDlng}), one-loop
resummed perturbation theory only says that for sufficiently small $g$,
where $O(g\ln(1/g))\gg O(g)$,
there is a {\em positive} correction to the Debye
mass of lowest-order perturbation theory following from
the pole definition (\ref{mDpoledef}), and that it is gauge
independent.

On the lattice, the static gluon propagator of pure SU(2)
gauge theory at high temperature \index{gluon propagator!lattice}
has been studied in various gauges \cite{Heller:1997nq,Cucchieri:2001tw}
with the result that the electrostatic propagator is exponentially
screened with a screening mass that indeed appears to be gauge independent
and which is about 60\% larger than the leading-order {Debye mass}
for temperatures $T/T_c$ up to about $10^4$.
Similar results have been obtained recently also for the
case of SU(3) \cite{Nakamura:2002ki,Nakamura:2003pu}.

In \cite{Rebhan:1994mx}, an estimate of the $O(g)$
contribution to (\ref{mDlng}) has been made using the crude
approximation of a simple massive propagator for the magnetostatic
one, which leads to
\be\label{mDlnm}
{\delta m^2_D\over \hat m^2_D}=
\frac{N}{2\pi}\sqrt{\frac6{2N+N_f}}\,g\,\left[\ln{2m_D\over m_m}-\frac12
\right].
\ee
On the lattice one finds
strong gauge dependences of the magnetostatic
screening function, but the data
are consistent with an over-all exponential
behaviour corresponding to $m_m \approx 0.5 g^2T$ in all gauges
\cite{Heller:1997nq,Cucchieri:2000cy}. Using this number
in a self-consistent evaluation of (\ref{mDlnm}) gives
an estimate for $m_D$ which is about 20\% larger than
the leading-order value for $T/T_c=10\ldots 10^4$.

This shows that there are strong non-perturbative contributions
to the Debye screening mass $m_D$ even at very high temperatures.
Assuming that these are predominantly of order $g^2T$, one-loop resummed
perturbation theory (which is as far as one can get)
is able to account for about 1/3 of
this inherently non-perturbative physics already, if one introduces a 
simple, purely phenomenological magnetic screening mass.

\subsubsection{Non-perturbative definitions of the Debye mass}

A different approach to studying Debye screening non-perturbatively
without the complication of gauge fixing is to consider spatial
correlation functions of appropriate gauge-invariant operators
such as those of the {Polyakov loop}
\be
L(\mathbf x)={1\over N} {\rm Tr}\, {\cal P} \exp \left\{
-\I g \int_0^\beta d\tau \, A_0(\tau,\mathbf x)\right\}.
\ee
The correlation of two such operators is related to the free energy
of a quark-antiquark pair \cite{McLerran:1981pb}. In lowest
order perturbation theory this is given by the square of
a Yukawa potential with screening mass $\hat m_D$
\cite{Nadkarni:1986cz}; at
one-loop order one can in fact identify contributions
of the form (\ref{mDlnm}) if one assumes magnetic screening
\cite{Braaten:1994pk,Rebhan:1994mx}, but there is the problem that 
through higher loop orders the
large-distance behaviour becomes 
dominated by the magnetostatic modes and their 
lightest bound states \cite{Braaten:1995qx}.

In \cite{Arnold:1995bh}, Arnold and Yaffe have proposed to use Euclidean
time reflection symmetry to distinguish electric and magnetic
contributions to screening, and have given a prescription to
compute the sublogarithmic contribution of order $g^2T$ to $m_D$
nonperturbatively. This has been carried out in 3-d lattice simulations
for SU(2) \cite{Kajantie:1997pd,Laine:1997nq} as well as for
SU(3) \cite{Laine:1999hh}. The Debye mass thus defined shows
even larger deviations from the lowest-order perturbative results
than that from gauge-fixed lattice propagators. E.g.,
in SU(2) at $T=10^4 T_c$ this deviation turns out to be over 100\%,
while in SU(3) the dominance of $g^2T$
contributions is even more pronounced.

Clearly, (resummed) perturbation theory is of no use here for
any temperature of practical interest. However, it should be
noted that the magnitude
of the contributions from the completely nonperturbative magnetostatic
sector depends strongly on the quantity considered. It is
significantly smaller in the definition of the {Debye mass} through
the exponential decay of gauge-fixed
gluon propagators, which
leads to smaller screening masses on the
lattice \cite{Heller:1997nq,Cucchieri:2001tw,Nakamura:2002ki,Nakamura:2003pu} 
and which also seem to provide a useful nonperturbative
description of Debye screening as they turn out to be gauge-independent
in accordance with the arguments of section \ref{gindproof}.


\subsection{Magnetostatic screening}
\label{sectmm}

The next-to-leading order correction to the
other structure function, $\Pi_{ii}(0,k)$,
can be derived in a calculation
analogous to the one leading to
\eref{dPi00QCD}, and is found to be \cite{Rebhan:1993az}
\begin{eqnarray}
\delta\Pi_{ii}(0,k)&=&g^2N\hat m_DT\biggl\{
\frac{(\alpha +1)^2+10}{16} \frac{k}{\hat m_D} \nonumber\\
&&+\frac1{4\pi}\left[2-\frac{k^2+4\hat m_D^2}{\hat m_Dk}\arctan\frac{k}{2\hat m_D}\right]
\biggr\}.
\label{piii}
\end{eqnarray}
Apparently, the magnetic permeability defined by
$1/\mu=1-\frac12\Pi_{ii}/k^2$ is a gauge-dependent quantity
beyond leading order. The gauge-dependent terms vanish
only at the location of the pole of the transverse gluon propagator,
which is at ${\bf k}=0$. There the correction term vanishes
completely,
which means that there is no magnetic mass squared of the
order $g^3T^2$.
The magnetic mass must therefore be $\ll g^{3/2}T$ at weak coupling.

For small $k\ll \hat m_D$, \eref{piii} has a linear behaviour
with gauge dependent, but positive definite coefficient.
The transverse propagator therefore has the form
$1/(k^2-c k)$ with $c\sim g^2T$. This corresponds to
a pole at space-like momentum with $k\sim g^2T$ \cite{Kalashnikov:1981ed}. 
However,
in this regime the inherently nonperturbative contributions
$\sim g^4T^2$ to $\Pi_{\mu\nu}$ become relevant and are expected
to remove this pathology by the generation of a magnetic mass
$m_m\sim g^2T$. This is completely analogous to the infrared
problem of the perturbative expansion of the pressure identified in
\cite{Linde:1980ts,Gross:1981br}, but while the latter sets in
at 4-loop order, in the transverse gauge-boson self-energy
with external momenta $\lesssim g^2 T$ this problem starts already
at 2-loop order.

In a HTL-resummed 2-loop calculation in general
covariant gauge, reference~\cite{Reinbach:1999bi}
has verified that such a magnetic mass receives contributions
exclusively from the ultrasoft momentum regime $k\sim g^2T$
where the relevant effective theory is 
three-dimensional Yang-Mills theory, which is confining
and inherently nonperturbative.
There exists in fact an elegant attempt towards an analytic
nonperturbative study of this theory
through its Schr\"odinger equation using results from
two-dimensional conformal field theory
\cite{Karabali:1996ps,Karabali:1997iu,Karabali:1997wk}.
This leads to the estimate $m_m=g^2NT/(2\pi)$, which
is reasonably
close to the results for magnetostatic propagators obtained
in the lattice calculations of references~\cite{Heller:1997nq,Cucchieri:2001tw}
quoted above.

The other approaches that have been tried to obtain an analytic estimate
for the magnetic mass mostly involve a reorganization of perturbation
theory by assuming a more or less complicated mass term for the
magnetostatic gluons, setting up a self-consistent gap equation
and solving it. However, the various possibilities lead to rather
contradictory results 
\cite{Alexanian:1995rp,Buchmuller:1995qy,Jackiw:1996nf,Jackiw:1997jg,Buchmuller:1997pp,Cornwall:1998dc}.
In fact, lattice calculations indicate that the
phenomenon of magnetic screening is not well described by a
simple pole mass \cite{Cucchieri:2001tw}.

In the Abelian case, one may expect that
magnetostatic fields are completely unscreened and it
can indeed be proved rigorously to
all orders of perturbation theory \cite{Fradkin:1965,Blaizot:1995kg}
that $\Pi_{ii}(0,k\to0)=O(k^2)$. In massless scalar
electrodynamics, an unresummed one-loop calculation
actually gives $\Pi_{ii}(0,k\to0)={1\08}e^2kT$ suggesting that
the magnetostatic propagator has the form $1/[k^2+e^2kT/8]$,
which would imply power-law 
screening. However, a
resummation of the thermal mass of the scalar particles removes this
unphysical behaviour
\cite{Kraemmer:1995az,Blaizot:1995kg}.



\subsection{Dynamical screening and damping at high temperature}
\label{subsecdds}

A logarithmic sensitivity to the nonperturbative physics
of the (chromo-) magneto\-static sector has in fact been encountered
early on also in the calculation of the damping rate for a heavy fermion 
\cite{Pisarski:1989vd}, and more generally of hard
particles \cite{Lebedev:1990ev,Lebedev:1991un,Burgess:1992wc,Rebhan:1992ca}.
It also turns out to occur for
soft quasi-particles as soon as they have nonvanishing (group) velocity
\cite{Pisarski:1993rf,Flechsig:1995sk}.

Because this logarithmic sensitivity arises only if one internal
line of (resummed) one-loop diagrams is static, the
coefficient of the resulting $g\ln(1/g)$-term is almost
as easy to obtain as in the case of the Debye mass, even though
the external line is non-static and soft, therefore requiring
HTL-resummed vertices (see figure~\ref{Figdpi}).

The infrared singularity arises (again) from the dressed one-loop
diagram with two propagators, one of which is magnetostatic and
thus massless in the HTL approximation, and the other of the same
type as the external one, so only the first diagram in figure~\ref{Figdpi}
is relevant. The dressed 3-vertices in it are needed only
in the limit of one leg being magnetostatic and having zero momentum.
Because of the gauge invariance of HTL's, these are determined
by the HTL self-energies through a differential Ward identity, e.g.\
\index{Ward identities}
\be
\hat\Gamma_{\mu\nu\varrho}(k;-k;0)=
-{\partial \over \partial k^\varrho}\hat\Pi_{\mu\nu}(k)
\ee
for the 3-gluon vertex (colour indices omitted).

Comparatively simple algebra gives \cite{Flechsig:1995sk}
\be
\delta\Pi_I(k) \simeq -g^2 N 4 \mathbf k^2 [1+\partial _{\mathbf k^2}\Pi_I(k)]^2 
{ \mathcal S_I(k)},\qquad {I=A,B}
\label{dPiI}
\ee
where
\be{ \mathcal S_I(k)} := T\int{d^3p\over (2\pi)^3}{ 1\over \mathbf p^2}
{-1\over (k-p)^2-\Pi_I(k-p)} \Big|_{k^2=\Pi_I(k), p^0=0}
\ee
and the logarithmic (mass-shell) singularity arises because
$(k-p)^2-\Pi_I(k-p) \to -\mathbf p^2+2\mathbf p\mathbf k-\Pi_I(k-p)
+\Pi_I(k)\, { \sim |\mathbf p|} $ as $k^2 \to \Pi_I(k)$.

The IR-singular part of ${\mathcal S}_I(k)$ is given by
\bea
{\mathcal S}_I(k)&=&T\int{d^3p\over (2\pi)^3}{1\over \mathbf p^2}
{1\over \mathbf p^2-2\mathbf p\mathbf k+\Pi_I(k-p)-\Pi_I(k)-\I \varepsilon}\nonumber\\
&\simeq& T [1+\partial _{\mathbf k^2}\Pi_I(k)]^{-1}\int{d^3p\over (2\pi)^3}
{1\over \mathbf p^2}{1\over \mathbf p^2-2\mathbf p\mathbf k-\I \varepsilon}\nonumber\\
&=&T[1+\partial _{\mathbf k^2}\Pi_I(k)]^{-1} \int_{ \lambda}^\infty
{dp\over p}{1\over 2|\mathbf k|}\ln{p+2|\mathbf k|-\I \varepsilon \over
  p-2|\mathbf k|-\I \varepsilon} \label{calSI}
\eea
where in the last line we have inserted an 
IR cutoff $\lambda\ll gT$ for the $p$-integral
in order to isolate the singular behaviour.

One finds that (\ref{calSI}) has a singular imaginary part
for 
propagating modes, and a singular real part
in screening situations
\be\fl
{\mathcal S}_I(k)\simeq {T\over 8\pi}[1+\partial _{\mathbf k^2}\Pi_I(k)]^{-1} 
\times\left\{
\begin{array}{ll}
\I|\mathbf k|^{-1}
\ln(|\mathbf k|/ \lambda) + O(\lambda^0) 
&\textrm{for $\mathbf k^2>0$}\\
\kappa^{-1} 
\ln(\kappa/ \lambda) + O(\lambda^0)
&\textrm{for $\mathbf k^2 = - \kappa^2<0$}
\end{array} \right.
\ee
The case $\mathbf k=0$, on the other hand,
is IR-safe, because (\ref{dPiI}) is proportional
to $\mathbf k^2$, while
\be\label{SIk0}\fl
{\mathcal S}_I(k)\longrightarrow {T\over 4\pi^2 \lambda}
[1+\partial _{\mathbf k^2}\Pi_I(k)]^{-1}\Big|_{\mathbf k=0} + O\({T|\mathbf k|\over \lambda^2}\)
\quad \mbox{for $\mathbf k \to 0$.}
\ee

This shows that there is a common origin for the infrared sensitivity
of screening and damping of HTL quasi-particles.
Provided that the scale where the logarithmic divergences are cut off
is the magnetic scale $g^2T$, the coefficients of the
leading
$g \ln(1/g)$-terms are determined and in fact lead to
beautifully simple results:
For the damping of moving quasi-particles one obtains
\cite{Pisarski:1993rf,Flechsig:1995sk}
\index{HTL quasi-particles!damping}
\be\label{Damping}\fl
\quad\gamma_I(|\mathbf k|) 
\simeq {g^2NT\over 4\pi}
{|\mathbf k|[1+\partial _{\mathbf k^2}\Pi_I(k)]\over 
\omega(|\mathbf k|)[1-\partial _{\omega^2}\Pi_I(k)]}
\ln{1\over g} 
\equiv{g^2NT\over 4\pi}v_I(|\mathbf k|) \ln{1\over g} 
\ee
where $v_I(|\mathbf k|)$ is the group
velocity of mode $I$ (which vanishes at $\mathbf k=0$). 
The IR-sensitive NLO correction to
screening takes its simplest form when formulated as \cite{Flechsig:1995sk}
\be
\quad\delta\kappa^2_I(\omega)={g^2NT\over 2\pi}\kappa_I(\omega)
\( \ln{1\over g}+ O(1) \)
\ee
where $\kappa_I(\omega)$ is the inverse screening length of mode $I$
at frequency $\omega<\omega_{\rm pl.}$ (which in the static limit
approaches the {Debye mass} and
perturbatively vanishing magnetic mass, resp., while
approaching zero for both modes as $\omega\to\omega_{\rm pl.}$).

A completely analogous calculation for the fermionic modes (for
which there are no screening masses) gives
\be\label{gpm}
\gamma_\pm(|\mathbf k|)={g^2C_FT\over 4\pi}\big|v_\pm(|\mathbf k|)\big| 
\( \ln{1\over g} + O(1) \)
\ee
for $|\mathbf k|>0$. 
The group velocity $v_\pm$ equals $\pm\frac13$ in the limit
$\mathbf k\to 0$, and increases monotonically towards $+1$ for
larger momenta (with a zero for the $(-)$-branch at $|\mathbf k_{\rm dip}|/\hat M
\approx 0.41$). 
For strictly $|\mathbf k|=0$, the IR sensitivity in fact disappears
because (\ref{gpm}) is no longer valid for $|\mathbf k|\ll\lambda$,
but one has $\gamma_\pm(|\mathbf k|)|_{\rm sing.}
\propto g^2 T |\mathbf k|/\lambda$ instead.
Thus $\gamma_\pm(0)$ is calculable at order $g^2T$ in HTL-resummed
perturbation theory, and has been calculated in 
\cite{Kobes:1992ys,Braaten:1992gd}. 
For nonvanishing $|\mathbf k|\sim \lambda \sim g^2T$ HTL-resummed
perturbation theory breaks down (the IR sensitivity of $\gamma_\pm$ for
small nonvanishing $\mathbf k$ has recently been displayed also
in \cite{Abada:2000hh}).
For $|\mathbf k|\gtrsim gT$, the damping is calculable in the
leading-log approximation, and even completely at 
$|\mathbf k|=|\mathbf k_{\rm dip}|$ in the case of $\gamma_-$, though
this has not been done yet.

The fermionic result (\ref{gpm}) applies in fact equally to QED, for
which one just needs to replace $g^2C_F\to e^2$.
This is somewhat disturbing as QED does not allow
a non-zero {magnetic mass} as IR cutoff, and
it has been conjectured that the damping $\gamma\sim g^2T$ or $e^2T$
itself might act as an effective IR cutoff 
\cite{Lebedev:1990ev,Lebedev:1991un,Pisarski:1993rf,Altherr:1993ti},
which however led to further difficulties \cite{Baier:1992bv,Peigne:1993ky}.
The solution for QED was finally
found in references~\cite{Takashiba:1996qa,%
Blaizot:1996hd,Blaizot:1997az,Blaizot:1997kw}
where it was shown 
that there the fermionic modes undergo over-exponential damping
in the form $\E^{-\gamma t} \to \exp({-{e^2\over 4\pi}
T t \ln( 
\omega_{\rm pl.}t)})$ (for $v\to1$),
so finite time 
is the actual IR cut-off (see also \cite{Wang:1999mb,Boyanovsky:1999cy,Wang:2000vi}).
The fermion propagator
has in fact no simple quasi-particle pole in momentum space
or any other singularity near the light-cone
\cite{Weldon:2003}, but nevertheless a sharply
peaked spectral density.

In non-Abelian
theories, on the other hand, one does expect static (chro\-mo-)\-magnetic
fields to have finite range, and lattice results do confirm
this, so the above leading-log results
for an (exponential) damping constant are
expected to be
applicable after all, at least for sufficiently weak coupling.

As will be discussed further in section \ref{secbey},
the damping rate of hard gluons 
$\gamma=\gamma_A(|\mathbf k|\sim T)\simeq {1\04\pi}
g^2N T\ln (1/g)$ defines
an important scale in the dynamics of nonabelian fields
that is parametrically larger than the magnetic mass scale $g^2 T$.
It sets the time scale for colour relaxation 
and determines, to leading order, the colour conductivity
through \cite{Selikhov:1993ns,Arnold:1999uy,MartinezResco:2000pz}
\be\label{sigmac}
\sigma_c=\hat\omega^2_{\rm pl}/\gamma.
\ee

\subsection{Damping of high-momentum fermions in a degenerate plasma}

\begin{figure}
\centerline{\includegraphics[bb=75 430 285 475,width=4.5cm]{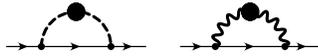}}
\caption{Next-to-leading order contributions to the self-energy
of a hard fermion. The dashed and wiggly lines with a blob
refer to the HTL/HDL resummed longitudinal and transverse
gauge boson propagators, respectively.
\label{figdSigmah}}
\end{figure}

At zero temperature and nonvanishing
chemical potential, the damping rate of fermionic
excitations is calculable in
perturbation theory, for in the absence
of Bose enhancement the dynamical screening
of the quasi-static transverse modes in the
HDL gauge boson propagator is sufficient to remove all
infrared divergences \cite{LeBellac:1997kr,Vanderheyden:1997bw,Manuel:2000mk}.
The leading-order
damping rate of a high energy/momentum mode with $E,p \gg g\mu$
is given by the diagrams in figure \ref{figdSigmah}, where
only the gauge boson propagator needs to be resummed. Explicitly, one has 
\bea\label{gfdeg}
\gamma_+(E)&={g^2 C_F\02v}\int {d^3q\0(2\pi)^3}\left[
\theta(q_0)-\theta(\mu-E+q_0)\right]\nonumber\\
&\times
\left[ \hat\rho_{\ell}(q_0,q)+v^2(1-\cos^2\xi)\hat\rho_t(q_0,q)\right]_{q_0=
qv\cos\xi}
\eea
where $v=p/E$ is the velocity of the fermion and $\hat\rho_{t,\ell}$
are defined in \eref{Deltat} and \eref{Deltaell}. 
(In the case of QED $g^2C_F$ reduces simply to $e^2$.)
For massless fermions
$v=1$ and  $\hat\rho_{t,\ell}$ are the HDL expressions given in
appendix \ref{App:A}, but \eref{gfdeg} remains applicable to
massive fermions and correspondingly modified spectral functions
for the gauge bosons \cite{Manuel:1996td,Manuel:2000mk}.

Since dynamical screening vanishes in the zero-frequency
limit according to \eref{dynscr}, there is a qualitative deviation
from the usual behaviour of the damping of fermionic excitations in the
vicinity of the Fermi surface, which for short-range interactions
vanishing like $(E-\mu)^2$ as $E\to\mu$ due to
phase-space restrictions in the fermion-fermion scattering
\cite{Luttinger:1961}.
The magnetic interactions instead give rise to a $|E-\mu|$ behaviour
characteristic of non-Fermi liquids
\cite{Holstein:1973,LeBellac:1997kr,Vanderheyden:1997bw,Manuel:2000mk}
which reads
\be\label{gfdegFs}
\gamma_+(E)={g^2C_F\024\pi}|E-\mu|+{g^2C_F\064\,v_F^2m_D}(E-\mu)^2+
\Or(|E-\mu|^3)
\ee
with $m_D^2=v_F\hat m_D^2$ and $\hat m_D^2$ given by the $T=0$ limit
of \eref{mD}. Here the first term is from the dynamically screened
quasi-static magnetic interactions and the second term from
Debye screened electrostatic ones. In a nonrelativistic situation
($v\ll 1$)
the latter is the dominant one (except very close to the Fermi surface).

Far away from the Fermi surface, the damping approaches
\cite{Vanderheyden:1997bw}
\be
\gamma_+(E)\sim 0.019\, g^2 C_F m_D,\quad E-\mu\gg\mu
\ee
which is also the damping relevant for hard anti-fermions \cite{Manuel:2000mk}
(which are auto\-matically far from their Fermi surface).

\subsection{Non-Fermi-liquid contributions to the real
part of the fermion self-energy}

The real part of the fermion self-energy at zero temperature
can equally be calculated in HDL perturbation theory, but
the most interesting aspect of it, a nonanalytic behaviour
in the vicinity of the Fermi surface, can be inferred from the
behaviour of its imaginary part through a Kramers-Kronig
dispersion relation, which implies
\cite{Holstein:1973,Brown:2000eh,Manuel:2000mk,Boyanovsky:2000bc}
\be\fl
{\rm Re}\,\Sigma_+(E,p)\simeq
{\rm Re}\,\Sigma_+(\mu,p)-{g^2 C_F\012\pi^2}(E-\mu)\ln{m_D\0|E-\mu|}
+\Or(|E-\mu|).
\label{ReSigmazerot}
\ee
This quantity is gauge independent on the mass shell of the hard
particle, to which \eref{ReSigmazerot} presents the leading correction.
The energy-independent part is governed
by the asymptotic fermionic mass,
\be
{\rm Re}\,\Sigma_+(\mu,p) = M_\infty^2/(2p),\quad M_\infty^2=2\hat M^2=
{1\04\pi^2}g^2 C_f\mu^2.
\ee
This corresponds to a correction to the Fermi momentum
defined by $\omega_+(p_F)=\mu$, which for effectively massless
fermions reads
\be
p_F/\mu=1-{g^2C_f\08\pi^2}+\ldots\,.
\ee
The logarithmic term in \eref{ReSigmazerot} leads to a correction
to the group velocity of the form
\be\label{vgnonfermi}
v_g={\6 E\0\6p}=1-{g^2C_f\012\pi^2}\ln{m_D\0|E-\mu|}+\ldots\,,
\ee
which dominates over the contribution from the asymptotic fermion mass
for $|E-\mu|\ll m_D$, and eventually spoils (HDL) perturbation theory
when $|E-\mu|/m_D \lesssim \exp(-C/g^2)$.
Evidently, the quasistatic magnetic interactions lead to
significant changes in the vicinity of the Fermi surface.
Such non-Fermi-liquid corrections can even be the dominant
effects in certain quantities, such as the entropy
or specific heat at low temperature $T\ll m_D\sim g\mu$, as
discussed in section \ref{seclowT}. 
In a colour superconductor, where the quasiparticle and quasiholes
at the Fermi surface develop a gap of order $b_1 \mu g^{-5} \exp(-c_1/g)$
as mentioned in section \ref{secCSC}, the non-Fermi-liquid corrections
to the real part of the fermion self-energy remain perturbative
and contribute at the level of the constant $b_1$, resulting
in a significant reduction of the magnitude of the gap
\cite{Brown:1999aq,Wang:2001aq} compared to previous results
which did not include non-Fermi-liquid contributions to the
quark self-energy \cite{Schafer:1999jg,Pisarski:1999bf,Pisarski:1999tv}.

One should note that
the nonanalytic behaviour of \eref{ReSigmazerot} is
indeed in the energy variable rather than the momentum, despite
the fact that the derivation has been on-shell where the two
are related, as can be verified by explicit calculation
of ${\rm Re}\,\6\Sigma_+/\6p_i$, which is indeed analytic on the
Fermi surface \cite{Brown:2000eh}. 
One consequence of this is that
similar logarithmic terms appear in
the quark-gluon vertex only in a special kinematic
regime, namely in
\be\fl
\lim_{E'\to E\approx\mu} \lim_{p'\to p\approx\mu}
\Lambda_\mu(E',\9p';E,\9p)=
{g^3 C_f\012\pi^2}\delta_\mu^0 \ln{m_D\0|E-\mu|}
+\Or(|E-\mu|),
\ee
where $(E',\9p')$ and $(E,\9p)$ are the four-momenta of the quarks,
but not when the order of the limits is interchanged.
Since magnetic interactions become (almost) unscreened only
in the regime where $\omega\ll k$, it can been argued that
non-Fermi-liquid corrections manifest themselves only
through corrections on the quark self-energy rather than the vertices
\cite{Brown:2000eh}.



\subsection{NLO corrections to real parts of dispersion laws
at high temperature}

At nonvanishing temperature, the real parts of the
dispersion laws of fermionic and gluonic quasi-particles
remain IR-safe in NLO HTL-resummed perturbation theory
(in contrast to the imaginary parts which are perturbatively
accessible only at $T=0$ or for $\mathbf k=0$ and
exceptional momenta where the group
velocity vanishes).
However, such calculations are 
exceedingly involved, and
only some partial results exist so far 
in QCD \cite{Flechsig:1995uz,Flechsig:1998mn}.

In the following, we shall restrict our attention to
the case $\mathbf k^2/\omega_{\rm pl.}^2 \gg 1 $ and consider
the different branches of the dispersion laws in turn.


\subsubsection{Energetic quarks and transverse gluons}

At large momenta only the normal branch of quark excitations and
the transverse gluons have nonnegligible spectral weight.
Their leading-order thermal masses become momentum independent
in this limit and are given in one-loop order
by their HTL values, $m_\infty^2=\hat m_D^2/2$ and
$M_\infty^2=2 \hat M^2$ with $\hat m_D^2$ and $\hat M^2$ given
by \eref{mD} and \eref{MF} respectively.
While the HTL self-energies are no longer accurate at large
momenta, their light-cone limit
is exact to one-loop order and thus still determines the leading-order
asymptotic masses \cite{Kraemmer:1990drA,Flechsig:1996ju}.

There are however nevertheless higher-order corrections
which in fact require HTL resummation as indicated in
\eref{dmas} and \eref{dMas} and which have complicated
momentum dependence that has not been evaluated so far.
Presently, only the particular averages \eref{deltamasav} and \eref{deltaMasav}
are known from their relation to the plasmon effect in
$\Phi$-derivable thermodynamics as discussed in Sect.~\ref{secBIR}.

\subsubsection{Longitudinal Plasmons}\label{secplasmonfate}

For momenta $\mathbf k^2 \gg \omega_{\rm pl.}^2$, the longitudinal
plasmon branch approaches the light-cone, as can be seen in
figure~\ref{figg}. From 
$k^2=\hat\Pi_B(k)$ and (\ref{PiB}) one finds
\be
\omega^2_B(|\mathbf k|) \to 
\mathbf k^2\(1+4{\rm e}^{-6\mathbf k^2/(e^2 T^2)-2}\)
\ee
with $e^2 = g^2(N+N_f/2)$ in QCD, so the light-cone is
approached exponentially as $|\mathbf k|$ is increased. 
If one also calculates the residue,
one finds that this goes to zero at the same time, and exponentially so, too.
A similar behaviour occurs in the ``plasmino'' branch of the
fermion propagator at momenta $\mathbf k^2 \gg \hat M^2$.

Instead of QCD, we shall consider the
analytically tractable case of massless {scalar electrodynamics}
as
a simple toy model
with at least some similarities to the vastly more complicated QCD case
in that in both theories there are
bosonic self-interactions.
There are however no HTL vertices in scalar electrodynamics, 
which makes it possible
to carry out a complete momentum-dependent NLO calculation 
\cite{Kraemmer:1995az}.

Comparing HTL values of and NLO corrections to 
$\Pi_{00}=-\mathbf k^2\Pi_B/k^2$, one finds
that as $k^2\to0$
there are collinear singularities in both:
\index{collinear singularities}
\be\hat\Pi_B(k)/k^2 \to {\hat m_D^2\over 2\mathbf k^2} 
{ \ln{\mathbf k^2\over k^2}}
\ee
diverges logarithmically\footnote{This is in fact the technical reason
why the longitudinal branch approaches the light-cone exponentially
when $\mathbf k^2 \gg \omega_{\rm pl.}^2$.}, whereas
\be\label{dPiBNLOlc}
\delta\Pi_B/k^2 \to -e{\hat m^2 \over |\mathbf k|{\sqrt{k^2}}}
\ee
(with $\hat m\propto e T$ the thermal mass of the scalar).
Because (\ref{dPiBNLOlc})
diverges stronger than logarithmically, one has
$ \delta\Pi_B > \hat\Pi_B$ eventually as $ k^2\to 0$.
Clearly, this leads to a breakdown of perturbation theory
in the immediate neighbourhood of the light-cone (${k^2/|\mathbf k|^2}
\lesssim (e/\ln{1\over e})^2$), which
this time is not caused by the massless magnetostatic modes, but
rather by the massless hard modes contained in the HTL's.

\begin{figure}
\centerline{\includegraphics[
viewport = 170 360 460 600,scale=0.64]{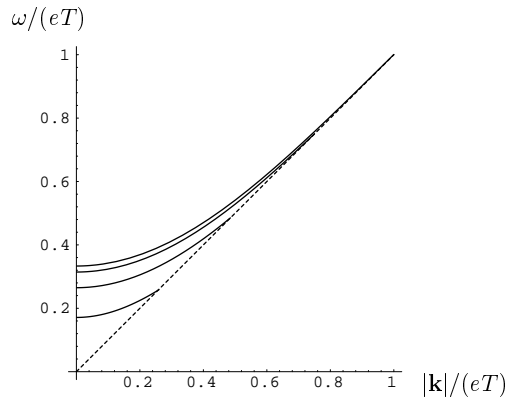}}

\caption{ The longitudinal {plasmon} branch of 
{scalar electrodynamics}
including NLO corrections to the HTL result. The upper of the
four lines gives the HTL result and the lines below correspond
to NLO corrections with $e=0.3$, 1, and 2, respectively.
The latter three lines cross the light-cone such that the
phase velocity starts to exceed 1, but with group velocity $<1$
throughout. In the space-like region, the plasmon modes are
damped by {Landau damping}, which is strong except in the immediate
neighbourhood of the light-cone, where it is suppressed by
a factor of $\exp\{-e\sqrt{\mathbf k/[8(\mathbf k-\omega)]}\}$
\label{Figpllc}}
\end{figure}

However, a self-consistent gap equation for the scalar thermal
mass implies that also the hard scalar modes have a thermal
mass $\sim eT$. Including this by extending the resummation
of the scalar thermal mass to hard internal lines
renders $\Pi_B/k^2$ 
finite in the light-cone limit, with the result
\bea\label{PiBk20}
\lim\limits_{k^2\to0}{\Pi_B^{\rm resum.}\over  k^2}
&= {e^2T^2\over  3\mathbf k^2}
  \Bigl[ 
 \ln{ 2T\over
  \hat m} 
+{1\over  2}
-\gamma_E+{\z'(2)\over  \z(2)} +O(e)\Bigr] \nonumber\\
&={e^2T^2\over  3\mathbf k^2}
  \Bigl[ \ln{2.094\ldots \over e}+O(e)\Bigr]
\eea
such that there is a solution to the dispersion law with $k^2=0$
at 
$\mathbf k^2/(e^2T^2)={1\over 3}\ln(2.094/ e)
+O(e)$. Because all {collinear singularities} have disappeared,
continuity implies that there are also solutions for
space-like momenta $k^2<0$, so the longitudinal plasmon branch pierces
the light-cone, having group velocity $v<1$ throughout, though,
as shown in figure~\ref{Figpllc}.
While at HTL level, the strong {Landau damping} at $k^2<0$
switches on discontinuously, it now does so smoothly
through an extra factor $\exp[-e\sqrt{|\mathbf k|/[8(|\mathbf k|-\omega)]}]$,
removing the longitudinal plasmons through over-damping
for $(|\mathbf k|-\omega)/|\mathbf k| \gtrsim e^2$.
This is in fact an essential singularity in the self-energy
now, which, as has been shown in reference~\cite{Weldon:2001vt}
occurs generally at $k_0=|\mathbf k|$ when the
hard degrees of freedom have nonzero mass. 

So the {collinear singularities} that spoil HTL-resummed perturbation
theory on the light-cone are associated with a slight but
nevertheless qualitative change of the spectrum of longitudinal
plasmons: instead of being time-like throughout and 
existing for higher momenta, albeit with
exponentially small and decreasing
residue and effective mass, they become space-like at
a particular point $|\mathbf k|_{\rm crit.} \sim eT\ln{1\over e}$.
For $|\mathbf k|>|\mathbf k|_{\rm crit.}$ Landau damping sets
in smoothly but rapidly, so that these modes soon become over-damped
as $|\mathbf k|$ is increased.

This phenomenon is in fact known to occur in non-ultrarelativistic 
($T<m_e$) QED \cite{Tsytovich:1961}, and has been considered in the case of
QCD by Silin and Ursov \cite{Silin:1988js},
who speculated that it may lead to Cherenkov-like phenomena in the
quark-gluon plasma.

In QCD, the situation is in fact much more complicated. Under the
assumption that the {collinear singularities} are removed solely by the
resummation of \index{asymptotic thermal mass}
asymptotic gluonic and fermionic thermal masses
in hard internal lines, the value of $|\mathbf k|$
where longitudinal plasmons turn space-like
has been calculated in \cite{Kraemmer:1995az}.
For a pure-glue plasma, it reads
\be\label{k2critQCD}
\mathbf k^2_{\rm crit.}=g^2T^2[\ln{1.48\ldots\over g}+\ldots].
\ee
Such an extended resummation can in fact be related to
an improved and still gauge-invariant version of the
HTL effective action \cite{Flechsig:1996ju}. However
it is likely that damping effects are of equal importance
here (in contrast to scalar electrodynamics), so that in
(\ref{k2critQCD})
only the coefficient of the logarithm is complete, but not
the constant under the log.

\section{Resummations beyond hard thermal loops}
\label{secbey}

In the previous section we have considered perturbative corrections
to dynamical quantities at soft scales $\sim gT$. As soon as there
is a sensitivity to ``ultrasoft'' scales $\sim g^2T$, (HTL) perturbation
theory breaks down and typically only leading logarithms can
be computed as we have discussed.
This is in particular the case when external momenta are
either ultrasoft or very close to the light-cone.

In the following we shall briefly discuss a few such cases where
resummations of the perturbative series have to include more than
the HTL diagrams and where important progress has been achieved recently.

\subsection{Ultrasoft amplitudes}

For ultrasoft external momenta, the infrared sensitivity 
may in fact become so large that HTL diagrams are no longer
of leading order. The diagrams of figure \ref{Figdpi} have
in fact been evaluated in \cite{Bodeker:1999ud} for $k_0,|\9k|\lesssim g^2T$
with the result
\be\label{usapila}
\fl
\delta\Pi_{\mu\nu}(k)=-i\hat m_D^2 N g^2T
k_0 \int {d\Omega_{\9v}\over 4\pi}\int {d\Omega_{\9v'}\over 4\pi}
{v_\mu v'_\nu \0 (v\cdot k)(v' \cdot k)}
\left[ I(\9v,\9v') \ln{gT\0\lambda} + {\rm finite} \right]
\ee
with $\lambda\ll gT$ and
\be\label{Ivvp}
I(\9v,\9v')=-\delta^{(S_2)}(\9v-\9v')+{1\0\pi^2}{(\9v\cdot \9v')^2\0
\sqrt{1-(\9v\cdot \9v')^2}}.
\ee
For $k_0,|\9k|\sim g^2T$ this is of the same order than the
HTL self-energy $\hat\Pi_{\mu\nu}\sim \hat m_D^2$ and even logarithmically
enhanced.

This strong infrared sensitivity is 
entirely due to the HTL parts of the dressed vertices in
figure \ref{Figdpi}, with contributions involving tree-level vertices being
suppressed by powers of $g$. By blowing up the
HTL vertices in figure \ref{Figdpi}, one may view the diagram
involving one 4-vertex
as one hard loop with one soft propagator insertion
so that the largeness of \eref{usapila} indeed means that
corrections within a HTL diagram cease to be perturbative.
This is in fact special to
nonabelian gauge fields and absent in Abelian theories
(where there are no HTL vertices involving exclusively
gauge fields to build the diagrams of figure \ref{Figdpi});
in Abelian theories there are cancellations between self-energy
and vertex corrections \cite{Lebedev:1992kt,Kraemmer:1995az,Carrington:1998at,Carrington:1998fx},
which however do not carry over to the nonabelian case.

\begin{figure}
\centerline{\includegraphics[width=8truecm]{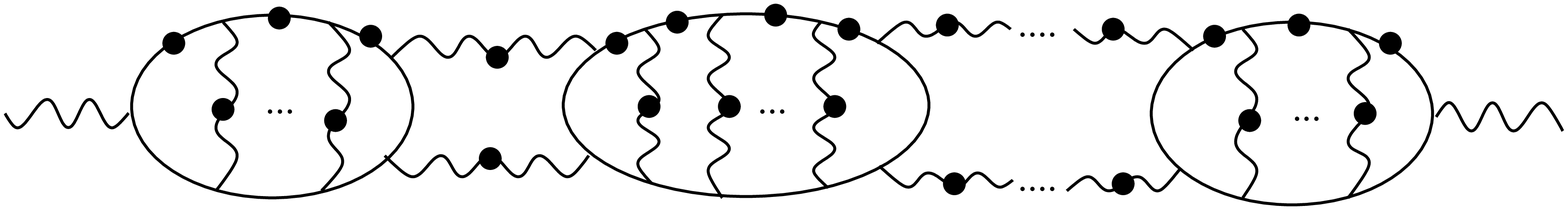}}
\caption{Generalized ladder diagrams contributing to ultrasoft
amplitudes.
\label{FigUSA}}
\end{figure}

Moreover, it turns out that there are infinitely many diagrams
that need to be taken into account in order to obtain the
leading order terms in ultrasoft amplitudes. In the case
of the gluon self-energy, such diagrams are generalized
ladder diagrams as shown in figure \ref{FigUSA}.

In order to characterize the leading contributions to ultrasoft
amplitudes, a diagrammatic approach is no longer practicable, but
it turns out that one can generalize the kinetic theory treatment
of HTL's to ultrasoft amplitudes. This requires the inclusion
of a collision term which is proportional to the colour relaxation
time $\tau_{col}\sim 1/\gamma$, where $\gamma$ is given
by \eref{Damping} in the large-momentum limit ($v_A=1$).

In leading-logarithmic approximation the 
required nonabelian Boltzmann equation
which generalizes the nonabelian Vlasov equation \eref{cDWvE} has the form
\cite{Bodeker:1998hm,Arnold:1998cy,Litim:1999ns,Blaizot:1999xk,Litim:1999id,Blaizot:1999fq}
\be
[v\cdot D,W(x,\9v)]^a=\9v\cdot \9E^a(x)
+\gamma \int{d\Omega_{\9v'}
} I(\9v,\9v') W(x,\9v'),
\ee
with $I(\9v,\9v')$ defined in \eref{Ivvp}.
Beyond the leading-log approximation
\cite{Blaizot:1999fq,Arnold:1999uz,Arnold:1999ux,Arnold:1999uy} 
the quantity $\gamma I(\9v,\9v')$
needs to be replaced by a collision operator that involves also
the HTL gauge-boson self-energies. The ultrasoft
vertex functions obtained formally in analogy to the expansion
\eref{jindexp} however still share many of the remarkable
properties of the hard thermal loops, namely gauge-fixing
independence and simple Ward identities \cite{Blaizot:1999fq}.

At still smaller scales, $\omega \ll |\9k| \ll \gamma$, it is
in fact possible to simplify the above effective
theory of ultrasoft modes and to construct a local effective theory 
(B\"odeker's effective theory)
\cite{Bodeker:1998hm,Arnold:1998cy,Arnold:1999jf}, because
on distances $|\9k|^{-1}$ much larger than the colour
coherence length $\gamma^{-1}$ there are no longer
propagating modes with definite colour. 
B\"odeker's  effective theory is a stochastic theory given by the
Langevin equation
\be\label{BET}\fl
\sigma_c \9E^a=(\9D\times \9B)^a+\bzeta^a,\quad
\langle\zeta^{ia}(x)\zeta^{jb}(x')  \rangle=2\sigma_c T \delta^{ij}
\delta^{ab}\delta^4(x-x'),
\ee
where $\bzeta$ is a Gaussian noise term and
$\sigma_c$ the colour conductivity given to leading
order by \eref{sigmac}; beyond leading-log order, this 
parameter has been
determined in reference \cite{Arnold:1999uy}.
The effective theory \eref{BET}
is also UV finite and
was put to use in the numerical calculation \cite{Moore:1998zk} of
the rate of baryon number violation \cite{Kuzmin:1985mm} in the hot
symmetry-restored phase of electroweak theory, which at leading
order is governed by nonperturbative nonabelian gauge field
fluctuations with spatial momenta $\sim g^2T$ and
frequencies $\sim g^4 T \ln(1/g)$.


\subsection{Light-like external momenta}

As we have seen in the example of NLO corrections to
the dispersion law of longitudinal plasmons in section \ref{secplasmonfate},
the HTL perturbation theory becomes insufficient not only
for ultrasoft momenta $|\9k|\lesssim g^2 T$, but also
when amplitudes involve harder momenta that are however nearly light-like,
which gives rise to collinear singularities.
This problem has surfaced in particular in the calculation of the
production rate of real (non-thermalized) photons in
a quark-gluon plasma 
from HTL-resummed perturbation
theory \cite{Baier:1994zb,Aurenche:1996is,Aurenche:1996sh}
and it turned out that damping effects on hard internal
lines have to be included as the dominant regulator
of collinear singularities
\cite{Gelis:2000dp}.

\begin{figure}
\centerline{\includegraphics[viewport=100 525 495 660,scale=0.3]{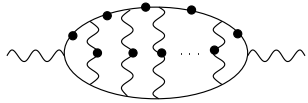}}
\caption{Ladder diagrams contributing to photon production
at leading order.
In a quark-gluon plasma, the external lines correspond to
on-shell photons, the internal lines to dressed quark and gluon
propagators. The quarks are hard but collinear with the
external photon and exchange soft gluons such as to maintain
collinearity.
\label{Figlad}}
\end{figure}

Moreover, for a complete leading-order calculation it turns
out to be necessary to resum all ladder diagrams built from
dressed propagators of the form shown in figure \ref{Figlad}.
This reflects the necessity to take into account
the physical phenomenon of Landau-Pomeranchuk-Migdal suppression\footnote{For
a review in the context of energy loss calculations see
e.g.\ reference \cite{Baier:2000mf}}
of photon emission
\cite{Aurenche:2000gf}. 
However, contrary to the conclusion arrived at
in reference \cite{Aurenche:1999tq}, thermal photon production
from a QCD plasma was shown in reference
\cite{Arnold:2001ba} not to be sensitive to ultrasoft nonperturbative
physics,
and a complete leading-order evaluation of the photon production
rate from a hot quark-gluon plasma 
was finally
accomplished in references \cite{Arnold:2001ms,Arnold:2002ja}. (The analogous
problem in dilepton production was recently solved in
reference \cite{Aurenche:2002wq}.)

\subsection{Transport coefficients}

Another important example where an expansion in number of loops breaks down
is in transport coefficients, which in linear response theory
are given by Kubo relations \cite{Hosoya:1984id}
through correlation functions
in the limit
of vanishing frequency and at zero spatial momentum of the form
\be
\lim_{\omega\to 0}{1\0\omega}\int d^4x e^{i\omega t}
\left\langle [\mathcal O(t,\9x), \mathcal O(0)] \right\rangle,
\ee
where $\mathcal O$ represents components of a conserved current or of the stress
tensor.

In scalar field theory, the required resummation 
(at leading order) has been worked out in
references \cite{Jeon:1995if,Jeon:1996zm,Wang:2002nb}. The relevant
diagrams turn out to be uncrossed ladder diagrams, where
each rung brings a pair of propagators with pinching
singularities cut off by their thermal width.

In gauge theory, transport coefficients turn out to depend
on physics at soft scales such as to require also hard-thermal-loop
resummation. For example, the leading term in the
shear viscosity has the parametric form
$\eta \propto T^3/(g^4\ln(1/g))$, where the logarithm
is due to the dynamical screening 
provided by the HTL gauge propagator
\cite{Baym:1990uj,Heiselberg:1994vy},
and an analogous leading logarithm appears in the various
diffusion constants \cite{Heiselberg:1994px}.
The initial calculations of these quantities in kinetic
theory however turned out to be incomplete already to leading-logarithmic
accuracy and have been completed only recently in \cite{Arnold:2000dr}.

In reference \cite{ValleBasagoiti:2002ir} a purely
diagrammatic rederivation 
of the relevant integral equation starting from the Kubo formula
has been given, which is however incomplete as far as
Ward identities are concerned; its completion in the Abelian
case has been discussed in
\cite{Aarts:2002tn,Boyanovsky:2002te}. To organize a diagrammatic calculation
of transport coefficients, the 2PI approximation scheme
mentioned in section \ref{subsecscappconv} is an efficient
means. It turns out that the 3-loop $\Phi$-derivable approximation
is necessary and sufficient to obtain the leading-order results
for shear viscosity
in a scalar theory \cite{Calzetta:1999ps,Aarts:2003bk},
and to leading-log order in QED \cite{Aarts:2003bk}.

Whereas at leading-log accuracy the diagrams that need to be
resummed are dressed uncrossed ladder diagrams like those
of figure \ref{Figlad}, the set of diagrams that contribute in a
complete leading-order
calculation is still much more complicated.
Their resummation has nevertheless been achieved recently
by means of a Boltzmann equation with
a collision term evaluated to
sufficient accuracy 
and with the resulting integral equation
treated by variational techniques. The results obtained 
within this effective kinetic theory \cite{Arnold:2002zm} 
so far comprise
shear viscosity, electrical conductivity, and flavour
diffusion constants \cite{Arnold:2003zc} (bulk viscosity, however,
turns out to be more difficult and has not yet been
determined even to leading-log accuracy).

Nonperturbative (all order in $g$) 
results for transport coefficients in a gauge theory
have been derived for the toy model of
large-flavour-number QED or QCD in reference \cite{Moore:2001fg}
allowing for a test of the quality of
HTL approximations, which turn out to work
remarkably well up to the point where
the renormalization-scale dependence becomes the dominant
uncertainty.

\section{Hard thermal loops and gravity}
\label{secgrav}

In ultrarelativistic field theory, the HTL approximation
to the polarization tensor of gauge bosons
determines the leading-order
spectrum of gauge-field quasi-particles and the linear response
of an ultrarelativistic plasma to external perturbations.

In the physics of the very early universe, which is filled
with a hot plasma of various elementary particles, the
gravitational polarization tensor is also of interest.
It describes the (linear) response to metric perturbations and
its infrared behaviour determines the evolution of
large-scale cosmological perturbations, which provide
the seeds for structure formation that are nowadays
being studied directly through the anisotropies
of the cosmic microwave background and which are being measured
with stunning accuracy \cite{
Jaffe:2000tx,Bennett:2003bz}.
The gravitational polarization tensor 
is also a central quantity in the theory of
stochastic (semiclassical) gravity \cite{Hu:1999mm,Hu:2003qn}, which
aims at a general self-consistent description of
quantum statistical fluctuations of matter in
a curved background geometry as a stepping-stone
towards a full quantum theory of gravity.

\subsection{HTL gravitational polarization tensor}

If $\Gamma$ denotes all contributions to the effective action besides
the classical Einstein-Hilbert action, the energy-momentum tensor
is given by the one-point 1PI vertex function
\begin{equation}
T_{\mu\nu}(x)={2\over\sqrt{-g(x)}}
{\delta\Gamma\over\delta g^{\mu\nu}(x)}
\end{equation}
and the gravitational polarization tensor by the two-point function
\begin{equation}
-\Pi_{\mu\nu\alpha\beta}(x,y)\equiv
{\delta^2\Gamma\over\delta g^{\mu\nu}(x)\delta g^{\alpha\beta}(y)}=
{1\over2}{\delta\left(\sqrt{-g(x)}T_{\mu\nu}(x)\right)
\over\delta g^{\alpha\beta}(y)}.
\end{equation}
{From} the last equality it is clear that 
$\Pi^{\alpha\beta\mu\nu}$ describes the response
of the (thermal) matter energy-momentum tensor
to perturbations in the metric. Equating $\Pi^{\alpha\beta\mu\nu}$
to the perturbation of the Einstein tensor gives self-consistent
equations for metric perturbations and, in particular, cosmological
perturbations.

In the ultrarelativistic limit where all bare masses can be
neglected and in the limit of temperature much larger than
spatial and temporal variations, the effective (thermal) action
is conformally
invariant, that is $\Gamma[g]=\Gamma[\Omega^2g]$ (the
conformal anomaly like other renormalization issues can be
neglected in the high-temperature domain).

This conformal invariance is crucial for the
application to cosmological perturbations for two
reasons. Firstly, it allows us to have matter 
in thermal equilibrium
despite a space-time dependent metric. As long as the latter is
conformally flat,
$ds^2=\sigma(\tau,{\bf x})[d\tau^2-d{\bf x}^2]$,
the local temperature on the curved background
is determined by the scale factor $\sigma$.
Secondly, the thermal correlation functions are simply given
by the conformal transforms of their counterparts on a flat 
background,
so that ordinary momentum-space techniques can be employed for their
evaluation.

The gravitational polarization tensor in the HTL approximation
has been first calculated fully in \cite{Rebhan:1991yr}, after
earlier work 
\cite{Gross:1982cv,Kikuchi:1984np,Nakazawa:1985zq,Gribosky:1989yk}
had attempted (in vain) to identify the Jeans mass (a negative Debye mass
squared signalling gravitational instability rather than screening
of static sources) from
the static limit of the momentum-space
quantity $\tilde\Pi_{\mu\nu\alpha\beta}(k)$,
$k_0=0$, $\mathbf k\to0$ in flat space.

Like the HTL self-energies of QED or QCD, $\Pi_{\mu\nu\alpha\beta}$
is an inherently nonlocal object. Because it is a tensor of rank 4,
and the local plasma rest frame singles out the time direction,
it has a much more complicated structure. From $\eta^{\mu\nu}$,
$u_\mu=\delta_\mu^0$ and $k_\mu$ one can build 14 tensors to 
form a basis for $\tilde\Pi_{\mu\nu\alpha\beta}(k)$. Its HTL
limit ($k_0,|\mathbf k| \ll T$), however, satisfies the Ward identity
\begin{equation}\label{Wardid}
4k^\mu\tilde\Pi_{\mu\nu\alpha\beta}(k)=
k_\nu T_{\alpha\beta}-k^\sigma \left(T_{\alpha\sigma}\eta_{\beta\nu}+
T_{\beta\sigma}\eta_{\alpha\nu}\right)
\end{equation}
corresponding to diffeomorphism invariance as well as a further
one corresponding to conformal invariance,
the ``Weyl identity''
\begin{equation}\label{Weylid}
\eta^{\mu\nu}\tilde\Pi_{\mu\nu\alpha\beta}(k)=-
{1\over2}T_{\alpha\beta},
\end{equation}
where $T_{\alpha\beta}=P_0(4\,\delta^0_\mu\delta^0_\nu-\eta_{\mu\nu})$
and $P_0$ the ideal-gas pressure $\propto T^4$.
These identities reduce the number of independent structure functions
to three, which may be chosen as
\begin{equation}\fl
\Pi_1(k)\equiv\tilde\Pi_{0000}(k)/\rho,\quad
\Pi_2(k)\equiv\tilde\Pi_{0\mu}{}^\mu{}_0(k)/\rho,\quad
\Pi_3(k)\equiv\tilde\Pi_{\mu\nu}{}^{\mu\nu}(k)/\rho,
\end{equation}
where $\rho=T_{00}=3 P_{SB}$.

The HTL limit is in fact universal: the thermal matter may be
composed of any form of ultrarelativistic matter such as
also gravitons, which are equally important as any other
thermalized matter if the graviton background has comparable temperature.
It reads\cite{Rebhan:1991yr}
\be\label{ABChat}
\hat \Pi_1(k)={k^0\02|{\bf k}|}\ln{k^0+|{\bf k}|\0k^0-|{\bf k}|}-{5\04},
\quad
\hat \Pi_2=-1, \quad \hat \Pi_3=0.
\ee

As in ordinary hot gauge theories, the entire HTL effective
action is determined by the Ward identities and has been
constructed in \cite{Brandt:1992qn,Frenkel:1995bg}. The 3-graviton HTL vertex
has moreover been worked out explicitly in 
\cite{Frenkel:1991dw,Brandt:1993dk}
and subleading corrections beyond the HTL 
approximation have been considered in \cite{deAlmeida:1994wy}.

{}From \eref{ABChat} one can formally derive an HTL-dressed
graviton propagator which contains three independent
transverse-traceless tensors. As the HTL propagators in gauge
theories, this describes collective phenomena in the form
of nontrivial quasi-particle dispersion laws and Landau damping
(from the imaginary part of $\hat A$). In the context of
cosmological perturbations, the latter
corresponds to the collisionless damping studied in \cite{Bond:1983hb},
as we shall further discuss below.

In the limit $k_0^2,\mathbf k^2\gg GT^4$, 
where $G$ is the gravitational coupling
constant, one can ignore any background curvature terms $G\rho\propto GT^4$,
and use the momentum-space propagator to read off
the dispersion relations for the three branches of
gravitational quasiparticles. One of these, the spatially
transverse-traceless branch, corresponds to the gravitons
of the vacuum theory. Denoting this branch by $\mathcal A$,
\eref{ABChat} implies an asymptotic thermal mass for the gravitons 
according to \cite{Rebhan:1991yr}
\be
\omega_{\mathcal A}^2\to k^2+m_{\mathcal A\infty}^2=k^2+{5\over9}
\times 16\pi G\rho.
\ee
While in a linear response theory branch $\mathcal A$ neither couples to perturbations
in the energy density $\delta T^{00}$ nor the energy flux $\delta T^{0i}$,
the two additional branches $\mathcal B$ and $\mathcal C$
both couple to energy flux, but only $\mathcal C$ to energy density.
The additional branches correspond to excitations that are
purely collective phenomena, and like the longitudinal plasmon
in gauge theories, they disappear from the spectrum
as $k^2/(GT^4)\to \infty$. In contrast to ordinary
gauge theories, however, they do so by acquiring effective thermal
masses that grow without bound, while at the same time
the residues of the corresponding poles disappear in a power-law
behaviour (instead of becoming massless and 
disappearing exponentially) \cite{Rebhan:1991yr}.
(Some subleading corrections to these asymptotic masses at one-loop
order have been determined in \cite{Brandt:1998hd}, but
these are gauge dependent and therefore evidently incomplete.)

For long wavelengths $\mathbf k^2\sim GT^4$ the momentum-space HTL graviton
propagator exhibits an instability, most prominently in mode $\mathcal C$,
which is reminiscent of the gravitational Jeans instability
and which is only to be expected since gravitational sources unlike charges
cannot be screened.
However, this instability occurs at momentum scales which
are comparable with the necessarily nonvanishing curvature
$R\sim GT^4$. The flat-space graviton propagator is therefore
no longer the appropriate quantity to analyse;
metric perturbations have to be studied in a curved
time-dependent background.

\subsection{Selfconsistent 
cosmological perturbations from thermal field theory}

The conformal covariance of the HTL gravitational polarization
tensor $\hat\Pi_{\mu\nu\alpha\beta}(x,y)$ as expressed by the
Weyl identity \eref{Weylid} determines this non-local quantity
in a curved but conformally flat space according to
\begin{equation}
\hat\Pi_{\mu\nu\alpha\beta}(x,y)\Big|_{g=\sigma\eta}=\sigma(x)\sigma(y)
\hat\Pi_{\mu\nu\alpha\beta}(x-y)\Big|_{g=\eta}.
\end{equation}

A closed set of equations for metric perturbations is obtained from
using this in the right-hand side of the Einstein
equation linearized around a conformally-flat background cosmological model
\cite{Rebhan:1991yr,Kraemmer:1991du,Rebhan:1992rt}
\bea\label{dGmn}\fl
\d G^{\mu\nu}&\equiv&{\d(R^{\mu\nu}-\2 g^{\mu\nu}R)\0\d g^{\a\b}}
\d g^{\a\b}=-8\pi G\,\d T^{\mu\nu}
=-8\pi G\, 
\int_{x^\prime} {\delta T^{\mu\nu}(x)\over
\delta g_{\alpha\beta}(x^{\prime})} \delta g_{\alpha\beta}(x^{\prime}) 
\nonumber \\ \fl
&=& 4 \pi G  \left[T^{\mu\nu} g^{\alpha\beta}
+ 2 T^{\mu\alpha} g^{\beta\nu} \right]\delta g_{\alpha\beta}(x)
+{16\pi G\over \sqrt{-g(x)}} \int_{x^\prime}
\Pi^{\mu\nu\alpha\beta}(x,x^\prime) \delta
g_{\alpha\beta}(x^{\prime}).
\eea
In a hydrodynamic approach, $\delta T^{\mu\nu}$ is usually determined
by certain equations of state together with covariant conservation, the
simplest case of which is that of a perfect fluid, which has been
studied in the pioneering work of Lifshitz 
\cite{Lifshitz:1946du,Lifshitz:1963ps}. Many generalizations have
since been worked out and cast into a gauge invariant
form by Bardeen \cite{Bardeen:1980kt} (see also 
\cite{Kodama:1985bj,Mukhanov:1992me}).

Relativistic and (nearly) collisionless matter, however, has more
complicated gravitational interactions than a perfect fluid. This
is usually studied using classical kinetic theory \cite{Ehlers:1971,Ste:NRKT}
and the case of purely collisonless matter has been worked out
in \cite{Stewart:1972} with some numerical solutions obtained
in \cite{Peebles:1980,Bond:1983hb,Schaefer:1991xn} for particular
gauge choices.

\subsubsection{Purely collisionless matter}
In \cite{Kraemmer:1991du}, the self-consistent equations for (scalar)
density perturbations of a radiation-dominated
Robertson-Walker-Friedmann model with collisionless matter
have been derived using the HTL gravitational
polarization tensor. By virtue of the diffeomorphism Ward
identity \eref{Wardid}, these equations turn out to be
automatically gauge independent.

For example, in the simple radiation-dominated and
spatially-flat Einstein-de Sitter model
\be\label{ds2}
ds^2=\sigma(\tau)(d\tau^2-d{\bf x}^2),\qquad
\sigma(\tau)={8\pi G\rho_0\03}\tau^2,
\ee
where $\rho_0$ is the energy density when $\sigma=1$ and
$\tau$ is the conformal time (which equals the size of
the Hubble horizon in comoving coordinates), the scalar
part of metric perturbations 
can be parametrized in terms of four
scalar functions
\be
\d g_{\mu\nu}^{(S)}=\s(\tau) 
\pmatrix{C & D_{,i} \cr D_{,j} & A\d_{ij}+B_{,ij}\cr} .
\ee
Of these, two can be gauged away by diffeomorphisms. But
in a gauge-invariant framework only gauge-invariant combinations
enter nontrivially. Only two independent gauge-invariant combinations
exist, which may be chosen as
\bea
\Phi&=& A+{\dot\s\0\s}(D-\2\dot B) \\
\Pi &=& \2(\ddot B+{\dot\s\0\s}\dot B+C-A)-\dot D-{\dot\s\0\s}D,
\eea
where a dot denotes differentiation with respect to the conformal time
variable $\tau$.

Each spatial Fourier mode with wave vector $\bf k$ is
related to perturbations in the energy density and anisotropic pressure
according to
\be\label{deltapianis}
\delta={1\03}x^2 \Phi,\qquad \pi_{\rm anis.}={1\03}x^2 \Pi,
\ee
where
\be\label{xdef}
x\equiv{k\tau}={R_H\0\lambda/(2\pi)},
\ee
which measures the (growing) size of the Hubble horizon over
the wavelength of a given mode (which is constant in comoving coordinates).
In \eref{deltapianis} 
energy density perturbations $\delta$ are defined with respect to
space-like hypersurfaces representing everywhere the local rest frame
of the full energy-momentum tensor, whereas $\pi_{\rm anis.}$ is an
unambiguous quantity, since there is no anisotropic pressure in
the background.

Correspondingly, when specifying to scalar perturbations, there are
just two independent equations contained in \eref{dGmn}. Because
of conformal invariance, the trace of \eref{dGmn} is particularly
simple and yields a finite-order differential equation in $x$,
\be
\Phi''+{4\0x}\Phi'+{1\03}\Phi={2\03}\Pi-{2\0x}\Pi' 
\ee
(a prime denotes differentiation with respect to the dimensionless
time variable $x$).
The other components, however, involve the nonlocalities of
the gravitational polarization tensor. These lead to an
integro-differential equation, which upon imposing retarded boundary
conditions reads \cite{Kraemmer:1991du}
\be\label{scp}\fl
(x^2-3)\Phi+3x\Phi'=6\Pi-12\int_{x_0}^x dx'\,
j_0(x-x')[\Phi'(x')+\Pi'(x')]+\varphi(x-x_0)
\ee
where $j_0(x)=\sin(x)/x$ arises as Fourier transform of
$\hat\Pi_1(\omega/k)$ in \eref{ABChat}.
$\varphi(x-x_0)$ encodes the initial conditions,
the simplest choice of which corresponds to $\varphi(x-x_0)\propto j_0(x-x_0)$.

Similar integro-differential equations have been obtained
from coupled Einstein-Vlasov equations in particular gauges, and
the above one can be shown to arise from a gauge-invariant
reformulation of classical kinetic theory \cite{Rebhan:1994zw}.
Usually, such equations are studied numerically,
with only some asymptotic behaviour having been analysed
analytically \cite{Zakharov:1979,Vishniac:1982}.
Remarkably enough,
they can be solved analytically \cite{Kraemmer:1991du} provided 
initial conditions
are formulated for $x_0\to0$. In this case 
a power series ansatz for $\Phi$ and $\Pi$
leads to solvable recursion relations for an
alternating series
that converges faster than trigonometric functions. 

This also holds true for the vector (rotational) and tensor 
perturbations and
when the more realistic case of a two-component system of a perfect
radiation fluid combined with a collisionless 
ultrarelativistic plasma is considered 
\cite{Rebhan:1992rt}. In the case of rotational perturbations
these studies led to novel solutions not 
considered before \cite{Rebhan:1992sr}.

\begin{figure}
\centerline{ 
\includegraphics[viewport=150 215 450 510,width=8cm]{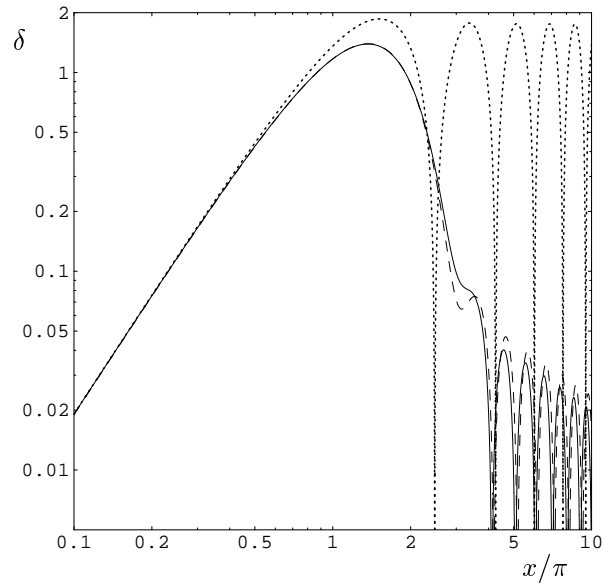} }
\caption{The energy-density contrast (arbitrary
normalization) as a function of $x/\pi$ for a collisionless
ultrarelativistic plasma (full line), a scalar plasma with
quartic self-interactions $\lambda\phi^4$ and $\lambda=1$ (dashed
line), and a perfect radiation fluid (dotted line). }
\label{plot}\end{figure}

In figure~\ref{plot}, the solution for the energy-density contrast
is given in a doubly-logarithmic plot (full line) and compared with
the perfect-fluid case (dotted line). In the latter, one has growth
of the energy-density contrast as long as the wavelength of the
perturbation exceeds the size of the Hubble radius ($x\ll1$).
After the Hubble horizon has grown such as to encompass about
one half wavelength ($x=\pi$), further growth of the perturbation
is stopped by the strong radiation pressure, turning it into
an (undamped) acoustic wave propagating with the speed of sound in radiation,
$v=1/\sqrt3$. The collisionless case is similar as concerns
the superhorizon-sized perturbations, but after horizon crossing,
there is strong damping $\sim1/x$, and the phase velocity is about 1.
This indeed reproduces the findings of the numerical studies
of reference~\cite{Bond:1983hb}. They can be understood as follows: a energy-density
perturbation consisting of collisionless particles propagates with
the speed of their constituents, which in the ultrarelativistic case
is the speed of light, and there is collisionless damping
in the form of directional dispersion.

Tensor perturbations, which correspond to primordial gravitational
waves, are also modified by a nearly collisionless background component
as worked out in references~\cite{Rebhan:1992rt,Rebhan:1994zw}.

\subsubsection{Weak self-interactions and HTL resummation}

The thermal-field-theory treat\-ment of the effects
of an ultrarelativistic plasma on the evolution of cosmological perturbations
can be used also to study the effect of weak self-interactions
in the plasma by calculating higher-order contributions
to the gravitational polarization tensor. A virtue of this approach
is that everything is formulated in purely geometrical terms,
without explicit recourse to perturbations in the (gauge variant)
distribution functions of a kinetic-theory treatment.

\begin{figure}
\centerline{
\includegraphics[width=5cm,bb=82.9728 385 385.027 488.732]{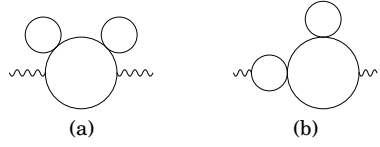} }
\caption{Two examples of infrared divergent contributions to
the gravitational polarization tensor in scalar $\phi^4$
theory beyond two-loop order.}
\label{f3}\end{figure}

In reference~\cite{Nachbagauer:1996wn}, 
the gravitational polarization tensor has been
calculated in a $\lambda\phi^4$ theory through order $\lambda^{3/2}$.
The next-to-leading order contributions to $\Pi_{\mu\nu\a\b}$
at order $\lambda^1$ are contained
in the high-temperature limit of two-loop diagrams and their evaluation is
straightforward. However, starting at three-loop order, there are
infrared divergences which signal a breakdown of the conventional
perturbative series. This is caused by the generation of a thermal
mass $\propto \sqrt{\lambda}T$ for
the hot scalars. If this is not resummed into a
correspondingly massive scalar propagator, repeated insertions
of scalar self-energy diagrams in a scalar line produces arbitrarily
high powers of massless scalar propagators all with the same momentum,
and thus increasingly singular infrared behaviour (figure~\ref{f3}a).

However, it is not sufficient to resum this thermal mass for the
hot scalars, as this would break conformal invariance (which in the
ultrarelativistic case is only broken by
the nonthermal conformal anomaly).
Indeed, there are also vertex subdiagrams $\propto\l T^2$ 
that have a similar
effect as a self-energy insertion, see figure~\ref{f3}b.
As in the hard-thermal-loop
resummation program for ordinary gauge theories, one has to resum also 
nonlocal vertex contributions.
Doing so, the result turns out to satisfy both the diffeomorphism
and conformal Ward identities. 

In the low-momentum limit of interest in the theory of
cosmological perturbations, the function $\Pi_1$ in \eref{ABChat} that
governs the evolution of scalar perturbations reads, to
order $\lambda^{3/2}$
\begin{eqnarray}\label{Ar}\fl
\Pi_1(\varpi) = \varpi\,{\rm artanh}{1\over \varpi}-\frac54
+{5\lambda\over 8\pi^2}\left[ 2\left(\varpi\,{\rm artanh}
{1\over \varpi}\right)^2
-\varpi\,{\rm artanh}{1\over \varpi}-
{\varpi^2\over {\varpi^2-1}} \right] \nonumber\\
\fl\qquad+{5\lambda^{3/2}\over 8\pi^3} \biggl[
3\left({\varpi^2-1}-\varpi\sqrt{\varpi^2-1}\right)\left(\varpi\,{\rm artanh}
{1\over \varpi}\right)^2 \nonumber\\
\fl\qquad\qquad+6 \left( \varpi\sqrt{\varpi^2-1}-\varpi^2-{\varpi\over \sqrt{\varpi^2-1}}
\right)\varpi
\,{\rm artanh}{1\over \varpi}\nonumber\\
\fl\qquad\qquad+{\varpi\over ({\varpi^2-1})^{3/2}}+3{\varpi^2\over {\varpi^2-1}}+
6{\varpi\over \sqrt{\varpi^2-1}}-3\varpi\sqrt{\varpi^2-1}+3\varpi^2
\biggr]
\end{eqnarray}
where $\varpi\equiv k_0/|\mathbf k|$,
and similarly complicated expressions arise for $\Pi_2$ and $\Pi_3$, which
in the collisionless limit \eref{ABChat} are simple constants.

The Fourier transform of this expression determines the kernel
in the convolution integral of \eref{scp}. At order $\lambda^1$,
it can still be expressed in terms of well-known special functions
\cite{Nachbagauer:1995jy}, whereas at order $\lambda^{3/2}$ this would involve
rather intractable integrals over Lommel functions. However, all
that is needed for finding analytical solutions is their power
series representations which are comparatively simple. 
Given them, it is as easy as before to solve the perturbation equations.
However, one finds that the asymptotic behaviour $x\gg1$ eventually
becomes sensitive to higher and higher loop orders. The reason for
this is that higher loop orders come with increasingly singular
contributions at $\varpi=\pm1$ to $\Pi(\varpi)$, and the large-$x$
behaviour is dominated by the latter. This could be cured by
a further resummation similar to the one considered in the
case of the dispersion relations of longitudinal plasmons 
in the vicinity of the light-cone in \eref{PiBk20}, but it turns out
that a particular Pad\'e-approximant based on the perturbative result 
reflects the effects of this further resummation quite well 
\cite{Nachbagauer:1996ec}.
The result for the density perturbations in a scalar plasma with
$\lambda\phi^4$-interactions and $\lambda=1$ are shown in figure~\ref{plot}
by the dashed line,
where it is compared with the collisionless case (full line) and the one of
a perfect radiation fluid (dotted line).
The effects of the self-interactions within the ultrarelativistic plasma
become important only for $x\gtrsim \pi$, where the strong collisionless
damping is somewhat reduced and the phase velocity is smaller than 1.

\section{Conclusion}

In this review we have concentrated on the progress made during
the last decade in calculating by analytic means static equilibrium and
dynamic near-equilibrium properties in ultrarelativistic gauge
theories. Even when the practical utility of the obtained results
is often open to debate, they are hopefully paving the way for
understanding the more complicated realistic situations to
be analysed in present or future heavy-ion experiments, or in
astrophysical problems.

Static quantities at high temperature and not too high chemical
potential can be efficiently investigated by lattice simulations,
whereas perturbative methods seemed to be too poorly convergent
to be of any predictive power in the applications of interest.
However, in the last few years there have been several different
investigations leading to the conclusion that a careful resummation
which emphasizes a weakly interacting quasiparticle picture,
or, in dimensional reduction, the effective-field-theory aspect,
is able to provide dramatic improvements, restoring predictivity
even down to a few times the deconfinement temperature in strongly
interacting QCD.

In nonabelian gauge theories the perturbative approach is
limited by the magnetic mass scale, or, more precisely, the
physics of confinement in the chromo-magnetostatic sector.
However, weak-coupling effective-field-theory methods can still
be used to combine analytical and (numerical) nonperturbative 
techniques to achieve further progress which is complementary to
or perhaps beyond
the capacities of a direct numerical approach in four dimensions.

In the case of dynamic properties, and also in the case
of static properties when dimensional reduction is not applicable
as in cold ultradegenerate plasmas,
the analytical approach is
even more important, and it is in fact here that the greatest
variety of phenomena are encountered. 
At weak coupling, there are several spatial
or temporal scales that need to be distinguished and are at the
root of the required resummations in a perturbative treatment.
In terms of frequencies and momenta,
the first important scale below the hard scale
of temperature or chemical potential is the soft scale
set by the Debye mass responsible for screening of electric fields
as well as for the frequency of long-wavelength plasma oscillations.
This is the realm of HTL resummations, which is however limited
by an eventual infrared sensitivity to ultrasoft scales occurring
at some (mostly very low) order of the expansion.

Resummations at zero temperature and high chemical potential
are not limited by the magnetic mass scale, but nearly
static magnetic modes lead to qualitative changes (non-Fermi-liquid
behaviour). There is furthermore the nonperturbative phenomenon
of colour superconductivity, which is however to some extent
accessible by weak-coupling methods, which strongly depend on HDL resummation
techniques.

At high temperature, there is,
between soft and ultrasoft scales, a further
scale set by the damping rate of the hard plasma constituents,
which in a nonabelian plasma determines a colour coherence
length. The corresponding energy scale is enhanced over the ultrasoft
scale by a logarithm in the inverse coupling, which allows for
novel systematic developments that we could only cursorily
describe.

In the final brief excursion to general relativity, we described
the role of the HTL contributions to the gravitational polarization
tensor in the theory of cosmological perturbations. There
the soft scale is given by the Jeans mass which is in a cosmological
situation comparable with the scale of the (inverse) Hubble horizon.

It should be needless to emphasize that many interesting
topics in thermal field theory have been covered
only cursorily or even not at all.
Some of the very recent developments in fact are about to
leave the arena of traditional thermal field theory towards a more
general, fully nonequilibrium field theory, which is only timely
in view of the wealth of experimental of data to be expected
from the modern relativistic heavy-ion colliders.
But (near-) equilibrium thermal field theory will certainly continue to be
an important theoretical laboratory where many physical
concepts are brought together for cross-fertilization.

\appendix
\def\thesection{\Alph{section}}
\section{Spectral representation of HTL/HDL propagators}

\subsection{Gauge boson propagator}
\label{App:A}

For the two nontrivial structure functions of the HTL/HDL gauge boson
propagator corresponding to the branches A and B it is convenient
to separate off the kinematical pole at $k^2=0$ in $\Delta_B$ and
to define
\bea\fl
\label{Deltat}
\Delta_t=-\Delta_A={-1\over k_0^2-\mathbf k^2 - \hat\Pi_A(k_0,|\mathbf k|)}
=\int_{-\infty}^\infty {dk_0'\over2\pi} 
{\hat\rho_t(k_0',|\mathbf k|)\over k_0'-k_0},\\ \fl
\label{Deltaell}
\Delta_\ell=-{k^2\0\mathbf k^2}\Delta_B={-1\over\mathbf k^2+\hat\Pi_B(k_0,|\mathbf k|)}
=\int_{-\infty}^\infty {dk_0'\over2\pi} 
{\hat\rho_\ell(k_0',|\mathbf k|)\over k_0'-k_0}-{1\0\mathbf k^2},
\eea
where $\hat\Pi_{A,B}$ are the HTL quantities given in \eref{PiA}, \eref{PiB}.
The spectral functions are given by
\bea\fl
\label{hatrhotl}
\hat\rho_{t,\ell}(k_0,|\mathbf k|)= 
{\rm Disc}\,\Delta_{t,\ell}(k_0,|\mathbf k|) 
= 2 \lim_{\epsilon\to0}
\Im \Delta_{t,\ell}(k_0+\I \epsilon,|\mathbf k|) \nonumber\\
\lo=2\pi\varepsilon(k_0)\,z_{t,\ell}(|\mathbf k|)\,\delta(k_0^2-\omega_{t,\ell}^2(
|\mathbf k|))+\beta_{t,\ell}(k_0,|\mathbf k|) \,\theta(-k^2)
\eea
with $\omega_{t,\ell}^2(|\mathbf k|)$ as shown in \fref{figg} 
(for $\mathbf k^2>0$). For small and large values of $\mathbf k^2$
they are approximated by \cite{Weldon:1982aq}
\bea\fl
\omega_t^2 \simeq \omega_{\rm pl.}^2+{6\05}\mathbf k^2,\quad
&\omega_\ell^2 \simeq \omega_{\rm pl.}^2+{3\05}\mathbf k^2,\qquad
&\mathbf k^2\ll\omega_{\rm pl.}^2
\\ \fl
\omega_t^2 \simeq \mathbf k^2+m_\infty^2,\quad
&\omega_\ell^2 \simeq \mathbf k^2+4 \mathbf k^2
\exp\left(-{\mathbf k^2\0m_\infty^2}-2\right),\quad
&\mathbf k^2 \gg m_\infty^2
\eea
where $\omega_{\rm pl.}^2=\hat m_D^2/3$ is the plasma frequency
common to both modes, and $m_\infty^2=\hat m_D^2/2$ is the asymptotic
mass of transverse quasiparticles. The effective thermal mass
of mode $\ell$ (or B) vanishes exponentially for large $\mathbf k^2$. 

The residues $z_{t,\ell}$ are defined by
\be
z_{t,\ell}^{-1}=\left[ {\partial\0\partial k_0^2} (-\Delta_{t,\ell})^{-1}
\right]\Big|_{\Delta_{t,\ell}^{-1}=0}
\ee
and explicitly read
\be\fl
z_t={2 k_0^2 k^2 \0\hat m_D^2 k_0^2-(k^2)^2}\Big|_{k_0=\omega_t(|\mathbf k|)},
\quad
z_\ell={2 k_0^2 k^2 \0 \mathbf k^2 (\hat m_D^2 - k^2)}
\Big|_{k_0=\omega_\ell(|\mathbf k|)}
\ee
with the following asymptotic limits \cite{Pisarski:1989cs}
\bea\fl
z_t \simeq 1-{4\mathbf k^2\05\omega_{\rm pl.}^2},\quad
&z_\ell \simeq {\omega_{\rm pl.}^2\0\mathbf k^2}\left(
1-{3\010}{\mathbf k^2\0\omega_{\rm pl.}^2}\right),\qquad
&\mathbf k^2\ll\omega_{\rm pl.}^2
\\ \fl
z_t \simeq 1-{m_\infty^2\02\mathbf k^2}\left(\ln{4\mathbf k^2\0m_\infty^2}
-2\right),\quad
&z_\ell \simeq {4\mathbf k^2\0m_\infty^2}
\exp\left(-{\mathbf k^2\0m_\infty^2}-2\right),\quad
&\mathbf k^2 \gg m_\infty^2
.\eea
The singular behaviour of $z_\ell$ for $\mathbf k^2\to0$
is in fact only due to the factor $k^2/\mathbf k^2$ in \eref{Deltaell};
the residue in $\Delta_B$ approaches $1$ in this limit.
For $\mathbf k^2 \gg m_\infty^2$, the residue in $\Delta_\ell$ 
vanishes exponentially, as mentioned in \Sref{sec:gfqplo}.

The Landau-damping functions $\beta_{t,\ell}$ are given by
\be\label{betatl}\fl
\beta_t(k_0,|\mathbf k|)=\pi \hat m_D^2 {k_0(-k^2)\02|\mathbf k|^3} 
|\Delta_t(k_0,|\mathbf k|)|^2,
\quad
\beta_\ell(k_0,|\mathbf k|)=\pi \hat m_D^2 {k_0\0|\mathbf k|} 
|\Delta_\ell(k_0,|\mathbf k|)|^2.
\ee
These are odd functions in $k_0$ which vanish at $k_0=0$
and at $k_0^2=\mathbf k^2$. For large $\mathbf k^2$ and
fixed ratio $k_0/|\mathbf k|$, $\beta_{t,\ell}$ decay
like $1/\mathbf k^4$.

The spectral functions $\rho_{t,\ell}$ satisfy certain sum
rules which can be obtained by a Taylor expansion of
\eref{Deltat} and \eref{Deltaell} in $k_0$
\cite{Braaten:1991dd,Schulz:1992zv,LeB:TFT}. A special,
particularly important case is obtained by putting $k_0=0$
in \eref{Deltat} and \eref{Deltaell} which yields
\be\fl
\int_{-\infty}^\infty {dk_0\over2\pi} 
{\hat\rho_t(k_0,|\mathbf k|)\over k_0}={1\0\mathbf k^2},\quad
\int_{-\infty}^\infty {dk_0\over2\pi} 
{\hat\rho_\ell(k_0,|\mathbf k|)\over k_0}={1\0\mathbf k^2}-{1\0\mathbf k^2+\hat m_D^2}.
\ee

More complicated sum rules have been found in applications
of HTL resummations. In \cite{Thoma:2000ne} it has been
shown that one can reexpress the integrals involving only the
continuous parts of the spectral functions appearing
in the energy loss formulae of heavy particles \cite{Braaten:1991we}
in terms of
generalized (Lorentz-transformed) Kramers-Kronig relations;
in \cite{Aurenche:2002pd} a sum rule involving also only the Landau
damping domain $|x|<1$ with $x\equiv k_0/|\mathbf k|$ has been
derived which appears 
in calculations of
the photon or dilepton production rates in a quark-gluon plasma
\cite{
Peitzmann:2001mz
}
\be\label{AGZsumrule}\fl
{1\0\pi}\int_0^1 {dx\0x} {2\,\Im \hat\Pi_i (x) \0
[t+\Re\hat\Pi_i (x)]^2+[\Im\hat\Pi_i (x)]^2}
={1\0t+\Re\hat\Pi_i (\infty)}-{1\0t+\Re\hat\Pi_i (0)}
\ee
for $t>0$ and $i=A,B$.

A peculiar sum rule has been encountered in \cite{Blaizot:2000fc},
\be\label{SHTLsumrule}
\int {d^4 k\0 k_0} \left\{ 2\Im \hat\Pi_A \Re \Delta_t - \Im \hat\Pi_B
\Re \hat \Delta_\ell \right\} = 0
\ee
which has so far only been shown to hold by numerical integrations.
Like \eref{AGZsumrule} this
sum rule receives contributions only 
from the Landau
damping domain $k^2<0$, but it
involves the two branches at the same time and it holds
only under both $k_0$ and $|\mathbf k|$ integrations.

\subsection{Fermion propagator}

The spectral representation of the two branches of
the HTL fermion propagator is given by
\be
\Delta_\pm={-1\over k_0 \mp (|\mathbf k| + \hat\Sigma_\pm(k_0,|\mathbf k|)}
=\int_{-\infty}^\infty {dk_0'\over2\pi} 
{\hat\rho_\pm(k_0',|\mathbf k|)\over k_0'-k_0}
\ee
where $\rho_\pm$ are defined in analogy to \eref{hatrhotl},
\be\fl
\label{hatrhopm}
\hat\rho_\pm(k_0,|\mathbf k|)= 
2\pi\varepsilon(k_0)\,z_\pm(|\mathbf k|)\,\delta(k_0^2-\omega_\pm^2(
|\mathbf k|))+\beta_\pm(k_0,|\mathbf k|) \,\theta(-k^2)
\ee
with $\omega_\pm^2(|\mathbf k|)$ as shown in \fref{figf}. 
For small and large values of $\mathbf k^2$
they are approximated by \cite{Pisarski:1989wb}
\bea\fl
\omega_+ \simeq \hat M + {|\mathbf k|\03},\quad
&\omega_- \simeq \hat M - {|\mathbf k|\03},\qquad
&\mathbf k^2\ll\hat M^2
\\ \fl
\omega_+^2 \simeq \mathbf k^2+M_\infty^2,\quad
&\omega_- \simeq |\mathbf k| \left(1+2
\exp\left(-{\mathbf 4k^2\0M_\infty^2}-1\right)\right),\quad
&\mathbf k^2 \gg M_\infty^2
\eea
where $M_\infty^2=2 \hat M^2$ is the asymptotic
mass of energetic fermions of the (+)-branch. The effective thermal mass
of the additional ($-$)-branch vanishes exponentially for large $\mathbf k^2$.

The residues $z_\pm$ are given by the simple expression
\be
z_\pm = {\omega_\pm^2-\mathbf k^2\0 2 \hat M^2}
\ee
with the following asymptotic limits \cite{Pisarski:1989cs}
\bea\fl
z_+ \simeq {1\02}+{|\mathbf k|\03\hat M},\quad
&z_- \simeq {1\02}-{|\mathbf k|\03\hat M},\qquad
&\mathbf k^2\ll\hat M^2
\\ \fl
z_+ \simeq 1-{M_\infty^2\04\mathbf k^2}\left(\ln{4\mathbf k^2\0M_\infty^2}
-1\right),\quad
&z_- \simeq {4\mathbf k^2\0M_\infty^2}
\exp\left(-{\mathbf 4k^2\0M_\infty^2}-1\right),\quad
&\mathbf k^2 \gg M_\infty^2
.\eea
For $\mathbf k^2 \gg M_\infty^2$, the residue in $\Delta_-$ 
vanishes exponentially.

The Landau-damping functions $\beta_\pm$ are given by
\be
\beta_\pm=\pi \hat M^2 {|\mathbf k| \mp k_0 \0 \mathbf k^2}
\left| \Delta_\pm(k_0,|\mathbf k|) \right|^2.
\ee
For large $\mathbf k^2$ and
fixed ratio $k_0/|\mathbf k|$, they decay
like $1/\mathbf k^2$.

In contrast to the gauge boson case, the spectral functions
are not odd functions in $k_0$ but rather obey
\be
\rho_+(-k_0,|\mathbf k|) = \rho_-(k_0,|\mathbf k|).
\ee
Sum rules for these spectral functions have been discussed
in detail in \cite{LeB:TFT}. A more complicated one that
plays a role in the HTL resummed calculation of the
hard photon production rate \cite{Kapusta:1991qp,Baier:1992em}
has been given recently in \cite{Thoma:2000ne}.

\ack

We dedicate this review to the memory of Tanguy Altherr and
express our gratitude to our other friends and collaborators
on topics covered in this report, in particular
Jean-Paul Blaizot, Fritjof Flechsig, 
Andreas Gerhold, 
Edmond Iancu, Andreas Ipp, Randy Kobes, Gabor Kunstatter,
Peter Landshoff, 
Herbert Nachbagauer, 
Paul Romatschke, Hermann Schulz, and Dominik Schwarz.
We also thank Guy Moore, Dirk Rischke, and Mike Strickland for
their comments on a first version of this review.

\section*{References}


\end{document}